\newif\ifdraft \drafttrue   
\newif\iffull \fulltrue     
\newif\ifscr  \scrfalse     
\newif\iflater\latertrue    
\newif\ifever \everfalse    
\newif\ifunfinished \unfinishedfalse   
\newif\ifproof \prooftrue   
\newif\iftechrep\techrepfalse 

\makeatletter \@input{texdirectives} \makeatother

\newif\ifproftables         
  \iffull
    \proftablestrue
  \else
    \proftablesfalse
  \fi

\iffull
\iftechrep
\documentclass[a4paper]{report}
\usepackage[
  margin=3.5cm,
  includefoot,
  footskip=30pt,
]{geometry}
\usepackage{times}
\usepackage{inconsolata}
\else
\documentclass{jfp}
\fi
\else
\documentclass{sigplanconf}
\fi

\usepackage{preamble}

\iffull
\fi

\iffull

\else

\fi

\begin{document}


\iffull\else
  \exclusivelicense
  \conferenceinfo{ICFP~'13}{September 25--27, 2013, Boston, MA, USA}
  \copyrightyear{2013}
  \copyrightdata{978-1-4503-2326-0/13/09}
  \doi{2500365.2500574}
\fi  

\title{\iftechrep\bf\huge\fi 
Testing Noninterference, Quickly
}
\newcommand{\authorcontent}{
  {\large C\u{A}T\u{A}LIN HRI\c{T}CU}\\
    {\em Inria, Paris-Rocquencourt, France}\\[2ex]
  {\large LEONIDAS LAMPROPOULOS}\\
    {\em University of Pennsylvania, Philadelphia, USA}\\[2ex]
  {\large ANTAL SPECTOR-ZABUSKY}\\
    {\em University of Pennsylvania, Philadelphia, USA}\\[2ex]
  {\large ARTHUR AZEVEDO DE AMORIM}\\
    {\em University of Pennsylvania, Philadelphia, USA}\\[2ex]
  {\large MAXIME D\'EN\`ES}\\
    {\em Inria, Paris-Rocquencourt, France}\\[2ex]
  {\large JOHN HUGHES}\\
    {\em Chalmers University, Gothenburg, Sweden}\\[2ex]
  {\large BENJAMIN C. PIERCE}\\
    {\em University of Pennsylvania, Philadelphia, USA}\\[2ex]
  {\large DIMITRIOS VYTINIOTIS}\\
    {\em Microsoft Research, Cambridge, UK}\\[2ex]
}
\iffull
\iftechrep
\author{\authorcontent}
\else
\author[C. Hritcu et al.]{\authorcontent}
\fi

\else
\authorinfo
    {
  C\u{a}t\u{a}lin Hri\c{t}cu\textsuperscript{1,4} \quad
  John Hughes\textsuperscript{2} \quad
  Benjamin C. Pierce\textsuperscript{1} \quad
  Antal Spector-Zabusky\textsuperscript{1} \quad
  Dimitrios Vytiniotis\textsuperscript{3} \\[1ex]
  Arthur Azevedo de Amorim\textsuperscript{1} \quad
  Leonidas Lampropoulos\textsuperscript{1} \\[1ex]
} {
  \textsuperscript{1}University of Pennsylvania\qquad
  \textsuperscript{2}Chalmers University\qquad
  \textsuperscript{3}Microsoft Research\qquad
  \textsuperscript{4}INRIA Paris\qquad
\vspace*{-1cm}
} {}
\fi
\maketitle


{
\let\section\OLDsection
\begin{abstract}
Information-flow control mechanisms are difficult both to design and
to prove correct.  To reduce the time wasted on 
doomed proof attempts due to
broken definitions, we advocate modern random testing techniques for finding
counterexamples during the design process.  We show how to use QuickCheck, a
property-based random-testing tool, to guide the design of
increasingly complex information-flow abstract machines, leading 
  up to a sophisticated register machine with
  a novel and highly permissive flow-sensitive dynamic
  enforcement mechanism that
\ifproof\chrev{is sound in the presence of}\fi{}
first-class public labels.
We find that both sophisticated
strategies for generating well-distributed random programs and readily
falsifiable formulations of noninterference properties are critically
important for efficient testing.
We propose several approaches and evaluate their effectiveness
on a collection of injected bugs of varying subtlety.  We also present an
effective technique for shrinking large counterexamples to minimal, easily
comprehensible ones. Taken together, our best methods enable us to quickly
and automatically generate simple counterexamples for
more than 45 bugs.
Moreover, we show how testing guides the discovery of the
  sophisticated invariants needed for
\ifproof\chrev{the}\fi{} noninterference proof of our
  most complex machine.
\end{abstract}
}

\iffull\else


\fi

\iftechrep
\renewcommand\thesection{\arabic{section}}
\tableofcontents
\newpage
\fi

\section{Introduction}

Secure information-flow control (IFC) is nearly impossible to achieve by
careful design alone.  The mechanisms involved
are intricate and easy to get wrong: static type systems
must impose numerous constraints that interact with other typing rules in
subtle ways~\cite{sabelfeld03:lang_based_security},
while dynamic mechanisms must appropriately propagate taints
and raise security exceptions when necessary~\cite{AustinF09, AustinF:2010,
  SabelfeldR09, Fenton74}.
In a dynamic setting, allowing IFC labels to vary dynamically
  (\IE performing flow-sensitive analysis) can lead to subtle information leaks
  through the labels themselves~\cite{RussoS10,
    ZhengM07}; these leaks are particularly hard to avoid if labels
  are observable inside the language~\cite{StefanRMM11, Exceptional}.
This intricacy makes it hard
to be confident in the correctness of such mechanisms without detailed
proofs; however, carrying out these proofs while designing the
mechanisms can be an exercise in frustration, with a great deal of time spent
attempting to verify broken definitions!  The question we
address in this paper is: Can we use modern \emph{testing} techniques to
discover bugs in IFC enforcement mechanisms quickly and effectively? If so,
then we can use testing 
to catch most errors during the design phase,
postponing proof attempts until we are reasonably confident that the design
is correct.

To answer this question, we undertake two case studies.
The first is aimed at
extending a simple abstract stack-and-pointer
machine to track dynamic information flow and enforce
termination-insensitive noninterference~\cite{sabelfeld03:lang_based_security}.
Although {this machine} is simple,
the exercise is nontrivial.
While even simpler notions of
dynamic \emph{taint tracking} are well studied for both high- and low-level
languages, it has only recently been shown~\cite{AustinF09, SabelfeldR09}
that dynamic checks are capable of soundly enforcing strong security
properties.  Moreover, until
recently~\cite{BichhawatRGH14,TestingNI,PicoCoq2013},
sound dynamic IFC has been studied only in the context of
high-level languages~\cite{AustinF09, StefanRMM11,
    Exceptional, SabelfeldR09, HedinS12, BichhawatRGH14b}; the
unstructured control flow of a low-level machine poses additional challenges.

We show how QuickCheck~\cite{ClaessenH00}, a popular property-based
testing tool, can be used to formulate and test noninterference
properties of our abstract machine,
quickly find a variety of missing-taint and
missing-exception bugs, and incrementally guide the design of a
correct version of the machine.  One significant challenge is
that both the strategy for generating random programs and the precise
formulation of the noninterference property have a dramatic impact on
the time required to discover bugs; we benchmark several variations
of each to identify the most effective choices.
In particular, we observe that checking the unwinding
conditions~\cite{GoguenM84} of our noninterference property can be much
more effective than directly testing the original property.
\ifever\bcp{We should probably say a bit more about our most important
  findings in the areas of program generation strategies and properties.}\fi

The second case study demonstrates the scalability of our techniques
by targeting the design of a novel and highly permissive flow-sensitive
dynamic IFC mechanism.
%
This experiment targets a more sophisticated 
register machine that is significantly more realistic than the first
and that includes
advanced features such as first-class public labels and
dynamically allocated memory with mutable labels.
We still quickly find all introduced flaws.  Moreover, we can use
testing to discover the sophisticated invariants required by a
\ifproof\chrev{complex}\fi{}
noninterference proof.
%

Our results should be of interest both to researchers in
language-based security, who can now add random testing to their tools
for debugging subtle IFC enforcement mechanisms 
\ifproof\chrev{and their noninterference proofs}\fi{};
and to the random-testing community,
where our techniques for generating and shrinking random programs may
be useful for checking other properties of abstract machines.
Our primary contributions are: (1) a
demonstration of the effectiveness of random testing for
discovering counterexamples to noninterference in low-level
information-flow machines;
(2) a range of program generation strategies
for finding such counterexamples; (3) an empirical comparison of how
effective combinations of these strategies and formulations of
noninterference are in finding counterexamples; (4) an effective
methodology for shrinking large counterexamples to smaller, more
readable ones;
(5) a demonstration that these techniques can speed the design
of a state-of-the-art flow-sensitive dynamic IFC mechanism that is
highly permissive \ifproof\chrev{and sound}\fi{} even though labels
are observable;
{} (6) 
\chrev{a demonstration that our techniques can aid in discovering}
the complex invariants
\ifproof\chrev{involved in the noninterference proofs for this}\fi{}
novel IFC mechanism\ifproof\chrev{; (7) a mechanized noninterference proof
for this mechanism}\fi.

Sections \ref{sec:basic} to \ref{sec:shrinking}
gradually introduce our testing methodology using
the simple stack machine as the running example.
Section~\ref{sec:regs} shows that our methodology scales up to the
more realistic register machine with advanced IFC features.
Section~\ref{sec:related} presents related work and
section~\ref{sec:conclusions} concludes and discusses future work.
Accompanying Haskell code associated to this paper and the Coq proofs mentioned
in Section~\ref{sec:regs} and Appendix~\ref{app:strengthening}
are available online at \url{https://github.com/QuickChick}.

\sloppy

A preliminary version of this work appeared in the
proceedings of the ICFP 2013 conference~\cite{TestingNI}.
Section~\ref{sec:regs} of this paper and the contributions therein
(points \chrev{5 to 7} above) are new.
The other sections have also been improved and extended with
additional counterexamples.

\fussy


\section{Basic IFC}
\label{sec:basic}


We begin by introducing the core of our abstract stack machine.
In \autoref{sec:cally} we will extend this
simple core with control flow (jumps and procedure calls), but the
presence of pointers already raises opportunities for some subtle
mistakes in information-flow control.

\ifever\asz{Not sure I like the superscript th.}\fi
\iffull Some notation: w\else W\fi{}e write $[\,]$ for the
empty list and $x:\ii{xs}$ for the list whose first element is $x$ and whose
tail is $\ii{xs}$; we also write $[x_0,x_1,\ldots,x_n]$ for the list $x_0 :
x_1 : \cdots : x_n : [\,]$.  If $\ii{xs}$ is a list and $0 \leq j < |\ii{xs}|$,
then $\ii{xs}(j)$ selects the $j^{\mbox{\tiny th}}$ element of $\ii{xs}$ and
$\update{\ii{xs}}{j}{x}$ produces the list that is like $\ii{xs}$ except that the
$j^{\mbox{\tiny th}}$ element is replaced by $x$.

\subsection[Bare stack machine]{Bare stack machine}
\label{sec:stack}


The most basic variant of our stack machine 
(without information-flow labels) has
seven instructions:
\[
\mathit{Instr} \mathrel{::=} 
  \ii{Push}\; n
  \;|\;
  \ii{Pop}
  \;|\;
  \ii{Load}
  \;|\;
  \ii{Store}
  \;|\;
  \ii{Add}
  \;|\;
  \ii{Noop}
  \;|\;
  \ii{Halt}
\]
The $n$ argument to \ii{Push} is an integer (an immediate constant).

A machine state $S$ is a 4-tuple consisting of a program counter $\pc$ (an
integer), a stack $s$ (a list of integers), a memory $m$ (another list of
integers), and an instruction memory $i$ (a list of instructions),
written $\MACHI{\pc}{s}{m}{i}$\,.
Since $i$ is cannot change during execution we will often write just
$\MACH{\pc}{s}{m}$ for the varying parts of the machine state.

The single-step reduction relation on machine states, written $S \step S'$,
is \iffull defined by the following rules: 

\typicallabel{Load}
\infrule[Bare-Noop]
  {i(\pc) = \ii{Noop}}
  {\MACH{\pc}{s}{m} \step \MACH{\pc\mathord{+}1}{s}{m}}
\infrule[Bare-Push]
  {i(\pc) = \ii{Push}\; n}
  {\MACH{\pc}{s}{m} \step \MACH{\pc\mathord{+}1}{n:s}{m}}
\infrule[Bare-Pop]
  {i(\pc) = \ii{Pop}}
  {\MACH{\pc}{n:s}{m} \step \MACH{\pc\mathord{+}1}{s}{m}}
\infrule[Bare-Load]
  {i(\pc) = \ii{Load} \andalso
   m(p) = n}
  {\MACH{\pc}{p:s}{m} \step \MACH{\pc\mathord{+}1}{n:s}{m}}
\infrule[Bare-Store]
  {i(\pc) = \ii{Store} \andalso
   m' = \update{m}{p}{n}}
  {\MACH{\pc}{p:n:s}{m} \step \MACH{\pc\mathord{+}1}{s}{m'}}
\infrule[Bare-Add]
  {i(\pc) = \ii{Add}}
  {\MACH{\pc}{n_1:n_2:s}{m} \step \MACH{\pc\mathord{+}1}{(n_1\mathord{+}n_2):s}{m}}
\else
straightforward to define; we elide it here, for brevity.
(It is included in a longer version of the paper, available from
\url{http://www.crash-safe.org/node/24}.) 
\fi
This relation is a partial function: it is
deterministic, but some 
machine states don't step to anything.
Such a stuck machine state is said to be \emph{halted} if $i(\pc) =
\ii{Halt}$ and \emph{failed} in all other cases (\EG if the machine is
trying to execute an \ii{Add} with an empty stack, or if the \pc{}
points outside the bounds of the instruction memory).
We write $\manystep$ for the reflexive, transitive closure of $\step$.  
When $S \manystep S'$ and $S'$ is a stuck state, we write $S \stuckswith S'$.

\subsection[Stack machine with labeled data]{Stack machine with labeled data}
\label{sec:basic-ifc}

In a (fine-grained) dynamic IFC system~\cite{AustinF09, StefanRMM11,
  Exceptional, SabelfeldR09, HedinS12, BichhawatRGH14b, PicoCoq2013} security
levels (called
labels) are attached to runtime values and propagated during
execution, enforcing the constraint that information derived from
secret data does not leak to untrusted processes or to the public
network.
Each value is protected by an individual IFC label representing a
security level (e.g., secret or public).
We now add labeled data to our simple stack machine.
Instead of bare integers,
the basic data items in the instruction and data memories and the stack are
now \emph{labeled integers}
of the form $\V{n}{\lab}$, where $n$ is an integer and $\lab$ is
\iffull
a \emph{label}:
\[
\lab \mathrel{::=} 
  \low
  \;|\;
  \high
\]
We read $\low$ as ``low'' (public) and $\high$ as ``high'' (secret).
\else
either $\low$ (public) or $\high$ (secret).
\fi
We order labels by $\low \flowsto \high$
\ifever\finish{and maybe: $\low
  \not\flowsto \high$}\fi 
and write $\lab_1 \join \lab_2$ for the 
\emph{join} (least upper bound) of $\lab_1$ and $\lab_2$.
%


The instructions are exactly the same except that the immediate argument to
\ii{Push} becomes a labeled integer\iffull:
\[
\mathit{Instr} \mathrel{::=} 
  \ii{Push}\; \V{n}{\lab}
  \;|\;
  \ii{Pop}
  \;|\;
  \ii{Load}
  \;|\;
  \ii{Store}
  \;|\;
  \ii{Add}
  \;|\;
  \ii{Noop}
  \;|\;
  \ii{Halt}
\]\else. \fi
Machine states have the same shape as the basic machine, with the stack and
memory now being lists of labeled integers.
The set of \emph{initial} states of this machine, $\Init$, contains states of the
form $\MACHI{0}{[\,]}{m_0}{i}$, where $m_0$ can be of any length and contains
only $\V{0}{\low}$. We use $\Halted$ to denote the set of halted
states of the machine, \IE $i(\pc) = \ii{Halt}$.

\subsection{Noninterference (EENI)}
\label{sec:eeni}

We define what it means for this basic IFC machine to be ``secure''
using a standard notion of
termination-insensitive
noninterference~\cite{sabelfeld03:lang_based_security, Exceptional,
  AustinF09, PicoCoq2013}; we call it \emph{end-to-end
  noninterference} (or \emph{EENI}) to distinguish it from the stronger
notions we will introduce in \autoref{sec:stronger}.
The main idea of EENI is to directly encode the intuition that secret inputs
should not influence public outputs.
By secret inputs we mean integers labeled $\high$ in the initial state;
because of the form of our initial states, such labeled integers
can appear only in instruction memories.
By secret outputs we mean integers labeled $\high$ in a halted state.
More precisely, EENI states that for any two executions starting from
initial states that are \emph{indistinguishable to a low observer} (or
just \emph{indistinguishable}) and ending in halted states $S_1$ and
$S_2$, the final states $S_1$ and $S_2$ are also indistinguishable.
Intuitively, two states are indistinguishable if they
differ only in integers labeled $\high$.
To make this formal, we define an equivalence relation on
states compositionally from equivalence relations over their
components.

\begin{defn}\label{def:indist} \
\iffull \begin{itemize}
\item \fi
Two labeled integers $\V{n_1}{\lab_1}$ and $\V{n_2}{\lab_2}$ are said to be
  \emph{indistinguishable},
  written $\V{n_1}{\lab_1} \indist \V{n_2}{\lab_2}$, if
  either $\lab_1 = \lab_2 = \high$ or else $n_1 = n_2$ and $\lab_1 = \lab_2 = \low$.
\iffull
\item\fi
 Two instructions $i_1$ and $i_2$ are indistinguishable if they are
the same,
  or if $i_1 = \ii{Push}\; \V{n_1}{\lab_1}$, and $i_2 =
  \ii{Push}\; \V{n_2}{\lab_2}$, and $\V{n_1}{\lab_1} \indist \V{n_2}{\lab_2}$.
\iffull\item\fi
 Two lists (memories, stacks, or instruction memories)
  $\ii{xs}$ and $\ii{ys}$ are indistinguishable if they have the same
  length and $\ii{xs}(i) \indist \ii{ys}(i)$ for all $i$ such that
  $0 \leq i < |\ii{xs}|$.
\iffull \end{itemize} \fi
\end{defn}

For machine states we have a choice as to how much of the state we
want to consider observable; we choose (somewhat arbitrarily) that the
observer can only see the data and instruction memories, but not the stack
or the $\pc$. (Other choices would give the observer either somewhat more
power---e.g., we could make the stack
observable---or somewhat
less---e.g., we could restrict the observer to some designated region of
``I/O memory,'' or extend the architecture with I/O instructions and only
observe the traces of inputs and outputs~\cite{PicoCoq2013}.%
%
)

\ifever
\bcp{This last point is actually really awkward---the more I try to write
  around it, the more I think we should really just bite the bullet and
  include an Output instruction...}  \fi

\ifever
\feedback{Delphine: could [our choice of what's observable] be justified? the
  "somewhat arbitrarily" does sound odd. looks like this makes sense
  for EENI. but for other defs?}
\ch{It also works for LLNI + StartInitial; there we strengthen the
  equivalence only to get better testing behavior. It does not
  work for StartQuasiInitial though.}
\ch{The choice is arbitrary because there are many variants that
  work and which are in the end equivalent for our particular machine.
  It's like for observational equivalence; where one makes the arbitrary
  choice of only observing termination, when many other equivalent choices 
  exist. This is, however, just some vague intuition we don't want to
  make formal here.}
\fi

\begin{defn} \
  Machine states $S_1 = \MACHI{\pc_1}{s_1}{m_1}{i_1}$ and
  $S_2 = \MACHI{\pc_2}{s_2}{m_2}{i_2}$ are \emph{indistinguishable
  with respect to memories}, written $S_1 \indistmem S_2$,
  if $m_1 \indist m_2$  and $i_1 \indist i_2$.
\end{defn}

\begin{defn}
A machine semantics is \emph{end-to-end noninterfering} with respect to some
sets of states $\Start$ and $\End$ and an indistinguishability relation
$\indist$, written EENI$_{\Start,\End,{\indist}}$, if for any $S_1, S_2 \in
\Start$ such that $S_1 \indist S_2$ and such that
$S_1 \stuckswith S'_1$, $S_2 \stuckswith S'_2$,
and $S'_1, S'_2 \in \End$,
we have $S'_1 \indist S'_2$.
\end{defn}

We take EENI$_{\Init,\Halted,{\indistmem}}$ as our baseline security
property; \IE we only consider executions starting in initial states
and ending in halted states, and we use indistinguishability with
respect to memories.
The EENI definition above is, however, more general, and we will consider other
instantiations of it later.
\iffull\ifever
\ch{We might want to mention somewhere that when instantiating EENI
  for our machine we need to take $\End$ be a subset of $\Halted$,
  otherwise the property does not hold. Show counterexample.}
\ch{here would be too early anyway -- maybe in the appendix}
\fi\fi

\subsection{Information-flow rule design with QuickCheck}
\label{sec:basic-ifc-rules}

Our next task is to enrich the rules for the step function to take
information-flow labels into account.  For most of the rules, there are
multiple plausible ways to do this, and some opportunities for subtle
mistakes even with these few instructions.
To illustrate the design methodology we hope to support, we first
propose a naive set of rules and then use counterexamples generated
using QuickCheck and our custom generation and shrinking techniques
(described in detail in the following sections) to identify and help
repair mistakes until no more can be found.
%
%
\infrule[Noop]
  {i(\pc) = \ii{Noop}}
  {\MACH{\pc}{s}{m} \step \MACH{\pc\mathord{+}1}{s}{m}}
\infrule[Push]
  {i(\pc) = \ii{Push}\; \V{n}{\lab}}
  {\MACH{\pc}{s}{m} \step \MACH{\pc\mathord{+}1}{\V{n}{\lab}:s}{m}}
\infrule[Pop]
  {i(\pc) = \ii{Pop}}
  {\MACH{\pc}{\V{n}{\lab}:s}{m} \step \MACH{\pc\mathord{+}1}{s}{m}}
\infrule[\IfcBugLoadNoTaint]
  {i(\pc) = \ii{Load} \andalso
   m(p) = \V{n}{\lab_n}}
  {\MACH{\pc}{\V{p}{\lab_p}:s}{m} \step 
    \MACH{\pc\mathord{+}1}{\V{n}{\lab_n}:s}{m}} 
\infrule[\IfcBugStoreNoPointerTaintANDAllowWriteDownThroughHighPtr]
  {i(\pc) = \ii{Store} \andalso
   m' = \update{m}{p}{\V{n}{\lab_n}}}
  {\MACH{\pc}{\V{p}{\lab_p}:\V{n}{\lab_n}:s}{m} \step 
    \MACH{\pc\mathord{+}1}{s}{m'}} 
\infrule[\IfcBugArithNoTaint]
  {i(\pc) = \ii{Add}}
  {\MACH{\pc}{\V{n_1}{\lab_1}:\V{n_2}{\lab_2}:s}{m} \step \iffull\else\\\fi
    \MACH{\pc\mathord{+}1}{\V{(n_1\mathord{+}n_2)}{\low}:s}{m}}
\ifever
\ch{I find it very strange to explain rules as diffs wrt
  rules we've never showed}
\fi
The \rn{Noop} rule is the same as in the unlabeled machine.
In the \rn{Push} and \rn{Pop}
rules, we simply change from bare to labeled integers;
luckily, this obvious adaptation happens to be correct.  But now our
luck runs out: the simple changes that we've made in the other rules
will all turn out to be wrong.  (We include a star in the names of
incorrect rules to indicate this.  The rule
\IfcBugStoreNoPointerTaintANDAllowWriteDownThroughHighPtr{}
actually contains \emph{two} bugs,
which we refer to as \rn{a} and \rn{b}; we will discuss them separately
later.)
Fortunately, QuickCheck can rapidly pinpoint the
problems, as we will see.

\autoref{fig:counter-basic2} shows the first counterexample that QuickCheck
gives us when we present it with the step function defined by the six rules
above and ask it to try to invalidate the EENI property.  (The \LaTeX\ source
for all the figures was generated automatically by our QuickCheck testing
infrastructure.)  The first line of the figure is the counterexample itself: a
pair of four-instruction programs, differing only in the constant argument of
the second \ii{Push}.  The first program pushes \V{0}{\high}, while the second
pushes \V{1}{\high}, and these two labeled integers are indistinguishable.
We display the two programs, and the other parts of the
two machine states, in a ``merged'' format.  Pieces of data that are the same between
the two machines are written just once; at any place where the two machines
differ, the value of the first machine is written above the value of the second
machine, separated by a horizontal line.  The rest of the figure shows what
happens when we run this program.  On the first step, the \pc{} starts out at
$0$; the memory, which has two locations, starts out as
$[\V{0}{\low},\V{0}{\low}]$; the stack starts out empty; and the next
instruction to be executed ($i(\pc)$) is $\ii{Push}\;\V{1}{\low}$.
%
%
On the next
step, this labeled integer has been pushed on the stack and the next instruction is either
$\ii{Push}\; \V{0}{\high}$ or $\ii{Push}\;\V{1}{\high}$; one or the other of
these labeled integers is pushed on the stack.  On the next, we \ii{Store} the second
stack element ($\V{1}{\low}$) into the location pointed to by the first (either
$\V{0}{\high}$ or $\V{1}{\high}$), so that now the memory contains $\V{1}{\low}$
in either location $0$ or location $1$ (the other location remains unchanged,
and contains $\V{0}{\low}$).  At this point, both machines halt.
This pair of execution
sequences shows that EENI fails: in the initial state, the two programs are
indistinguishable to a low observer (their only difference is labeled $\high$),
but in the final states the memories contain different integers at the same
location, both of which are labeled~$\low$.


Thinking about this counterexample, it soon becomes apparent what went wrong
with the \ii{Store} instruction: since pointers labeled $\high$ are allowed to
vary between the two runs, it is not safe to store a low integer through a high
pointer.  One simple but draconian fix is simply to stop the machine if it tries
to perform such a store (\IE we could add the side-condition $\lab_p = \low$ to the
rule).  A more permissive option is to allow the store to take place, but
require it to taint the stored value with the label on the pointer:

\infrule[\IfcBugAllowWriteDownThroughHighPtr]
  {i(\pc) = \ii{Store} \andalso
  m' = \update{m}{p}{\V{n}{(\lab_n\mathord\join \lab_p)}}}
  {\MACH{\pc}{\V{p}{\lab_p}:\V{n}{\lab_n}:s}{m} \step \iffull\else\\\fi
    \MACH{\pc\mathord{+}1}{s}{m'}}

Unfortunately, QuickCheck's next counterexample
(\autoref{fig:counter-basic3}) shows that this rule is still not quite
good enough.  This counterexample is quite similar to the first one, but it
illustrates a more subtle point: our definition of noninterference allows
the observer to distinguish between final memory states that differ only in their
\emph{labels}.%
\footnote{See the first clause of
  Definition~\ref{def:indist}.  One might imagine that this could be fixed
  easily by changing the definition so that whether a label is high or low
  is not
  observable---\IE $\V{n}{\low}\indist \V{n}{\high}$ for any $n$.
  Sadly, this is known not to
  work~\cite{RussoS10, Fenton74}. (QuickCheck can also find a counterexample,
  which we present in \autoref{sec:counter-observable}. The
  counterexample relies on control flow, which is only
  introduced in \autoref{sec:cally}.)
\ifever
\feedback{Deian: I would suggest inlining this and maybe adding some
  intuition vs. just citing [..].}
\feedback{Delphine: agree strongly with Deian on this footnote}
\ch{Yes, but we don't have space. Maybe if we drop the boring
  discussion about the Add bug? \IE start without the Add bug}
\fi
}
\ifever
\feedback{Delphine: it might be counter-intuitive to call that program
  interferent.  the candidate indist. relation for value given in the
  footnote is larger than "only differing in the label" because values
  can also differ}
\ch{Indeed, I think we're being a bit handweavy here in order to save space}
\fi
  Since the
\rn{\IfcBugAllowWriteDownThroughHighPtr} rule taints the label of the stored
integer with the label of the pointer, the fact that the \ii{Store} changes
different locations is visible in the fact that a label changes from \low{}
to \high{} on a different memory location in each run.  To avoid this issue,
we adopt the ``no sensitive upgrades'' rule~\cite{zdancewic02:thesis,AustinF09}, which demands that the label on the current contents of a
memory location being stored into are above the label of the
pointer used for the store
---\IE it is illegal to overwrite a low value via a high pointer
(and trying to do so results in a fatal failure).
Adding this side condition brings us to a correct version of the \rn{Store}
rule.
\infrule[Store]
  {i(\pc) = \ii{Store} \andalso
   m(p) = \V{n'}{\lab_n'} \andalso
   \lab_p \flowsto \lab_n' \andalso
   m' = \update{m}{p}{\V{n}{(\lab_n\mathord\join\lab_p)}}}
  {\MACH{\pc}{\V{p}{\lab_p}:\V{n}{\lab_n}:s}{m} \step \iffull\else\\\fi
    \MACH{\pc\mathord{+}1}{s}{m'}} 


\counterexample{basic2}{Counterexample to \IfcBugStoreNoPointerTaintANDAllowWriteDownThroughHighPtr
\ifever
  \feedback{Deian: Maybe make the 0/1 different colors? Might make
    some of the later figures easier to read.}
\fi
}
\counterexample{basic3}{Counterexample
to \IfcBugAllowWriteDownThroughHighPtr
}
\counterexample*{basic4}{Counterexample to \IfcBugArithNoTaint}

The next counterexample found by QuickCheck
(\autoref{fig:counter-basic4}) points out a straightforward problem in the
\IfcBugArithNoTaint{} rule: adding \V{0}{\low} to \V{0}{\high} yields
\V{0}{\low}.
\iffull\else (We elide the detailed execution trace for this
example and most of the ones that we will see later, for brevity.  They can
be found in 
the long version.)\fi{}
The problem is that the taints on the arguments to \ii{Add} are not
propagated to its result. 
The \ii{Store} is needed in order to make the difference observable.
The easy (and standard) fix is to use the join of the
argument labels as the label of the result:

\infrule[Add]
  {i(\pc) = \ii{Add}}
  {\MACH{\pc}{\V{n_1}{\lab_1}:\V{n_2}{\lab_2}:s}{m} \step \iffull\else\\\fi
   \MACH{\pc\mathord{+}1}{\V{(n_1\mathord{+}n_2)}{(\lab_1\mathord\join \lab_2)}:s}{m}}

\counterexample*{basic5}{Counterexample to \IfcBugLoadNoTaint}

The final counterexample \iffull found by QuickCheck \fi{}
(\autoref{fig:counter-basic5}) alerts us to the
fact that the \IfcBugLoadNoTaint{} rule contains a similar
mistake to the original
\IfcBugStoreNoPointerTaintANDAllowWriteDownThroughHighPtr{} rule: loading
a low value through a high pointer should taint the loaded value.  The
program in \autoref{fig:counter-basic5} is a little longer than the one
in \autoref{fig:counter-basic2} because it needs to do a little work at
the beginning to set up a memory state containing two different low values.
It then pushes a high address pointing to one or the other of those cells onto
the stack; loads (different, low addresses)
through that pointer; and finally stores $\V{0}{\low}$ to the resulting 
address in memory and halts.  In this case, we can make the same
change to \rn{\IfcBugLoadNoTaint} as we did to
\IfcBugStoreNoPointerTaintANDAllowWriteDownThroughHighPtr{}: we taint the
loaded integer with the join of its label and the address's label.  This time
(unlike the case of \ii{Store}, where the fact that we were changing the
memory gave us additional opportunities for bugs), this change gives us the
correct rule for \ii{Load},
\infrule[Load]
  {i(\pc) = \ii{Load} \andalso
   m(p) = \V{n}{\lab_n}}
  {\MACH{\pc}{\V{p}{\lab_p}:s}{m} \step
    \MACH{\pc\mathord{+}1}{\V{n}{(\lab_n \join \lab_p)}:s}{m}} 
and QuickCheck is unable to find any further counterexamples.

\subsection{More bugs}

\ifever
\ch{Alternatively, we could consider having a separate appendix
  listing all bugs of this and the cally machine}
\fi

The original IFC version
of the step rules illustrate one set of mistakes
that we might plausibly have made, but 
there are more possible ones:
\infrule[\IfcBugPushNoTaint]
  {i(\pc) = \ii{Push}\; \V{n}{\lab}}
  {\MACH{\pc}{s}{m} \step \MACH{\pc\mathord{+}1}{\V{n}{\low}:s}{m}}
\infrule[\IfcBugStoreNoValueTaint]
  {i(\pc) = \ii{Store} \andalso
   m' = \update{m}{p}{\V{y}{\low}}}
  {\MACH{\pc}{\V{p}{\lab_p}:\V{n}{\lab_n}:s}{m} \step \iffull\else\\\fi
    \MACH{\pc\mathord{+}1}{s}{m'}} 
Although it is unlikely that we'd write these rather silly rules by accident, 
it is worth including them in our experiments because they can be
invalidated by short counterexamples and thus provide useful data
points for less effective testing strategies.

We will also gather statistics for a partially fixed but still wrong
rule for \ii{Store}, in which the no-sensitive-upgrades check is
performed but the result is not properly tainted:
\infrule[\IfcBugStoreNoPointerTaint]
  {i(\pc) = \ii{Store} \andalso
   m(p) = \V{n'}{\lab_n'} \andalso
   \lab_p \flowsto \lab_n' \andalso
   m' = \update{m}{p}{\V{n}{\lab_n}}}
  {\MACH{\pc}{\V{p}{\lab_p}:\V{n}{\lab_n}:s}{m} \step 
    \MACH{\pc\mathord{+}1}{s}{m'}} 

\section{QuickCheck}
\label{sec:quick-check}

We test noninterference using QuickCheck~\cite{ClaessenH00}, a
tool that tests properties expressed in Haskell.
Often, QuickCheck is used to test properties that should hold for all
inhabitants of a certain type.  
QuickCheck
repeatedly generates random values of 
the desired type, instantiates the property with them, and checks it directly
by evaluating it to a Boolean.  This process continues until either a
counterexample is found or a specified timeout is reached.
QuickCheck supplies default 
test data generators
for many standard types.  Additionally, the user can supply
custom generators for their own types.  In order to test EENI, for
example, we needed to define custom generators for 
labeled integers, instructions, and machine states (each of which depends on the
previous generator: machine states contain instructions, some of which
contain labeled integers).  The effectiveness of testing (\IE mean time to
discover bugs) depends on the sophistication of these generators, a topic we
explore in detail in~\autoref{sec:generation}.

QuickCheck properties may also be guarded by \emph{preconditions};
EENI is an example of why this is necessary, as it only applies to
pairs of indistinguishable initial machine states that both
successfully execute to halted states.
%
%
Testing a property with a precondition proceeds
similarly: a sequence of random values are generated and tested, up to a
user-specified maximum.
The difference is that if there is a precondition, it is instantiated with the
random value first.  If the precondition does not hold, this random value is
summarily discarded.
If the precondition does hold, then the rest of the property is checked
just as before.  Although preconditions are very useful, too high a proportion
of discards can lead to {very} {ineffective} testing or
a badly skewed distribution of test cases 
(since some kinds of test case may be discarded much more often than others).
\ifever
\feedback{Michal: I agree with the first part of the statement [in the
  previous phrase], but it is hard for me to find a justification for
  the second part.}\ch{what more can we say?}
\fi
To help diagnose such problems, QuickCheck can collect statistics
about the tests it tried.

When a test fails, the failing test case is often
large, containing many irrelevant details. QuickCheck then tries to
\emph{shrink} the test case, by searching for a similar but smaller test case
that also fails. To do this, it greedily explores a number of ``shrinking
candidates'': modifications of the original failing test case that are
``smaller'' in some sense.  The property is tested for each of these, and as
soon as a failure is found, that candidate becomes the starting point for a new
shrinking search (and the other candidates are discarded). Eventually this
process terminates in a failing test case that is \emph{locally minimal}: none
of its shrinking candidates fails. This failing case is then reported to
the user. It is often very much smaller than the original randomly
generated test case, and it is thus easy to use it to diagnose the failure
because it (hopefully) contains no irrelevant details. Just like generation
strategies, shrinking strategies are type dependent; they are defined by
QuickCheck for standard types, and by the user for other types. We discuss the
custom shrinking strategies we use for machine states
in \autoref{sec:shrinking}.

\ifever
\feedback{Andrew: p. 4 Section 3/4.  You never actually say that (much
  less how) you've encoded the machines into Haskell!}
\ch{isn't this obvious?}
\feedback{Delphine: I agree with Andrew that it would help to
  "instantiate" section 3 with your abstract machine}
\bcp{This seems low priority.}
\fi

\section{State Generation Strategies}
\label{sec:generation}

\combinedtable{basic}{
  Basic-EENI-Mem-Initial-Naive-False,
  Basic-EENI-Mem-Initial-Weighted-False,
  Basic-EENI-Mem-Initial-Sequence-False,
  Basic-EENI-Mem-Initial-Sequence-True,
  Basic-EENI-Mem-Initial-ByExec-True }{
  Comparison of generation strategies for the basic machine. The first
  part of the table shows the mean time to find a failing test case
  (MTTF) in milliseconds for each bug. The second part lists
  the arithmetic and
  geometric mean for the MTTF over all bugs. The third part shows the
  number of tests per second and the proportion of test cases that
  were discarded because they did not satisfy some precondition.}

We are ready now to begin exploring ways to generate potential
counterexamples.  At the outset, we need to address one fundamental issue.  
Noninterference properties quantify over a \emph{pair} of indistinguishable
starting states: $\forall S_1, S_2 \in \Start.\; S_1 \indist S_2
\Longrightarrow \dots$. This is a very strong precondition, which is
{extremely} unlikely to be satisfied for independently
generated states.  Instead, we
generate \emph{indistinguishable pairs} of states together. The first state
is generated randomly using one of the techniques described later in this
section. The second is obtained by randomly varying the ``high parts'' of the 
first. We refer to the second state as the {\em variation} of the first.
The resulting pair thus satisfies indistinguishability by construction.
Note that when implemented correctly this does not compromise
completeness: by generating a random state and randomly varying 
we still guarantee that it is possible to generate all pairs of indistinguishable
states. Naturally, the resulting distributions will depend on the specifics of 
the generation and variation methods used, as we shall see. 
%
%

Since EENI considers only executions that start at initial states, we only
need to randomly generate the contents of the instruction memory (the
program that the machine executes) together with the \emph{size} of the data
memory (in initial states, the contents of the memory are fixed and the stack is
guaranteed to be empty).


\autoref{fig:combinedtable-basic} offers an empirical comparison of all the
generation strategies described in this section. 
For
a given test-generation strategy, we inject the bugs from \autoref{sec:basic}
one at a time into the machine
definition and measure the time spent on average until that bug is
found (\emph{mean time to failure}, or MTTF).
Tests were run one at a time on seven identical machines,
each with 4$\times$ 2.4~GHz
Intel processors and 11.7~GB of RAM; they were running
{Fedora~18 and GHC~7.4.1, and using QuickCheck~2.7.6}.
%
We run each test for 5~minutes (300~seconds) or until 4000~counterexamples
are found, whichever comes first.
\ifever
\asz{How does this change for 1000 seconds?}
\fi

\subsection{Naive instruction generation}

The simplest way to generate programs is by choosing a sequence of
instructions \emph{independently} and \emph{uniformly}.
We generate
individual instructions by selecting an instruction type uniformly
(\IE \(\ii{Noop}\), \(\ii{Push}\), \ETC) and then filling in its
fields using QuickCheck's built-in generators. Labels are also
chosen uniformly. We then build the instruction memory by sampling
a number (currently a random number between 20 and 50)
\ifever
\dv{Can we just read this param from the tmu flags?}
\fi
of instructions from this generator.

The first column of \autoref{fig:combinedtable-basic}
\ifever\aaa{TODO: is there any better way of referencing this?}\fi
shows how this strategy performs on the bugs from \autoref{sec:basic}.
Disappointingly, but perhaps not too surprisingly, naive instruction
generation can only find four of the six bugs within 5 minutes.
\ifever
\ch{TODO: Make the 5 minutes be a macro that's generated from Haskell!
  together with the tables. Same for the 3100,
  which will be 4000 on next regeneration!}
\fi
\ifever\finish{Check these magic constants, once we've generated the final set of
  tables.}\fi
How can we do better? 


\proftable[t]{Initial-Naive-False}{
  Execution statistics for naive instruction
  generation.  Executions fail early, and the main reason for failure is
  stack underflow.
}

One obvious weakness is that the 
discard rate is quite high, indicating that one or both machines often fail
to reach a halted state. 
By asking QuickCheck to collect statistics on the execution traces of test
cases (\autoref{fig:prof-Initial-Naive-False}),
we can also see a second problem:
 the average execution length
is only \begingroup
  \onlylentrue
\ifonlylen%
0.47%
\else%
\begin{tabular}{rl}
\multicolumn{2}{l}{Average number of execution steps: 0.47}\\\midrule
74\% & stack underflow \\
21\% & halt \\
4\% & load or store out of range \\
\end{tabular}
\fi%
\endgroup steps!
Such short runs are not useful for finding counterexamples to EENI (at a
minimum, any counterexample must include a \ii{Store} instruction to put bad
values into the memory and a \ii{Halt} so that the run terminates, plus
whatever other instructions are needed to produce the bad states).  So our
next step is to vary the distribution of instructions so as to generate
programs that run for longer and thus have a chance to get into more
interesting states.

\subsection{Weighted distribution on instructions}

\autoref{fig:prof-Initial-Naive-False} shows that by far the most common
reason for early termination is a stack underflow.  After a bit of thought,
this makes perfect sense: the stack is initially empty, so if the first
instruction that we generate is anything but a \ii{Push}, \ii{Halt}, or
\ii{Noop}, we will fail immediately.  Instead of a uniform distribution on
instructions, we can do better by increasing the weights of \ii{Push} and
\ii{Halt}---\ii{Push} to reduce the number of stack underflows, and
\ii{Halt} because each execution must reach a halted state to satisfy EENI's
precondition. The results after this change are shown in the second column of
\autoref{fig:combinedtable-basic}.
Although this strategy still fails to find the \IfcBugLoadNoTaint{}
and \IfcBugStoreNoPointerTaint{} bugs in the allocated time, there is
a significant improvement on both discard rates and the MTTF for the
other bugs.
Run length is also better, averaging %
\begingroup
  \onlylentrue
\ifonlylen%
2.69%
\else%
\begin{tabular}{rl}
\multicolumn{2}{l}{Average number of execution steps: 2.69}\\\midrule
38\% & halt \\
35\% & stack underflow \\
25\% & load or store out of range \\
\end{tabular}
\fi%
\endgroup steps. As %
\autoref{fig:prof-Initial-Weighted-False}
shows, executing \ii{Halt} is now the main reason for termination,
with stack underflows and out-of-range accesses close behind.

\proftable[t]{Initial-Weighted-False}{Execution statistics when generating
  instructions with a weighted distribution.  The main reason for failure is
  now \ii{Halt}, followed by stack underflow.}

\subsection{Generating useful instruction sequences more often}
\label{sec:gen-seq}

To further reduce stack underflows we additionally generate \emph{sequences} of
instructions that make sense together. For instance, in addition to
generating single \ii{Store} instructions, we also generate sequences
of the form \([\ii{Push }\V{n}{\lab}, \ii{Push }\V{ma}{\lab'}, \ii{Store}]\),
where $\ii{ma}$ is a valid memory address. We also generate such
sequences for the other two instructions that use stack elements:
\([\ii{Push }\V{ma}{\lab}, \ii{Load}]\)
where $\ii{ma}$ is a valid memory address, and
\([\ii{Push }\V{n_1}{\lab_1}, \ii{Push }\V{n_2}{\lab_2}, \ii{Add}]\).
%
%
The results are shown in the third column
of \autoref{fig:combinedtable-basic}.
With sequence generation we can now find all bugs, faster than before.
Programs run for slightly longer
(\begingroup
  \onlylentrue
\ifonlylen%
3.86%
\else%
\begin{tabular}{rl}
\multicolumn{2}{l}{Average number of execution steps: 3.86}\\\midrule
37\% & halt \\
28\% & load or store out of range \\
20\% & stack underflow \\
13\% & sensitive upgrade \\
\end{tabular}
\fi%
\endgroup steps on average).
As expected, stack underflows are less common than before
(\autoref{fig:prof-Initial-Sequence-False}) and out-of-range
addresses are now the second biggest reason for termination.
 
\proftable[t]{Initial-Sequence-False}{
  Execution statistics when generating sequences of instructions.  Out-of-range
  addresses are now the biggest reason for termination.}

\subsection{Smart integers: generating addresses more often}

To reduce the number of errors caused by out-of-range
addresses, we additionally give preference to \emph{valid} memory addresses, \IE
integers within (data and instruction) memory bounds, when generating integers.
We do this not only when generating the state of the first machine, but also
when \emph{varying} it, since both machines need to halt successfully
in order to satisfy the precondition of EENI. Column four of
\autoref{fig:combinedtable-basic} shows the
results after making this improvement to the previous generator.
We see an improvement on the MTTF. The average run length is now
\begingroup
  \onlylentrue
\ifonlylen%
4.22%
\else%
\begin{tabular}{rl}
\multicolumn{2}{l}{Average number of execution steps: 4.22}\\\midrule
41\% & halt \\
21\% & stack underflow \\
21\% & load or store out of range \\
15\% & sensitive upgrade \\
\end{tabular}
\fi%
\endgroup steps, and
the percentage of address-out-of-range errors is decreased
(\autoref{fig:prof-Initial-Sequence-True}).

\proftable[t]{Initial-Sequence-True}{
   Execution statistics when using smart integers with sequences of
   instructions.  The percentage of address out of range errors has halved.}

\subsection{Generation by execution}\label{sec:gen-by-exec-basic}
We can go even further.
In addition to weighted distributions, sequences, and smart integers, we 
try to generate instructions that
\emph{do not cause a crash}.
In general (for more interesting machines) deciding whether
an arbitrary instruction sequence causes a crash is undecidable.
In particular we cannot know in advance all possible states
in which an instruction will be executed.
We can only make a guess---a very accurate one for this simple machine.
This leads us to the following \emph{generation by execution} strategy: We
generate a single instruction or a small sequence of instructions, as
before, except that now we restrict generation to instructions that do not
cause the machine to crash in the current state.  When we find one, we
execute it to reach a new state and then repeat the process to generate
further instructions.  We continue until we have generated a reasonably
sized instruction stream (currently, randomly chosen between 20--50
instructions).
We discuss how this idea generalizes to machines with
nontrivial control flow in \autoref{sec:gen-by-exec-cally}. 

As we generate more instructions, we
increase the probability
of generating a $\ii{Halt}$ instruction, to reduce the chances of the
machine running off the end of the instruction stream. As a result, (i)~we
maintain low discard ratios for EENI since we increase the probability that
executions finish with a $\ii{Halt}$ instruction, and (ii)~we avoid
{extremely long} executions whose long time to generate and run could be more
fruitfully used for other test cases.

\ifever
\dv{An alternative plan would be to generate a \ii{halt} instruction in the end 
of the instruction stream but that would fail to tackle problem (ii) and would
be less effective in the presence of interesting control flow, a point that 
we return to in \autoref{sec:cally}.}
\iflater\dv{Someone might say: oh why not then just create a final
  halt if you think you are running off the instruction stream? This has two
  disadvantages: executions will be long, and second it's not generalizable
  to the cally machine that really benefits from the increasing
  halting probability.  I am not fully convinced that the argument belongs
  here, but it's an interesting discussion point to be made I believe and
  some readers may raise this point.}\bcp{Yes, I wondered about this.}\fi
\fi

The MTTF (last column of \autoref{fig:combinedtable-basic}) is now
significantly lower than in any previous generation method, although
this strategy runs fewer tests per second than the previous ones
(because both test case generation and execution take longer).
\autoref{fig:profvar-Initial-ByExec-True} shows that 94\% of the pairs
both successfully halt, which is in line with the very low discard
rate of \autoref{fig:combinedtable-basic}, and that programs run for
much longer.
Happily, varying a machine that successfully halts
has a high probability of generating a machine that also halts.

\profvartable[t]{Initial-ByExec-True}{
  Execution statistics for generation by execution, broken down for the variations.}

\section{Control Flow}
\label{sec:cally}

\sloppy

Up to this point, we've seen how varying the program generation strategy can
make orders-of-magnitude difference in the speed at which counterexamples
are found for a very simple---almost trivial---information-flow stack machine.
Now we are ready to make the machine more interesting and see how these
techniques perform on the new bugs that arise, as well as how their
performance changes on the bugs we've already seen.
In this section, we add \ii{Jump}, \ii{Call}, and \ii{Return}
instructions---and, with them, the possibility that information can leak via
the program's control flow.

\fussy

\subsection{Jumps, implicit flows, and the \pc{} label}
\label{sec:jumps}

We first add a new \ii{Jump} instruction that takes the first element from
the stack and sets the \pc{} to that address:
\infrule[\IfcBugJumpNoRaisePcANDJumpLowerPc]
  {i(\pc) = \ii{Jump}}
  {\MACH{\pc}{\V{n}{\lab_n}:s}{m} \step \MACH{n}{s}{m}}
(The jump target may be an invalid address.  In this case, the machine will
be stuck on the next instruction.)

\counterexample*{jumpy1}{Counterexample to \IfcBugJumpNoRaisePcANDJumpLowerPc: A textbook example of an implicit flow }

Note that this rule simply ignores the label on the jump target on the
stack.  This is unsound, and QuickCheck easily finds the counterexample in
\autoref{fig:counter-jumpy1}---a textbook case of an
\emph{implicit flow}~\cite{sabelfeld03:lang_based_security}.  A
secret is used as the target of a jump, which causes the instructions that
are executed afterwards to differ between the two machines; one of the
machines halts immediately, whereas the other one does a \ii{Store} to a low
location and only then halts, causing the final memories to be
distinguishable.

\ShowCounterexamplePCLabels{}

The standard way to prevent implicit flows is to label the \pc{}---\IE to
make it a labeled integer, not a bare integer.
Initial states have $\V{pc}{\lab_\pc} = \V{0}{\low}$, and 
after a jump to a secret address the label of the \pc{} becomes
$\high$:
\infrule[\IfcBugJumpLowerPc]
  {i(\pc) = \ii{Jump}}
  {\MACH{\V{\pc}{\lab_\pc}}{\V{n}{\lab_n}:s}{m} \step \MACH{\V{n}{\lab_n}}{s}{m}}
While the \pc{} is (labeled) high, the two machines may be
  executing different instructions, and so we cannot expect the
  machine states to correspond. We therefore extend the definition of
  $\indistmem$ so that \emph{all} high machine states are deemed
  equivalent.
(We call a state ``high'' if the \pc{} is labeled $\high$,
and ``low'' otherwise.)
\ifever
\feedback{Deian: This is minor, but maybe explain why this notion of
  equivalence is okay. [even if you just refresh the reader's mind of
  what the attacker model is]}
\ch{will be showing counterexample to this in the appendix}
\fi

\begin{defn} \
  Machine states $S_1 = \MACHI{\V{\pc_1}{\lab_{\pc_1}}}{s_1}{m_1}{i_1}$ and
  $S_2 = \MACHI{\V{\pc_2}{\lab_{\pc_2}}}{s_2}{m_2}{i_2}$ are indistinguishable
  with respect to memories, written $S_1 \indistmem S_2$,
  if either $\labelof{\pc_1} = \labelof{\pc_2} = \high$
     or else $\labelof{\pc_1} = \labelof{\pc_2} = \low$ and
        $m_1 \indist m_2$  and $i_1 \indist i_2$.
\ifever
\bcp{@Catalin: pc's should also be equal, right?}
\ch{@Benjamin: Not really necessary for EENI + StartInitial or
  StartQuasiInitial in the same way it wasn't necessary for the Basic
  variant of $\indistmem$ (it would be necessary for EENI +
  StartArbitrary, but we don't really look at that). This is
  also what our current implementation does.}
\bcp{OK.  Alejandro was confused by this, so maybe it needs a comment at
  some point.}
\fi
\end{defn}

\counterexample{jumpy2}{Counterexample to \IfcBugJumpLowerPc: \ii{Jump} should not lower the \pc{} label
%
}

The \IfcBugJumpLowerPc{} rule is still wrong, however, since it not only
raises the \pc{} label when jumping to a high address but also lowers it
when jumping to a low address. The counterexample in 
\autoref{fig:counter-jumpy2} illustrates that the latter behavior is
problematic.
%
%
The fix is to label the \pc{} after a jump with the \emph{join} of the
current \pc{} label and the label of the target address.
\infrule[Jump]
  {i(\pc) = \ii{Jump}}
  {\MACH{\V{\pc}{\lab_\pc}}{\V{n}{\lab_n}:s}{m} \step \MACH{\V{n}{(\lab_n \join \lab_\pc)}}{s}{m}}

With this rule for jumps QuickCheck no longer finds any
counterexamples.
%
Some readers may find this odd: In order to fully address implicit
flows, it is usually necessary to modify the rules for memory stores
to handle the case where the \pc{} is labeled high~\cite{AustinF09,
  RussoS10}.
The current machine doesn't require this, but the reason is subtle:
here, the \pc{} can go from \low{} to \high{} when we jump to a secret
address, but it never goes from \high{} to \low!
It doesn't matter what the machine does when the \pc{} is high, because none
of its actions will ever be observable---all high machine states are
indistinguishable. 

To make things more interesting, we need to enrich the machine with some
mechanism that allows the \pc{} to safely return to \low{} after it has
become \high.  One way to achieve this is to add \ii{Call} and \ii{Return}
instructions, a task we turn to next.

\subsection{Restoring the \pc{} label with calls and returns}
\label{sec:calls-and-returns}

\ReturnControlsReturn

IFC systems (both static and dynamic) generally rely on control flow \emph{merge points} (\IE post-dominators of the branch point in the control flow
graph where the control was tainted by a secret) to detect when the
influence of secrets on control flow is no longer relevant and the \pc{}
label can safely be restored.
%
Control flow merge points are, however, much more evident for structured
control features such as conditionals than they are for jumps
(as long as we don't have exceptions~\cite{Exceptional, BichhawatRGH14}).
Moreover, since we are doing purely dynamic IFC we cannot
distinguish between safe uses of jumps and unsafe ones (\EG the one
in \autoref{fig:counter-jumpy2}).
So we keep jumps as they are (only raising the \pc{} label) and add
support for structured programming and restoring the \pc{} label
in the form of \ii{Call} and \ii{Return} instructions,
which are of course useful in their own right.

To support these instructions, we need some way of representing stack
frames.  We choose a straightforward representation, in which each stack
element $e$ can now be either a labeled integer $\V{n}{\ell}$ (as before)
or a \emph{return address},
marked $\retsym$, recording the \pc{} (including its label!) from which the
corresponding \ii{Call} was made.
We also extend the indistinguishability
relation on stack elements so that return addresses are only equivalent to
other return addresses and $\badret{n_1}{\lab_1} \indist \badret{n_2}{\lab_2}$ if
either $\lab_1 = \lab_2 = \high$ or else $n_1 = n_2$ and $\lab_1 = \lab_2 = \low$ (this
is the same as for labeled integers). (High return addresses and high
integers need to be distinguishable to a low observer, as we discovered
when QuickCheck generated an unexpected counterexample,
which we list in \autoref{sec:counter-stack}---understanding it requires
reading the rest of this section.)

We also need a way to pass arguments to
and return results from a called procedure.  For this, we annotate the
\ii{Call} and \ii{Return} instructions with a positive integer indicating how many
integers should be passed or returned (0 or 1 in the case of
\ii{Return}\ifever\bcp{awkward wording}\fi).
Formally,
$\ii{Call}\; k$ expects an address $\V{n}{\lab_n}$ followed by $k$
integers $\ii{ns}$ on the stack.  It
sets the \pc{} to $n$, labels this new \pc{}
by the join of $\lab_n$ and the current \pc{} label (as we did for
\ii{Jump}---we're eliding the step of getting this bit wrong at first and
letting QuickCheck find a counterexample), and adds 
the return address frame to the stack \emph{under} the $k$ arguments.  
\infrule[\IfcBugValueOrVoidOnReturnCall]
  {i(\pc) = \ii{Call}\; k \andalso
   \ii{ns} = \V{n_1}{\lab_1}:\ldots:\V{n_k}{\lab_k}}
  {\MACH{\V{\pc}{\lab_\pc}}{\V{n}{\lab_n}:\ii{ns}:s}{m} \step
   \MACH{\V{n}{(\lab_n \join \lab_\pc)}}{\ii{ns}:\badret{(\pc\mathord{+}1)}{\lab_\pc}:s}{m}}

$\ii{Return}\; k'$\ifever\bcp{why $k'$ rather than just $n$?}
\ch{It was because it will later be moved to the call
  and there we already have an $n$. Feel free to change this though.}\fi{}
traverses the stack until it finds the first 
return address and jumps to it. Moreover it restores the \pc{} label
to the label stored in that $\retsym$ entry, and preserves the first
$k'$ elements on the stack as return values, discarding all other
elements in this stack frame (like $\ii{ns}$, $\ii{ns}'$ stands
for a list of labeled integers; in particular it cannot contain return addresses).

\ifever\bcp{Why only 0 or 1 return
  values?  Explain.}\ch{That's what our implementation does; there
  is no deeper reason for it. This 0-1 thing did confuse 2 of our
  feedback-givers, so maybe just swipe this condition under the
  carpet?}\bcp{But it will still be there in the definition, waiting to
  confuse readers that look carefully.  Let's just explain that this was an
  arbitrary choice that we made in the design.}
\ch{I was proposing removing it from the definition,
  not just from the explanation. It's fine either way.}
\bcp{@Catalin: This sounds like the right thing to do, though I do remember
  arguing with John about whether fixing it to be 1 would make finding
  counterexamples very much harder---but in any case not before
  submission: no time to regenerate tables.}
\fi  

\infrule[\IfcBugReturnNoTaintANDIfcBugValueOrVoidOnReturnReturn]
  {i(\pc) = \ii{Return}\; k' \quad k' \in \{0,1\} \quad
  \ii{ns} = \V{n_1}{\lab_1}:\ldots:\V{n_{k'}}{\lab_{k'}}}
  {\MACH{\V{\pc}{\lab_\pc}}{\ii{ns}:\ii{ns}':\badret{n}{\lab_n}:s}{m} \step
   \MACH{\V{n}{\lab_n}}{\ii{ns}:s}{m}}

Finally, we observe that we cannot expect the current EENI instantiation to
hold for 
this changed machine, since now one machine can halt in a high state
while the other can continue, return to a low state, and only then 
halt.
\ifever\ch{will show counterexample to this in the
  appendix}\bcp{it would be much more fun to do this paragraph socratically:
show the counterexample (here, not in an appendix!), comment that it is
actually spurious, and then fix the property.  This will work best / be most
truthful if this is the \emph{first} bug that QC finds---is it??}
\feedback{Deian: Same as before, explaining why this is intuitively
  okay would be helpful}
\ch{It would be nice, but we're not doing well on space}
\fi
Since we cannot equate high and low states
\iftrue (see \autoref{sec:counter-eeni-low} for a counterexample;
  again understanding it requires reading the rest of this section)\fi,
we need to change the EENI
instance we use to EENI$_{\Init,\Halted\cap\Low,{\indistmem}}$, where
 $\Low$ denotes the set of states with $\lab_\pc=\low$. Thus, we
only consider executions that end in a low halting state.

After these changes, we can turn QuickCheck loose and start finding more
bugs. 
The first one, listed in \autoref{fig:counter-cally1}, is essentially
another instance of the implicit flow bug, which is not surprising
given the discussion at the end of the previous subsection.
%
We adapted the Store rule trivially to the new setting,
but that is clearly not enough:
\infrule[Store*de]
  {i(\pc) = \ii{Store} \quad
   m(p) = \V{n'}{\lab_n'} \quad
   \lab_p \flowsto \lab_n' \quad
   m' = \update{m}{p}{\V{n}{(\lab_n\mathord\join\lab_p)}}}
  {\MACH{\V{\pc}{\lab_\pc}}{\V{p}{\lab_p}:\V{n}{\lab_n}:s}{m} \step \iffull\else\\\fi
    \MACH{\V{(\pc\mathord{+}1)}{\lab_\pc}}{s}{m'}} 
We need to change this rule so that the value written in memory
is tainted with the current \pc{} label:
\infrule[Store*e]
  {i(\pc) = \ii{Store} \quad
   m(p) = \V{n'}{\lab_n'} \quad
   \lab_p \flowsto \lab_n' \quad
   m' = \update{m}{p}{\V{n}{(\lab_n\mathord\join\lab_p\mathord\join\lab_\pc)}}}
  {\MACH{\V{\pc}{\lab_\pc}}{\V{p}{\lab_p}:\V{n}{\lab_n}:s}{m} \step \iffull\else\\\fi
    \MACH{\V{(\pc\mathord{+}1)}{\lab_\pc}}{s}{m'}} 
This eliminates the current counterexample; QuickCheck then
finds a very similar one in which the \emph{labels} of values in the
memories differ between the two
machines\iffull~(\autoref{fig:counter-cally2})\fi.
The usual way to prevent this problem is to extend the no-sensitive-upgrades
check so that low-labeled data cannot be overwritten in a high
context~\cite{zdancewic02:thesis,AustinF09}. This leads to the 
correct rule for stores:
\ifever\bcp{Do we have two rules named
  \rn{Store}?  Maybe the earlier one needs to be renamed, with a note that
  it is correct in the basic machine but will need to be fixed again
  later?}
\feedback{Michal: You keep calling the changed rule STORE, which is a
  little confusing}
\feedback{Andrew also found this confusing}
\ch{How about repeating the old rule for Store here giving it a
  different name, Store*3+4, before presenting the new rule?
  would that help?}
\bcp{No space!}
\fi

\counterexample*{cally1}{Counterexample to \rn{Store*de} rule:
  Raising \pc{} label is not enough to prevent
  implicit flows.  Once we have a mechanism (like \ii{Return}) for restoring
  the \pc{} label, we need to be more careful about stores in high
  contexts.
}

\iffull\counterexample*{cally2}{Counterexample to \rn{Store*d}:
  Implicit flow via labels.}\fi

\infrule[Store]
  {i(\pc) = \ii{Store} \quad
   m(p) = \V{n'}{\lab_n'} \quad
   \lab_p\mathord\join\lab_\pc \flowsto \lab_n' \quad
   m' = \update{m}{p}{\V{n}{(\lab_n\mathord\join\lab_p\mathord\join\lab_\pc)}}}
  {\MACH{\V{\pc}{\lab_\pc}}{\V{p}{\lab_p}:\V{n}{\lab_n}:s}{m} \step \iffull\else\\\fi
    \MACH{\V{(\pc\mathord{+}1)}{\lab_\pc}}{s}{m'}} 

The next counterexample found by QuickCheck
(\autoref{fig:counter-cally3})
shows that
returning values from a high context to a low one is unsound if we do not
label those values as secrets.
%
To fix this, we taint all the returned values with the pre-return \pc{}
label.

\counterexample*{cally3}{Counterexample
to \IfcBugReturnNoTaintANDIfcBugValueOrVoidOnReturnReturn: \ii{Return} needs to
taint the returned values.
\ifever\ch{It would be good to merge the two
    executions back in if possible}\fi}

\infrule[\IfcBugValueOrVoidOnReturnReturn]
  {i(\pc) = \ii{Return}\; k' \quad k' \in \{0,1\} \quad
  \ii{ns} = \V{n_1}{\lab_1}:\ldots:\V{n_{k'}}{\lab_{k'}} \\
  \ii{ns}_\pc = \V{n_1}{(\lab_1\mathord\join\lab_\pc)}
     :\ldots:\V{n_{k'}}{(\lab_{k'}\mathord\join\lab_\pc)}
  }
  {\MACH{\V{\pc}{\lab_\pc}}{\ii{ns}:\ii{ns}':\badret{n}{\lab_n}:s}{m} \step
   \MACH{\V{n}{\lab_n}}{\ii{ns}_\pc:s}{m}}

The next counterexample, listed in \autoref{fig:counter-cally4}, shows
(maybe somewhat surprisingly)
that it is unsound to specify the number of results to return in
the \ii{Return} instruction, because then the number of results returned
may depend on secret flows of control.
To restore soundness, we need to pre-declare at each \ii{Call} whether the
corresponding \ii{Return} will return a value---\IE the \ii{Call}
instruction should be annotated with \emph{two} integers, one for parameters
and the other for results. Stack elements $e$ are accordingly
either labeled values $\V{n}{\lab}$ or (labelled) pairs of a return
address $n$ and the number of return values $k$.
\[
e ::= \V{n}{\lab} \;|\; \ret{n}{k}{\lab}
\]
These changes lead us to the correct rules:

\counterexample*{cally4}{Counterexample to
  \IfcBugValueOrVoidOnReturnCall{} and
  \IfcBugValueOrVoidOnReturnReturn: It is unsound to choose how many
  results to return on \ii{Return}.}

\CallControlsReturn

\infrule[Call]
  {i(\pc) = \ii{Call}\; k\;k' \andalso k' \in \{0,1\} \andalso
   \ii{ns} = \V{n_1}{\lab_1}:\ldots:\V{n_k}{\lab_k}}
  {\MACH{\V{\pc}{\lab_\pc}}{\V{n}{\lab_n}:\ii{ns}:s}{m} \step
   \MACH{\V{n}{(\lab_n \join \lab_\pc)}}{\ii{ns}:\ret{\pc\mathord{+}1}{k'}{\lab_\pc}:s}{m}}

\infrule[Return]
  {i(\pc) = \ii{Return} \andalso
  \ii{ns} = \V{n_1}{\lab_1}:\ldots:\V{n_{k'}}{\lab_{k'}} \\
  \ii{ns}_\pc = \V{n_1}{(\lab_1\mathord\join\lab_\pc)}
     :\ldots:\V{n_{k'}}{(\lab_{k'}\mathord\join\lab_\pc)}
  }
  {\MACH{\V{\pc}{\lab_\pc}}{\ii{ns}:\ii{ns}':\ret{n}{k'}{\lab}:s}{m} \step
   \MACH{\V{n}{\lab}}{\ii{ns}_\pc:s}{m}}

The final counterexample found by QuickCheck is quite a bit longer
\iffull
(see \autoref{fig:counter-cally5})\else; we omit the details for 
brevity\fi.
It shows that we cannot allow instructions like \ii{Pop} to remove
return addresses from the stack, as does the following broken rule
(recall that $e$ denotes an arbitrary stack entry):
\ifever
\bcp{it took me a while to
  realize that this was not strictly the same as the earlier \rn{Pop} rule.
Couldn't we fudge the difference?  One problem is that we presented the
earlier rule as correct, but now it's become incorrect.  There's a similar
problem with the Store rule.  I suggest removing \IfcBugPopPopsReturns.}
\ch{I think \IfcBugPopPopsReturns{} is the most interesting bug we have,
  so we should keep it.
  If we look at it purely syntactically then the correct \rn{Pop}
  is the same as the correct \rn{Pop} from before -- but that's just
  in rule format, in Haskell it wouldn't be this way.
  If naming is the problem then let's fix naming.}\bcp{Or, perhaps easier at
this point, just say explicitly that we're re-using the same names even
though some types have changed so the rules mean something different.}
\fi
\infrule[\IfcBugPopPopsReturns]
  {i(\pc) = \ii{Pop}}
  {\MACH{\V{\pc}{\lab_\pc}}{e:s}{m} \step \MACH{\V{(\pc\mathord{+}1)}{\lab_\pc}}{s}{m}}
To protect the call frames on the stack, we change this rule to only
pop integers (all the other rules can already only operate on integers).
\infrule[Pop]
  {i(\pc) = \ii{Pop}}
  {\MACH{\V{\pc}{\lab_\pc}}{\V{n}{\lab_n}:s}{m} \step \MACH{\V{(\pc\mathord{+}1)}{\lab_\pc}}{s}{m}}

\iffull\cewide\fi
\counterexample*{cally5}{%
  Counterexample to \IfcBugPopPopsReturns: It is unsound not to protect the call
  stack.}



\ifever
\ch{We're not listing all the wrong rules for the cally machine}
\ch{I've added a few more, to me it seems that all things
  not explicitly listed can easily be inferred from the rest.}
\fi

\subsection{Generation by execution and control flow}
\label{sec:gen-by-exec-cally}

Generation by execution is still applicable in the presence of interesting
control flow but we have to make small modifications to the original algorithm.
We still generate a single instruction or sequence%
\footnote{In addition to the instruction sequences from
  \autoref{sec:gen-seq} we use two new sequences for \ii{Jump} and
  \ii{Call}: \([\ii{Push \V{ia}{\lab}}, \ii{Jump}]\) and
  \([\ii{Push }\V{n_k}{\lab_k}, \dots, \ii{Push }\V{n_1}{\lab_1},
     \ii{Push \V{ia}{\lab}}, \ii{Call\;k\;k'}]\),
  where $\ii{ia}$ is a valid memory address.}
that does not crash,
as before, and we execute it
to compute a new state. However, unlike before, while executing this newly generated sequence 
of instructions, it is possible to ``land'' in a position in the instruction stream where we have already 
generated an instruction (e.g. via a backward jump).
If this happens then we keep executing the 
already generated instructions. If the machine halts (or we reach a loop-avoiding 
cutoff) then we stop the process and return the so-far generated instruction stream.
If there are no more instructions to execute then we go on to generate more instructions.
There is one more possibility though: the machine may {\em crash} while executing an 
already generated instruction. To address this issue, 
we make sure that we never generate an instruction (\EG a jump)
that causes the machine to crash in a certain number of steps.
We refer to this number of steps as the \emph{lookahead} parameter and in our experiments we use 
a lookahead of 2 steps. If we cannot generate any instruction satisfying
this constraint, we retry with a smaller lookahead, until we succeed.

Since it now becomes possible to generate instruction streams that 
cause the machine to crash in some number of steps, one might be worried about
the discard ratio for EENI.
However, the ever increasing probability of
generating a \ii{Halt} (discussed in \autoref{sec:gen-by-exec-basic})
counterbalances this issue.

%


\subsection{Finding the bugs}\label{sec:finding-bugs-cally}

\combinedtable{cally}{
  Cally-EENI-Mem-Initial-ByExec2-True,
  Cally-EENI-Low-Initial-ByExec2-True,
  Cally-EENI-Low-QuasiInitial-ByExec2-True,
  Cally-LLNI-Low-QuasiInitial-ByExec2-True,
  Cally-SSNI-Full-Arbitrary-Naive-True,
  Cally-SSNI-Full-Arbitrary-TinySSNI-True}
{Experiments for control flow machine. MTTF given in
  milliseconds.
}

We experimentally evaluated the effectiveness of testing for this new
version of the stack machine,
by adding the bugs discussed in this section
to the ones applicable for the previous machine.
The results of generation by execution with lookahead for this machine
are shown in the first column of \autoref{fig:combinedtable-cally}.
As we can see, all old bugs are still found relatively fast.
It takes longer to find them when compared to the previous
machine, but this is to be expected: when we extend the machine, we
are also increasing the state space to be explored.
%
The new control-flow-specific bugs are all found, with the exception
of \IfcBugPopPopsReturns{}, which requires a larger timeout.
%
Discard rates are much higher compared to generation by execution in
\autoref{fig:combinedtable-basic}, for two reasons. First, control flow can
cause loops, so we discard machines  
that run for more than 50 steps without halting.
Detailed profiling
revealed that 18\% of the pairs of machines both loop, and loopy machines
push the average number of execution steps to 22.
Second, as described previously, generation by execution in the presence of control flow is much less accurate.

\subsection{Alternative generation strategies}

Generation by execution has proved
effective in
finding bugs. Even this method, however, required some tuning, driven by experimental evaluation. For instance, 
our first implementations did not involve
gradually increasing the probability of \ii{Halt} instructions.
We also experimented 
with different lookahead values. Larger lookaheads introduce significant overheads 
in generation as every generated instruction costs many steps of execution, and
the payoff of lower discard rates was not worth the increased generation cost.

We have also explored (and dismissed) several other generation
strategies, and we outline two of these below:
\begin{itemize}
  \item {\em Generation by forward execution}. Generation by execution 
  fills in the
  instruction stream in patches, due to generated jumps. It is hence
  possible for the instruction stream to contain ``holes'' filled with \ii{Noop}
  instructions. An alternative strategy is to generate instructions 
  in a forward style only: if we generate a branch then we {\em save} the current 
  state along with the branch target, but keep generating instructions as if the branch was 
  not taken. 
  If we ever reach the target of the branch we may use the 
  saved state 
as a potentially more accurate state that we can use to generate more instructions.
  This strategy delivered similar results as generation by execution, but due to
  its more complicated nature we dismissed it and used the basic design instead.

  \item {\em Variational generation by execution}.
  In this strategy, we first generate a machine with generation by execution.
  We then {\em vary} the machine and run generation by execution for the resulting machine, in the
  hope that we can fill in the holes in the originally generated instruction stream with 
  instructions from a variational execution. As before, we did not find
  that the results justified the generation overheads and 
  complexity of this strategy.
\end{itemize}


\section{Strengthening the Tested Property}
\label{sec:stronger}

The last few counterexamples in \autoref{sec:calls-and-returns}
are fairly long and quite difficult for
QuickCheck to find, even with the best test-generation strategy.
In this section we explore a different approach: strengthening
the {\em property} we
are testing so that counterexamples become shorter and easier to
find. \autoref{fig:combinedtable-cally} in
\autoref{sec:finding-bugs-cally} summarizes the variants of
noninterference that we consider and how they affect test performance.

\ifever
\ch{Could try to use the average counterexample length (number of
  executed instructions? size of instruction memory? one of them)
  after shrinking as a metric for how well we are doing here. For SSNI
  this metric is 1, but for the original EENI I expect it to be quite
  big.}
\ch{This would depend on our shrinking being really good; optimal?}
\ch{Otherwise use not average counterexample length, but minimal
  counterexample length. Although it would be very interesting to see
  what our distribution is on shrunk counterexamples, since we might
  be able to use that to improve shrinking. Would minimum + median +
  (arithmetic) average shrunk counterexample length be enough for
  this?}
\fi

\subsection{Making entire low states observable}

Every counterexample that we've seen involves pushing an address, executing
a \ii{Store} 
instruction, and halting.  These steps are all necessary because of the
choice we made in \autoref{sec:eeni} to ignore the stack when defining
indistinguishability on machine states.
A counterexample that leaks a secret onto the stack must continue by
storing it into memory; similarly, a counterexample that leaks a
secret into the \pc{} must execute \ii{Store} at least twice.  This
suggests that we can get shorter counterexamples by redefining
indistinguishability as follows:
\begin{defn} \
  Machine states $S_1 = \MACHI{\pc_1}{s_1}{m_1}{i_1}$ and $S_2 =
  \MACHI{\pc_2}{s_2}{m_2}{i_2}$ are indistinguishable with respect to
  entire low states,
  written $S_1 \indistlow S_2$,
  if either $\labelof{\pc_1} = \labelof{\pc_2} = \high$
     or else
        $\labelof{\pc_1} = \labelof{\pc_2} = \low$,
        $m_1 \indist m_2$, 
        $i_1 \indist i_2$, 
        $s_1 \indist s_2$, and
        $\pc_1 \indist \pc_2$.
\end{defn}

\sloppy

We now strengthen EENI$_{\Init,\Halted\cap\Low,{\indistmem}}$, the property we
have been testing so far, to EENI$_{\Init,\Halted\cap\Low,{\indistlow}}$; this is
stronger because $\indistmem$ and $\indistlow$ agree on initial states, while
for halted states ${\indistlow} \subset {\indistmem}$.  Indeed, for this
stronger property, QuickCheck finds bugs faster (compare the first two columns
of \autoref{fig:combinedtable-cally}).

\fussy

\ifever
\ch{TODO: We could also consider a completeness property.  Given a
  counterexample to EENI$_{\Init,\Halted\cap\Low,{\indistlow}}$ we can always
  construct a counterexample for EENI$_{\Init,\Halted\cap\Low,{\indistmem}}$.
  But then this means that the two properties are actually
  equivalent? No! This completeness property is not general, it only
  holds for our specific machine! We could also call this ``no
  spurious counterexamples'' (meta)property}
\ch{While for this step we can probably prove completeness in a
  machine-dependent but property independent way (probably just
  something about traces), this is not the case for quasi-initial
  states and the Cally machine.  There the lack of spurious
  counterexamples crucially depends on the property (noninterference).}
\asz{This all seems very interesting, but if it's relevant anywhere, it's
     probably only relevant for the full paper.}
\fi

\subsection{Quasi-initial states}

Many counterexamples begin by pushing values onto the stack and storing values
into memory. This is necessary because each test starts with an empty stack and
low, zeroed memory. We can make counterexamples easier to find by allowing
the two machines to start with 
arbitrary (indistinguishable) stacks and memories; we call such
states \emph{quasi-initial}.  Formally, the set \QInit{} of quasi-initial states
contains all states of the form $\MACHI{\V{0}{\low}}{s}{m}{i}$, for arbitrary
$s$, $m$, and $i$.


The advantage of generating more varied start states is that parts of
the state space may be difficult to reach by running generated
code from an initial state; for example, to get two return
addresses on the stack, we must successfully execute two \ii{Call}
instructions (see \EG \autoref{fig:counter-cally5}).
Thus, bugs that are only manifested in these
hard-to-reach states may be discovered very slowly or not at
all. Generating ``intermediate'' states directly gives us better
control over their distribution, which can help eliminate such 
blind spots in testing. The
disadvantage of this approach is that
a quasi-initial state may not be \emph{reachable} from any initial state, so in
principle QuickCheck may report spurious problems that cannot actually
arise in any real execution.
In general, we
could address such problems by carefully formulating the important
invariants of reachable states and ensuring that we generate
quasi-initial states satisfying them. In practice, though, for this
extremely simple machine we have not
encountered any spurious counterexamples, even with
quasi-initial states.
(This is different for the more complex register machine from
\autoref{sec:regs}; in that setting a generator for non-initial states
needs to produce only states satisfying strong invariants associated
with reachable states.)


Instantiating EENI with \QInit{}, we obtain a stronger property
EENI$_{\QInit,\Halted\cap\Low,{\indistlow}}$ (stronger because
${\Init} \subset {\QInit}$) that finds bugs faster, as column~3
of \autoref{fig:combinedtable-cally} shows.





\subsection{LLNI: Low-lockstep noninterference}\label{sec:llni}


While making the full state observable and starting from quasi-initial
states significantly improves EENI, we can get even better results
by moving to a yet stronger noninterference property.
The intuition is that EENI generates machines and runs them
for a long time, but it only compares the final states, and
only when both machines successfully halt; these preconditions lead to
rather large discard rates. Why not compare \emph{intermediate} states
as well, and report a bug as soon as intermediate states are
distinguishable? While the \pc{} is high, the two
machines may be executing different instructions, so their states will
naturally differ; we therefore ignore these states and
require only that low execution states are pointwise
indistinguishable. We call this new property
\emph{low-lockstep noninterference} (or \emph{LLNI}).
We write $S \manystep_t$ when an execution from state $S$ produces trace $t$
(a list of states).
Since this is just a state-collecting variant of the reflexive
transitive closure of $\step$, we allow partial executions and in
particular do not require that the last state in the trace is stuck or
halting.



\newcommand{\llnibody}{%
  with $S_1 \indist S_2$, \chrev{$S_1 \manystep_{t_1}$, and $S_2
    \manystep_{t_2}$}, 
  we have $t_1 \indist^* t_2$,
  where indistinguishability on traces $\indist^*$
  is defined inductively by the following rules:
\typicallabel{Lockstep End}
\infrule[Low Lockstep]
  {S_1, S_2 \in \Low \quad S_1 \indist S_2 \quad t_1 \indist^* t_2}
  {(S_1 : t_1) \indist^* (S_2 : t_2)}
\infrule[High Filter]
  {S_1 \not\in \Low \quad t_1 \indist^* t_2}
  {(S_1 : t_1) \indist^* t_2}
\chrev{
\infrule[End]
  {}
  {t \indist^* [\,]}
}
\infrule[Symmetry]
  {t_1 \indist^* t_2}
  {t_2 \indist^* t_1}
}
\begin{defn}
  A machine semantics is \emph{low-lockstep noninterfering} with
  respect to the indistinguishability relation $\indist$ (written
  $\text{LLNI}_{\indist}$) if, for any quasi-initial states 
  $S_1$ and $S_2$ \strikeout{if} \llnibody{}
\end{defn}

\noindent 
The rule \rn{Low Lockstep} requires low states in the two traces to be
pointwise indistinguishable, while \rn{High Filter} (together with
\rn{Symmetry}) simply filters out high states from either trace.
The remaining rule is about termination: because we are working with
termination-insensitive noninterference, we allow one of the traces to
continue (maybe forever) even if the other has terminated.
%
We implement these rules in Haskell as a recursive predicate over
finite traces.

In general, LLNI implies EENI (see Appendix~\ref{app:strengthening}),
but not vice versa. However,
the correct version of our machine does satisfy LLNI, and we have not
observed any cases where QuickChecking a buggy machine with
LLNI finds a bug that is not also a bug with regard to EENI.
%
Testing LLNI instead of EENI leads to significant improvement in the
bug detection rate for all bugs, as the results in the fourth column
\autoref{fig:combinedtable-cally} show.
In these experiments no generated machine states are discarded, since
LLNI applies to both successful (halting) executions
and failing executions.
The generation strategies described in \autoref{sec:generation}
apply to LLNI without much change; also, as for EENI, generation by
execution (with lookahead of 2 steps)
performs better than the more basic strategies, so we don't consider
those for LLNI.



\subsection{SSNI: Single-step noninterference}\label{sec:ssni}

Until now, we have focused on using sophisticated (and potentially slow)
techniques for generating long-running initial (or quasi-initial)
machine states, and then checking
equivalence for low halting states (EENI) or at every low step (LLNI).  An
alternative is to define a stronger property that talks about \emph{all}
possible single steps of execution starting from two indistinguishable
states.

Proofs of noninterference usually go by induction on a pair of
execution traces; to preserve the corresponding
invariant, the proof needs to consider how each execution step affects the
indistinguishability relation. This gives rise to properties known as
``unwinding conditions''~\cite{GoguenM84}; the corresponding
conditions for our machine form a property we call \emph{single-step
noninterference} (\emph{SSNI}).


We start by observing that LLNI implies that, if two low states are
indistinguishable and each takes a step to another low state, then
the resulting states are also indistinguishable.
%
However, this alone is not a strong enough inductive invariant to guarantee the
indistinguishability of whole traces.
In particular, if the two machines perform a \ii{Return} from a high state to a
low state, we would need to conclude that the two low states are equivalent
without  knowing anything about the original high states. This indicates
that, for SSNI, we can no longer consider all high states indistinguishable.
The indistinguishability relation on high states needs to be strong
enough to ensure that if both machines return to low states, those
low states are also indistinguishable.
Moreover, we need to ensure that if one of the machines takes a step from a high
state to another high state, then the old and new high states are equivalent.
The following definition captures all these constraints formally.

\ifever
\ch{It would be great to show some diagrams here!}
\fi

\newcommand{\ssnilabel}{} 
\newcommand{\ssnibody}[1]{%
the following 
conditions are satisfied:
\chrev{
\begin{enumerate}
\item\label{ssni:low-step\ssnilabel}
For all #1$S_1, S_2 \in \Low$,
if $S_1 \indist S_2$, $S_1 \step S_1'$, and $S_2 \step S_2'$,
then $S_1' \indist S_2'$;
\item\label{ssni:high-step\ssnilabel}
For all #1$S \not\in \Low$ if $S \step S'$ and $S' \not\in \Low$,
then $S \indist S'$;
\item\label{ssni:return-step\ssnilabel}
For all #1$S_1, S_2 \not\in \Low$,
if $S_1 \indist S_2$, $S_1 \step S_1'$, $S_2 \step S_2'$,
and $S_1', S_2' \in \Low$,
then $S_1' \indist S_2'$.
\end{enumerate}
}
}
\begin{defn}\label{def:SSNI}
A machine semantics is \emph{single-step noninterfering}
with respect to the indistinguishability relation $\indist$
(written SSNI$_{\indist}$) if \ssnibody{}
\end{defn}
Note that SSNI talks about completely arbitrary
states, not just (quasi-)initial ones.

\ifever
\ch{At some point I had the impression that condition 2 is too strong,
  all we need for LLNI/EENI is a weaker triangle property. Is that
  really the case? The triangle property is equivalent by transitivity
  to 2 in the context of proving LLNI.}
\fi



The definition of SSNI is parametric in the indistinguishability
relation used, and it can take some work to find the right relation.
As discussed above, $\indistlow$ is too weak and QuickCheck can
easily find counterexamples to condition~\ref{ssni:return-step}, \EG
by choosing two indistinguishable machine states with
$i = \left[\ii{Return}\ReturnRet{\True}\right]$,
$\pc = \CEPC{0 \labelsym \high}$, and
$s = \left[\HSret{\variation{0}{1}}{\False}{\low}\right]$;
after a single step the two machines have distinguishable $\pc$s
$\CEPC{0 \labelsym \low}$ and $\CEPC{1 \labelsym \low}$, respectively.
On the other hand, treating high states exactly like low states in
the indistinguishability relation is too strong.
In this case QuickCheck finds counterexamples to
condition~\ref{ssni:high-step}, \EG a single machine state with
$i = \left[\begin{array}{l}\ii{Pop}\end{array}\right]$,
$\pc = \CEPC{0 \labelsym \high}$, and
$s = \left[0 \labelsym \low\right]$ steps to a state
with $s = \left[\right]$, which would not be considered indistinguishable.
%
%
These counterexamples
show that indistinguishable high
states can have different \pc{}s and can have completely different
stack frames at the top of the stack.
So all we can require for two high states to be equivalent is
that their memories and instruction memories agree and that the parts
of the stacks below the topmost low return address are equivalent.
This is strong enough to ensure condition~\ref{ssni:return-step}.

\begin{defn}\label{def:indistfull} \
  Machine states $S_1 = \MACHI{\pc_1}{s_1}{m_1}{i_1}$ and
  $S_2 = \MACHI{\pc_2}{s_2}{m_2}{i_2}$ are \emph{indistinguishable
  with respect to whole machine states}, written $S_1 \indistfull S_2$,
  if $m_1 \indist m_2$, $i_1 \indist i_2$, $\labelof{\pc_1} = \labelof{\pc_2}$,
  and additionally
  \begin{itemize}
  \item if $\labelof{\pc_1} = \low$ then $s_1 \indist s_2$ and $\pc_1 \indist \pc_2$, and
  \item if $\labelof{\pc_1} = \high$ then $\cropStack~s_1 \indist
  \cropStack~s_2$. 
  \end{itemize}
\end{defn}
The $\cropStack$ helper function takes a stack and removes elements from
the top until it reaches the first low return address (or until all
elements are removed). (As most of our definitions so far, this
definition is tailored to a 2-element lattice; we will generalize
it to an arbitrary lattice in \autoref{sec:invariant}.)

The fifth column of \autoref{fig:combinedtable-cally}
shows that, even with arbitrary starting states generated completely
naively, SSNI$_{\indistfull}$ performs very well.
If we tweak the weights a bit and additionally observe that since
we only execute the generated
machine for only one step, we can begin with very small states (\EG the
instruction memory can be of size~2), then
we can find all bugs very quickly.
As the last column of \autoref{fig:combinedtable-cally} illustrates,
each bug is found in under 20 milliseconds.
%
(This last optimization is a bit risky, since we need to make
sure that these very small states are still large enough to exercise all bugs we
might have---\EG an instruction memory of size 1 is not
enough to exhibit the \IfcBugValueOrVoidOnReturn{} bug using SSNI.)
Compared to other properties, QuickCheck executes many more tests per
second with SSNI for both generation strategies.

\subsection{Discussion}
\label{sec:strengthening-discussion}

In this section we have seen that strengthening the noninterference
property is a very effective way of improving the
effectiveness of random testing our IFC machine. It is not without
costs, though.
Changing the security property required some expertise and, in the
case of LLNI and SSNI, manual proofs showing that the new property
implies EENI, the baseline security property (see
Appendix~\ref{app:strengthening}).
In the case of \chrev{LLNI and }SSNI
we used additional invariants of our machine
(\EG captured by $\indistfull$)
and finding these invariants 
is
the most creative part of doing a full security proof.
While we could use the counterexamples provided by QuickCheck to guide
our search for the right invariants, for more realistic
machines the process of interpreting the counterexamples and manually
refining the invariants {is} significantly harder than for our very
simple machine \chrev{(see \autoref{sec:regs})}.

The potential users of our techniques will have to choose a point in
the continuum between testing and proving that best matches the
characteristics of their practical application.
At one end, we present ways of testing the original EENI property
without changing it in any way, by putting all the smarts in clever
generation strategies.
At the other end, one can imagine using random testing just as the
first step towards a full proof of a stronger property such as SSNI.
\ifever
\bcp{The rest of the paragraph doesn't flow very well.}%
\asz{Tweaked---what do people think?}%
\fi
For a variant of our simple stack machine,
\citet{PicoCoq2013} did in fact prove recently in Coq
that
SSNI holds, and did not find any bugs that had escaped our testing.
Moreover, we proved in Coq that under reasonable assumptions
SSNI implies LLNI and LLNI implies EENI
(see Appendix~\ref{app:strengthening}).



\section{Shrinking Strategies}
\label{sec:shrinking}

\ifever
\feedback{Johannes: The shrinking discussion requires some knowledge of
  quickcheck to appreciate, I do not understand why a and bk should
  have the same structure and why that would cause the first shrinking
  attempts from bk to fail.}
\fi

The counterexamples presented in this paper are not the initial
randomly generated machine states;
they are the result of QuickCheck shrinking these to
minimal counterexamples. For example, randomly generated counterexamples
to \rn{EENI} for the \IfcBugPushNoTaint{} bug usually consist of
20--40 instructions; the minimal counterexample uses just 4
(see \autoref{fig:counter-basic3}).
In this section we describe the shrinking strategies we used.

\subsection{Shrinking labeled values and instructions}

By default, QuickCheck already implements a shrinking strategy for integers. For
labels, we shrink $\high$ to $\low$, because we prefer to see counterexamples in
which labels are only $\high$ if this is essential to the failure. Values are
shrunk by shrinking either the label or the contents. If we need to shrink
\emph{both} the label and the contents, then this is achieved in two separate
shrinking steps.

We allow any instruction to shrink to \ii{Noop}, which preserves a
counterexample if the instruction was unnecessary; or to \ii{Halt}, which
preserves a counterexample if the bug had already manifested by the time that
instruction was reached. To avoid an infinite shrinking loop, we do not
shrink \ii{Noop} at all, while \ii{Halt} can shrink only
to \ii{Noop}. Instructions of the form \ii{Push \V{n}{\lab}}
are also shrunk by
shrinking \(\V{n}{\lab}\).
Finally, instructions of the form $\ii{Call}~n~n'$ are shrunk
by shrinking \(n\) or \(n'\), or by being replaced with \(\ii{Jump}\).

\subsection{Shrinking machine states}

Machine states contain a data memory, a stack, an instruction memory
and the \pc{}.  For data memories, we can simply shrink the elements
using the techniques of the previous subsection. In addition, we allow
shrinking to remove arbitrary elements of the data memory completely.
However, the first element that we try to remove is the last one:
removing other elements changes all subsequent memory addresses,
potentialy invalidating the counterexample. Stacks can be shrunk
similarly: we can shrink their data elements or remove them
completely. We need to be extra clever in shrinking return
addresses---otherwise, it is very easy to obtain crashing states. This
is elaborated in the next subsection.

In the case of the instruction memory, we only try to
remove \(\ii{Noop}\) instructions, since removing other instructions
is likely to change the stack or the control flow fairly drastically,
and is thus likely to invalidate any counterexample. Other
instructions can still be removed in two stages, by first shrinking
them to a \ii{Noop}.

Finally, we choose not to shrink the \pc{}. Generation by execution
works by generating valid instructions starting from the
initial \pc{}. Shrinking its address will most likely lead to
immediate failure. One strategy we considered is {\em shrinking by
execution}, where we shrink by taking a step in the machine. However,
we didn't get a lot of benefit from such an approach. Even worse, if
the indistinguishability relation is too coarse grained, then
shrinking by execution can lead our states past the point where they
become distinct, but are still considered equivalent; such a
counterexample is not useful for debugging!

\subsection{Shrinking variations}

One difficulty that arises when shrinking noninterference counterexamples
is that the test cases must be pairs of \emph{indistinguishable}
machines. Shrinking each machine state independently will most likely
yield distinguishable pairs, which are invalid test cases, since they
fail to satisfy the precondition of the property we are testing. In
order to shrink effectively, we need to shrink both states of a
variation \emph{simultaneously}, and in the same way.

For instance, if we shrink one machine state by deleting a \ii{Noop} in the
middle of its instruction memory, then we must delete the same instruction in
the corresponding variation. %
\ifever
\asz{Should we explain why?  This replaced the less-interesting sentence ``For
instance, if we shrink one machine state by removing the last element of its
memory, then we should also remove the last element of the memory of the other
state in the variation.''}%
\fi
Similarly, if a
particular element gets shrunk in a memory location, then the same
location should be shrunk in the other state of the variation, and
only in ways that produce indistinguishable states. We have
implemented all of the shrinking strategies described above as
operations on \emph{pairs} of indistinguishable states, and ensured that they
generate only shrinking candidates that are also indistinguishable.

When we use the full state equivalence $\indistfull$, we can shrink stacks slightly
differently: we only need to synchronize shrinking steps on
the \emph{low} parts of the stacks. Since the equivalence relation
ignores the high half of the stacks, we are free to shrink those parts
of the two states independently, provided that high return addresses
don't get transformed into low ones.

\subsection{Optimizing shrinking}

\ifever
\feedback{Ben: The "Optimizing Shrinking" section sounds mostly like an
  application of existing generic techniques (smart shrinking, double
  shrinking), and the more novel aspects are left un-described: What
  sorts of programs needed double-shrinking? How close to minimal were
  the non-minimal counterexamples after various shrinking techniques
  were applied? Were they 10\% too big or 1000\% too big? What sorts
  of counterexamples remained non-minimal after all the hammers were
  pulled out?}
\ch{TODO: for now see if we can answer any of these questions without
  putting significant effort into obtaining numbers about shrinking}
\fi

We applied a number of optimizations to make the shrinking process
faster and more effective. One way we sped up shrinking was by
turning on QuickCheck's ``smart shrinking,'' which optimizes the order
in which shrinking candidates are tried. If a counterexample $a$ can
be shrunk to any $b_i$, but the first $k$ of these are not
counterexamples, then it is likely that the first $k$ shrinking
candidates for $b_{k+1}$ will not be counterexamples either, because
$a$ and $b_{k+1}$ are likely to be similar in structure and so to have
similar lists of shrinking candidates. Smart shrinking just changes
the order in which these candidates are tried: it defers the first
$k$ shrinking candidates for $b_{k+1}$ until after more likely
ones have been tried. This sped up shrinking dramatically in our
tests.
\ifever\ch{It would be nice to show some numbers}\fi

We also observed that many reported counterexamples contained \ii{Noop}
instructions---in some cases many of them---even though we
implemented \ii{Noop} removal as a shrinking step. On examining these
counterexamples, we discovered that they could not be shrunk because
removing a \ii{Noop} changes the addresses of subsequent instructions,
at least one of which was the target of a \ii{Jump} or \ii{Call}
instruction. So to preserve the behavior of the counterexample, we
needed to remove the \ii{Noop} instruction \emph{and adjust the target
of a control transfer} in the same shrinking step. Since control
transfer targets are taken off the stack, and such addresses can be
generated during the test in many different ways, we simply
allowed \ii{Noop} removal to be combined with any other shrinking
step---which might, for example, decrement any integer on the initial
stack, or any integer stored in the initial memory, or any constant in
a \ii{Push} instruction. This combined shrinking step was much more 
effective in removing unnecessary \ii{Noop}s.

Occasionally, we observed shrunk counterexamples containing two or more
unnecessary \ii{Noop}s, but where removing just one \ii{Noop} led to a
non-counterexample. We therefore used QuickCheck's \emph{double shrinking}, which allows a counterexample to shrink in two
steps to another counterexample, even if the intermediate value is not
a counterexample. With this technique, QuickCheck could remove all
unnecessary \ii{Noop}s, albeit at a cost in shrinking time.

We also observed that some reported test cases contained unnecessary
\emph{sequences} of instructions, which could be elided together, but
not one by one. We added a shrinking step that can replace any two
instructions by \ii{Noop}s simultaneously (and thus, thanks to double shrinking,
up to four), which solved this problem.

With this combination of methods, almost all counterexamples we found
shrink to minimal ones, from which no instruction, stack element,
or memory element could be removed without invalidating the
counterexample.

\section{Information-Flow Register Machine}
\label{sec:regs}

\ch{\bf TODO: make sure that all the rules in this section
  match both the Haskell and the Coq implementations.
  Same thing for the definitions (\EG indistinguishability)}

\rjmh{I did feel that the connection between the design presented in
  that section, and testing, isn't made as strongly as it could
  be. The first four sections present the design, which is quite
  intricate, without really any reference to testing--and then the
  fifth presents the test results. The claim in the introduction that
  we "use testing to discover the sophisticated invariants" isn't
  really justified in the current text. It begs the question: HOW did
  we use testing to discover them? It would be nice to show that,
  perhaps with counterexamples to some earlier incorrect-but-plausible
  formulations. There is a counterexample presented in section 8.2,
  but it isn't stated whether it was found by QuickCheck or not, or
  even whether it can be found by QuickCheck.}
\ch{I think we all are aware of this and think it would be nice to do
  better. We just didn't have enough time and/or manpower to make it
  happen for the JFP submission, but we can still work on this for
  the final version.}
\bcp{Even without restructuring the section, we can add a bit more text here
  about how we used testing to discover invariants.  I think even a short
  paragraph would help a lot.}
\ch{Agreed, we that testing was helped us find the right rules and
  invariants. For the invariants, did we get the well-stampedness
  invariant wrong in any interesting ways so that we can fix
  definitions and iterate the testing?}
\ch{I'm afraid we have no time for this once again}

\sloppy

We demonstrate the scalability of the techniques above by studying a
more realistic information-flow register machine.
%
%
Beyond registers (\autoref{sec:registers}),
this machine includes advanced features such as
first-class public labels (\autoref{sec:labels})
and dynamically allocated memory with mutable labels
(\autoref{sec:memory}).
The combination of these features makes the design of
{sound} IFC rules highly non-trivial, and thus discovering flaws
early by testing even more crucial.
%
The counterexamples produced by testing have guided us
in designing a novel and highly permissive
 flow-sensitive dynamic enforcement mechanism
and in the discovery of the sophisticated invariants needed for
\ifproof\chrev{the final}\fi{} noninterference
proof (\autoref{sec:invariant}).
Most importantly for the purpose of this paper,
we experimentally evaluate the scalability
of our best generation strategy (generation by execution) and
readily falsifiable formulations of noninterference by testing this more
complex machine (\autoref{sec:regs-testing-results}).

\fussy

\subsection{The core of the register machine}\label{sec:registers}

The core instructions of the new machine are very similar to those of
the simple stack machine from the previous sections:
\[
\begin{array}{l@{\hspace{1ex}}l}
\mathit{Instr} \mathrel{::=} &
  \ii{Put}~n~r_d
  \;|\;
  \ii{Mov}~r_s~r_d
  \;|\;
  \ii{Load}~r_p~r_d
  \;|\;
  \ii{Store}~r_p~r_s
  \;|\;
  \ii{Add}~r_1~r_2~r_d
  \;|\;
  \ii{Mult}~r_1~r_2~r_d
  \;|\;
  \\
&  \ii{Noop}
  \;|\;
  \ii{Halt}
  \;|\;
 \ii{Jump}~r
  \;|\;
  \ii{BranchNZ}~n~r
  \;|\;
  \ii{Call}~r_1~r_2~r_3
  \;|\;
  \ii{Return}
\end{array}
\]
The main difference is that the instructions now take their arguments
from and store their result into registers.
We use the meta-variable $r$ to range over a finite set of register identifiers.
The register machine has no $\ii{Pop}$ instruction and the
$\ii{Push}~\V{n}{\lab}$ instruction of the stack machine is replaced by
$\ii{Put}~n~r_d$, which stores the integer constant $n$ into
the destination register $r_d$.
\ifever
\chfoot{\ii{Push} took a labeled integer as argument, while
  \ii{Put} only takes an integer \ldots can this be rationally explained
  or should be sweep it under the rug?}
\fi
$\ii{Mov}~r_s~r_d$ copies the contents of register $r_s$ into register
$r_d$.
$\ii{Load}~r_p~r_d$ and $\ii{Store}~r_p~r_s$ take the pointer from
register $r_p$, and load the result into $r_d$ or store the
value of $r_s$ into memory.
On top of the familiar \ii{Add} we also add a multiplication
instruction \ii{Mult}.
\ii{Noop}, \ii{Halt}, and \ii{Jump} work as before, and we
additionally add a $\ii{BranchNZ}~n~r$ (``branch not zero'') instruction
that performs a conditional relative jump by adding the integer $n$ to
the \pc{} if the register $r$ contains a non-zero integer value;
otherwise the \pc{} is simply incremented.

While most instructions of the new machine only work with registers, the
\ii{Call} and \ii{Return} instructions also use a (protected) call stack.
We hard-code a simple calling convention in which the values of all
registers are passed from the callee to the caller, but only one
register is used for passing back a result value, while all the other
registers are automatically restored to the values before the call.
Similarly to the simple stack machine, where to enforce
noninterference it was necessary to specify the number of returned
values on calls, on the register machine the register in which
the result is returned and the label of the result are both specified
on \ii{Call}, saved on the stack, and used on the corresponding
\ii{Return}.

The states of the register machine have the form
$\RMACHI{\V{\pc}{\lab_\pc}}{\ii{rf}}{\ii{cs}}{m}{i}$ and include the register file
($\ii{rf}$, mapping register identifiers and to their values) and the
call stack ($\ii{cs}$, a list of stack frames). The labeled $\pc$, the data
memory $m$, and the instruction memory $i$ (omitted by convention
below) are familiar from the stack machine. The stepping rules
for most instructions above are simple; for instance:
%
%
\infrule[Mult]
  {i(\pc) = \ii{Mult}~r_1~r_2~r_d \andalso
   \ii{rf\,}[r_1] = \V{n_1}{\lab_1} \andalso
   \ii{rf\,}[r_2] = \V{n_2}{\lab_2} \\
   \ii{rf\,}' = \ii{rf\,}[r_d := \V{(n_1 \times n_2)}{(\lab_1\mathord\join \lab_2)}]}
  {\RMACH{\V{\pc}{\lab_\pc}}{\ii{rf}}{\ii{cs}}{m} \step
   \RMACH{\V{(\pc\mathord{+}1)}{\lab_\pc}}{\ii{rf\,}'}{\ii{cs}}{m}}
The interaction between public labels and \ii{Call} and \ii{Return},
however, is complex; we discuss it in detail in \autoref{sec:labels}.
The precise structure of the data memory and the rules for \ii{Load}
and \ii{Store} are discussed in \autoref{sec:memory}.







\subsection{First-class public labels}
\label{sec:labels}

First-class public labels are an important feature of several recent
IFC systems for functional programming languages~\cite{Exceptional,
  StefanRMM11}.
They support the development of realistic
applications in which new 
principals and labels are created dynamically~\cite{giffin:2012:hails},
and they are a key ingredient in recently proposed mechanisms for
{soundly} recovering from IFC violations~\cite{Exceptional,
  StefanRMM12jfp}.
While for simplicity we consider neither dynamic labels nor recoverable
exceptions, our register machine does have first-class public labels.
Enforcing noninterference for public labels is highly non-trivial,
especially in the presence of varying memory labels
(\autoref{sec:memory}).

Even before that, adding a $\ii{LabelOf}~r_s~r_d$ instruction that puts in
$r_d$ the label of the value in $r_s$
as a public first-class value
is unsound for a label lattice with at least three elements and a standard
mechanism for restoring the \pc{} on control flow merge point (\EG the
\ii{Call}s and \ii{Return}s of the stack machine
in~\autoref{sec:calls-and-returns}).
In a functional language we would write the counterexample as follows:
\[
\ii{LabelOf}~(\textsf{if }m\textsf{ then }\V{0}{\midl}
                           \textsf{ else }\V{0}{\high})
\]
This encodes the secret bit $m$ protected by label $\midl$ (where
$\low \flowsto \midl \flowsto \high$) by varying the {\em label} of
the conditional's result ($\midl$ on the if branch and $\high$ on the
else one), and then uses \ii{LabelOf} to expose that label.
Note that the counterexample involves two
  different labels (in our case $\midl$ and $\high$) that are higher
  or equal in lattice order ($\flowsto$) than the label of a secret
  ($\midl$); a two-point lattice is not rich enough.

To express the same counterexample on our low-level machine,
consider two indistinguishable states
$\RMACH{\V{0}{\low}}{\ii{rf}_1}{[\,]}{[\,]}$ and 
$\RMACH{\V{0}{\low}}{\ii{rf}_2}{[\,]}{[\,]}$ differing only in the value of
the $r_0$ register, which contains a secret bit labeled $\midl$:
$\ii{rf}_1[r_0] = \V{0}{\midl}$ while
$\ii{rf}_2[r_0] = \V{1}{\midl}$.
We use three more registers with the same value in both states:
$\ii{rf}_1[r_1] = \ii{rf}_2[r_1] = \V{3}{\low}$,
$\ii{rf}_1[r_2] = \ii{rf}_2[r_2] = \V{0}{\midl}$, and
$\ii{rf}_1[r_3] = \ii{rf}_2[r_3] = \V{0}{\high}$.
%
The instruction memory of both machines contains the following
program:
%
%
\[ [
\ii{Call}~r_1~r_3,\
\ii{LabelOf}~r_3~r_0,\
\ii{Halt},\
\ii{BranchNZ}~2~r_0,\
\ii{Mov}~r_2~r_3,\
\ii{Return}
] \]
The \ii{Call} transfers control to the $\ii{BranchNZ}$ instruction and
specifies $r_3$ as the return register.
Together with the \ii{Return}, this ensures that the $\pc$ label is
restored to $\low$ after branching on the secret bit via $\ii{BranchNZ}$.
The $\ii{BranchNZ}$ ensures that the $\ii{Mov}$ is executed
when $r_0$ is $\V{0}{\midl}$ (the first execution) and
skipped when $r_0$ is $\V{1}{\midl}$ (the second one).
The $\pc$ is raised to $\midl$ on the $\ii{BranchNZ}$, but that cannot
have any effect on the \ii{Mov} since the labels of the values in
$r_2$ and $r_3$ are anyway higher or equal than $\midl$.
Similarly, the $\ii{Return}$ could potentially join the $\pc$ label
$\midl$ to the label of the returned register $r_3$, but that would
again have no effect.
After the \ii{Return}, the \pc{} is again labeled $\low$ and the
register $r_3$ stores $\V{0}{\midl}$ in the first execution and
$\V{0}{\high}$ in the second.
Executing $\ii{LabelOf}$ exposes this label difference to the value level.
In the end, the first machine halts with $\V{\midl}{\low}$ in register
$r_3$, while the second one halts with $\V{\high}{\low}$ in $r_3$, a
distinguishable difference.
\ifever
\chfoot{Should try to run this example in Haskell (and if the
  infrastructure is in place at some point produce a nice figure
  containing all the intermediate states in the execution).}
\fi

Following \citet{Exceptional} and
\citet{StefanRMM11}, we can solve this problem by separating the choice of
label (which needs to 
be done in a low context) from the computation of the labeled data
(which happens in a high context).
Concretely, we require the programmer to specify the label of the
result of each procedure as part of the \ii{Call}
(in our simple instruction language the first-class label is first put
 in a register with \ii{PutLabel}, and only then passed to the \ii{Call}).
In the example above the \ii{Return} succeeds on both branches
only if the \ii{Call} is
annotated with $\high$, \IE a label that is more secure than the label
of the result register on either branch:
\[ [
\ii{PutLabel}~\high~r_4,\
\ii{Call}~r_1~r_3~r_4,\
\ii{LabelOf}~r_3~r_0,\
\ii{Halt},\
\ii{BranchNZ}~2~r_0,\
\ii{Mov}~r_2~r_3,\
\ii{Return}
] \]
Regardless of which branch is chosen, the procedure will return
$0@\high$ in register $r_3$, thus preventing $m$ from being leaked via
the labels.
Concretely, the new \ii{PutLabel} instruction loads the label $\high$
into register $r_4$, and then this register is passed as a third
argument to the \ii{Call}.
The \ii{Call} saves the $\high$ label in the new stack frame, and the
\ii{Return} checks that the $\pc{}$ label and the label of the result
in $r_3$ are both below or equal to $\high$ (which is trivially true
on both branches) and then {\em raises} the label of the result to $\high$.
The check performed on \ii{Return} would fail
at least on one of the branches 
if the \ii{Call} were wrongly annotated with $\low$ or $\midl$.

The rest of this subsection explains the part of our machine
definition concerning
first-class public labels, culminating with the stepping rules for
\ii{Call} and \ii{Return}.
For the sake of brevity,
in the rest of this and the following subsections, as opposed to
\autoref{sec:basic} and \autoref{sec:cally}, we \chrev{do not}
list wrong rules and the QuickCheck-found counterexamples that
have guided our search for the right formulations.
\ifever
\bcpfoot{Like John, I'm losing the testing story at
  this point.  How much extra space / work would it take to actually show
  the wrong version of the machine, let QC {\em find} the counterexample
  instead of just showing it, and then fix the machine?\ch{a lot and
  clearly more than what we have now (Aug 31)}\bcp{OK, but if we've going
  ahead with the current structure, then at least the reader needs to be
  explicitly alerted to the change of approach.  We could say, for instance,
  that we switch to a less ``socratic'' style for brevity, but that in fact
  we used testing heavily in formulating the machine we're about to
  describe.  (We do say most of this now.)  But a little rewording right
  here would also help: opening with ``we explain our formalization...''
  makes it seem as though the formalization itself is the main point.}
\ch{Tried to add something above.}}
\fi
For testing, we consider labels drawn from a four-element diamond
lattice:
\[
\lab \mathrel{::=} 
  \low
  \;|\;
  \mone
  \;|\;
  \mtwo
  \;|\;
  \high
\]
where $\low \flowsto \mone$, $\low \flowsto \mtwo$,
$\mone \flowsto \high$, and $\mtwo \flowsto \high$.
The labels $M_1$ and $M_2$ are incomparable.
This four-element lattice was rich enough for finding all bugs
introduced in the experiments from \autoref{sec:regs-testing-results}.
With this richer lattice our definition of ``low'' and ``high'' becomes
relative to an arbitrary observer label $\lab$: we call $\lab_1$ low
with respect to $\lab$ if $\lab_1 \flowsto \lab$ and high otherwise.
\chrev{The noninterference proof from \autoref{sec:invariant} is
  parameterized over an arbitrary finite lattice.}


The register machine works with labeled values $\V{v}{\lab}$,
where $v$ is an integer $n$ or a first-class label $\lab$ or a
pointer (more on pointers in \autoref{sec:memory}).
Besides \ii{PutLabel} and \ii{LabelOf}, which we have seen above, we
add three more new instructions that work with first-class labels:
\[
\mathit{Instr} \mathrel{::=}
\ldots
  \;|\;
  \ii{PutLabel}~\lab~r_d
  \;|\;
  \ii{LabelOf}~r_s~r_d
  \;|\;
  \ii{PcLabel}~r_d
  \;|\;
  \ii{Join}~r_1~r_2~r_d
  \;|\;
  \ii{FlowsTo}~r_1~r_2~r_d
\]
\ii{PcLabel} returns the label of the \pc{}, while \ii{Join} and
\ii{FlowsTo} compute $\join$ and $\flowsto$ on
first-class labels.
The stepping rules for these new instructions are all very simple:
\infrule[PutLabel]
  {i(\pc) = \ii{PutLabel}~\lab~r_d \andalso
   \ii{rf\,}' = \ii{rf\,}[r_d := \V{\lab}{\low}]}
  {\RMACH{\V{\pc}{\lab_\pc}}{\ii{rf}}{\ii{cs}}{m} \step
   \RMACH{\V{(\pc\mathord{+}1)}{\lab_\pc}}{\ii{rf\,}'}{\ii{cs}}{m}}
\infrule[LabelOf]
  {i(\pc) = \ii{LabelOf}~r_s~r_d \andalso
   \ii{rf\,}[r_s] = \V{n}{\lab} \andalso
   \ii{rf\,}' = \ii{rf\,}[r_d := \V{\lab}{\low}]}
  {\RMACH{\V{\pc}{\lab_\pc}}{\ii{rf}}{\ii{cs}}{m} \step
   \RMACH{\V{(\pc\mathord{+}1)}{\lab_\pc}}{\ii{rf\,}'}{\ii{cs}}{m}}
\infrule[PcLabel]
  {i(\pc) = \ii{PcLabel}~r_d \andalso
   \ii{rf\,}' = \ii{rf\,}[r_d := \V{\lab_\pc}{\low}]}
  {\RMACH{\V{\pc}{\lab_\pc}}{\ii{rf}}{\ii{cs}}{m} \step
   \RMACH{\V{(\pc\mathord{+}1)}{\lab_\pc}}{\ii{rf\,}'}{\ii{cs}}{m}}
\infrule[Join]
  {i(\pc) = \ii{Join}~r_1~r_2~r_d \andalso
   \ii{rf\,}[r_1] = \V{\lab_1}{\lab_1'} \andalso
   \ii{rf\,}[r_2] = \V{\lab_2}{\lab_2'} \\
   \ii{rf\,}' = \ii{rf\,}[r_d := \V{(\lab_1 \join \lab_2)}{(\lab_1' \join \lab_2')}]}
  {\RMACH{\V{\pc}{\lab_\pc}}{\ii{rf}}{\ii{cs}}{m} \step
   \RMACH{\V{(\pc\mathord{+}1)}{\lab_\pc}}{\ii{rf\,}'}{\ii{cs}}{m}}
\infrule[FlowsTo]
  {i(\pc) = \ii{FlowsTo}~r_1~r_2~r_d \andalso
   \ii{rf\,}[r_1] = \V{\lab_1}{\lab_1'} \andalso
   \ii{rf\,}[r_2] = \V{\lab_2}{\lab_2'} \\
   n = \text{if } \lab_1 \flowsto \lab_2 \text{ then } 1 \text{ else } 0 \andalso
   \ii{rf\,}' = \ii{rf\,}[r_d := \V{n}{(\lab_1' \join \lab_2')}]}
  {\RMACH{\V{\pc}{\lab_\pc}}{\ii{rf}}{\ii{cs}}{m} \step
   \RMACH{\V{(\pc\mathord{+}1)}{\lab_\pc}}{\ii{rf\,}'}{\ii{cs}}{m}}

Note that result of the \ii{LabelOf} instruction is a ``label value''
that is itself labeled $\low$.  So in a low context the labels
of values in registers---even labels on secret data---are public
information.

The stepping rules for \ii{Call} and \ii{Return} are more complex.
The $\ii{Call}~r_1~r_2~r_3$ instruction has 3 register arguments:
$r_1$ stores the address of a procedure,
$r_2$ is marked as a result register and is not restored on return, and
$r_3$ stores a first-class label that is
used to label the result value on return.
On executing $\ii{Call}~r_1~r_2~r_3$ the return address ($\pc\mathord{+}1$),
the contents of the whole register file (\ii{rf}),
the return register identifier $r_2$,
and the label in $r_3$ ($\lab$),
are all saved in a new stack frame,
and control is passed to the address in $r_1$ ($n$):
\infrule[Call]
  {i(\pc) = \ii{Call}~r_1~r_2~r_3 \andalso
   \ii{rf\,}[r_1] = \V{n}{\lab_n} \andalso
   \ii{rf\,}[r_3] = \V{\lab}{\lab'}}
  {\RMACH{\V{\pc}{\lab_\pc}}{\ii{rf}}{\ii{cs}}{m} \step
   \RMACH{\V{n}{(\lab_\pc \join \lab_n)}}
     {\ii{rf}}{(\V{(\pc\mathord{+}1)}{(\lab_\pc\join\lab')},\ii{rf},r_2,\lab) : \ii{cs}}{m}}
As was the case in \autoref{sec:calls-and-returns}, the address of the called
procedure ($n$) can be influenced by secrets, so we join its label
($\lab_n$) to the $\pc$ label ($\lab_\pc$).
Finally, since the label used to annotate the call ($\lab$) is first
class, it has itself a protecting label ($\lab'$), which we join to
the label of the return address ($\pc\mathord{+}1$).

On a \ii{Return} the top frame on the call stack is popped, the saved
\pc{} and register file are restored to their previous values, with
the exception of the return register whose content is preserved,
so that a value is passed from the callee to the caller in this register.
\infrule[Return]
  {i(\pc) = \ii{Return} \andalso
  \ii{rf\,}[r] = \V{v}{\lab} \andalso
  \lab \join \lab_\pc \flowsto \lab' \join \lab_\pc' \andalso
  \ii{rf\,}'' = \ii{rf\,}'[r := \V{v}{\lab'}]}
  {\RMACH{\V{\pc}{\lab_\pc}}{\ii{rf}}
      {(\V{n}{\lab_\pc'},\ii{rf\,}',r,\lab') : \ii{cs}}{m} \step
   \RMACH{\V{n}{\lab_\pc'}}{\ii{rf\,}''}{\ii{cs}}{m}}
The label check considers the total protection of a
value as the join of its explicit label and the \pc{}
label~\cite{Exceptional}.
This check ensures that the total protection of the returned
value ($v$) after the \ii{Return} (label $\lab' \join \lab_\pc'$) is at
least as strong as its protection before the return ($\lab \join
\lab_\pc$); in other words, it prevents \ii{Return} from declassifying
the result, which would break noninterference.
%

\subsection{Permissive flow-sensitive memory updates}
\label{sec:memory}



Testing has helped us more easily explore the intricate space of IFC
mechanisms and design a new one that addresses a current research
challenge in an interesting way.
The challenge we address is allowing the labels of the values in
memory to vary at runtime yet still be observable.\ifever\chfoot{Terminology:
  public labels vs observable labels}\fi{}
IFC systems that allow labels to change during execution are usually
called {\em flow-sensitive}, and are generally more permissive than
flow-insensitive systems that require labels to be fixed once and
for all.
Devising flow-sensitive dynamic IFC systems is, however, challenging.
%
The first solutions
proposed in the literature used static analysis to soundly approximate
the effects of branches not taken on IFC labels~\cite{GuernicBJS06,
  Guernic07, RussoS10}. Later, sound purely-dynamic
flow-sensitive monitors were proposed, based on dynamic checks called
no-sensitive-upgrades~\cite{zdancewic02:thesis, AustinF09} (which we
used for the stack machine in \autoref{sec:calls-and-returns})
and permissive upgrades~\cite{AustinF:2010, BichhawatRGH14b}.
These checks are, however, not sound when labels are observable (\EG
via the \ii{LabelOf} instruction we introduced in
\autoref{sec:labels}); intuitively they allow secret information to
leak into the labels of the values in memory and ensure soundness by
preventing these labels from being observed.\ifever\chfoot{This is my
  intuition but should make sure this is correct, by having some
  counterexamples at hand.}\fi{}
We are aware of only one flow-sensitive IFC system featuring
public (\IE observable) labels: the one recently proposed by \citet{BuirasSR14}.
Our solution is similar but in many ways more permissive than the one
by \citet{BuirasSR14}; on the other hand, their technique extends well
to concurrency, while here we only study a sequential setting.
The precise connection to \citet{BuirasSR14} is discussed throughout
this subsection.

The main idea is simple: we associate a label with each memory block
and this label protects not only the values inside, but also their
individual labels (\citet{BuirasSR14} don't store values in blocks,
but use a similar concept called ``the label on the reference label'').
This block label is chosen by the programmer at allocation time and is
fixed throughout execution, while the label on the values in the block
can vary more or less arbitrarily.
The only restriction we impose is that label updates can only happen
in contexts that are less classified than the label of the memory
block containing the updated label.
The rest of this subsection presents the technical details of our solution.

The register machine features a block-based memory
model~\cite{PicoCoq2013, Leroy-Blazy-memory-model,
  Leroy-Appel-Blazy-Stewart-memory-v2}.
As mentioned above, values include integers, first-class labels, and pointers:
\[
v \mathrel{::=}
  n
  \;|\;
  \lab
  \;|\;
  (b, o)
\]
A pointer is a pair $(b, o)$ of a block identifier $b$
and an integer offset $o$.
The memory $m$ is a partial function from a block identifier to a labeled
list of labeled values $\V{\ii{vs}}{\lab_b}$; we call $\lab_b$ the block label.
%
The stepping rule for $\ii{Load}~r_p~r_d$ looks up the value of $r_p$
in the register file, and proceeds only if it is a pointer $(b,o)$
labeled $\lab_p$:
\infrule[Load]
  {i(\pc) = \ii{Load}~r_p~r_d \andalso
   \ii{rf\,}[r_p] = \V{(b, o)}{\lab_p} \\
   m[b] = \V{\ii{vs}}{\lab_b} \andalso
   \ii{vs}[o] = \V{v}{\lab_v} \andalso
   \ii{rf\,}' = \ii{rf\,}[r_d := \V{v}{\lab_v}]}
  {\RMACH{\V{\pc}{\lab_\pc}}{\ii{rf}}{\ii{cs}}{m} \step
   \RMACH{\V{(\pc\mathord{+}1)}{(\lab_\pc \join \lab_p \join \lab_b)}}
         {\ii{rf\,}'}{\ii{cs}}{m}}
It looks up the block identifier $b$ in the memory $m$ and if the
block is allocated it obtains a list of values \ii{vs} labeled 
by the block label $\lab_b$.
The result of the \ii{Load} is a labeled value $\V{v}{\lab_v}$
obtained by looking up at offset $o$ in \ii{vs}.
The most interesting part is that the resulting $\pc$ label is the
join of the previous $\pc$ label $\lab_\pc$, the pointer label 
$\lab_p$, and the block label $\lab_b$.
Intuitively, before the load, the labels $\lab_b$ and $\lab_p$ protect
the value $v$ as well as its label $\lab_v$.
After the load, we could have protected the value $v$ by joining
$\lab_b$ and $\lab_p$ to $\lab_v$ instead of the \pc{} label and that
would have been more permissive.
However, this would have left the label $\lab_v$ unprotected, and
directly accessible via $\ii{LabelOf}$, breaking
noninterference.

The stepping rule for \ii{Store} takes a labeled value $\V{v}{\lab_v}$
and writes it to memory, overwriting the previous value at that
location as well as its label.
\infrule[Store]
  {i(\pc) = \ii{Store}~r_p~r_s \andalso
   \ii{rf\,}[r_p] = \V{(b, o)}{\lab_p} \andalso
   \ii{rf\,}[r_s] = \V{v}{\lab_v} \\
   m[b] = \V{\ii{vs}}{\lab_b} \quad
   (\lab_\pc \join \lab_p) \flowsto \lab_b \quad
   \ii{vs}' = \ii{vs}[o := \V{v}{\lab_v}] \quad
   m' = m[b := \V{\ii{vs}'}{\lab_b}]}
  {\RMACH{\V{\pc}{\lab_\pc}}{\ii{rf}}{\ii{cs}}{m} \step
   \RMACH{\V{(\pc\mathord{+}1)}{\lab_\pc}}{\ii{rf}}{\ii{cs}}{m'}}
Because the previous value is overwritten its label doesn't need to
be related in any way with the label of the new value.
This allows for {\em arbitrary} label changes in memory and is thus
more permissive than previous work based on upgrade operations that
can only raise the label of a value in
memory~\cite{BuirasSR14,HedinS12}.
The label check $(\lab_\pc \join \lab_p) \flowsto \lab_b$ ensures that
the label $\lab_b$ that will protect $\V{v}{\lab_v}$ after the store
is high enough to prevent revealing information about the context in
which or the pointer through which this store happened.
This ensures that no program can branch on a secret and based on this
change a labeled value in a memory block with a public block label, since
this would be observable via \ii{Load} as soon as the branching ends
and the \pc{} label is restored.
Similarly, this ensures that no program can vary a pointer based on
secrets and then use that pointer to do a store to a block with a
public block label, since that block can potentially also be
accessible via public pointers that can observe the stored value or
its label.
This is analogous to one of the checks performed for the upgrade
operation of \citet{BuirasSR14} (upgrade is further discussed below);
perhaps surprisingly this it the {\em only} check we need for our
\ii{Store} instruction.

Beyond \ii{Load} and \ii{Store} we have 8 other instructions that deal
with pointers and memory:
\[
\begin{array}{l@{\hspace{1ex}}l}
\mathit{Instr} \mathrel{::=} \ldots &
  \;|\;
  \ii{Alloc}~r_n~r_l~r_d
  \;|\;
  \ii{Write}~r_p~r_s
  \;|\;
  \ii{Upgrade}~r_p~r_l
  \;|\;
  \ii{Eq}~r_1~r_2~r_d
  \;|\;
  \ii{GetOffset}~r_p~r_d
  \\
& \;|\;
  \ii{SetOffset}~r_p~r_o~r_d
  \;|\;
  \ii{GetBlockSize}~r_p~r_d
  \;|\;
  \ii{GetBlockLabel}~r_p~r_d
\end{array}
\]

The $\ii{Alloc}~r_n~r_l~r_d$ instruction allocates a fresh block of
size $r_n$ with block label $r_l$ and stores in $r_d$ a pointer to the first
position in this block. The block is initially filled with $\V{0}{\low}$:
\infrule[Alloc]
  {i(\pc) = \ii{Alloc}~r_n~r_l~r_d \andalso
   \ii{rf\,}[r_n] = \V{n}{\lab_n} \andalso
   n > 0 \andalso
   \ii{rf\,}[r_l] = \V{\lab}{\lab'} \\
   \ii{fresh}~m~(\lab_\pc \join \lab_n \join \lab') = b \andalso
   m' = m[b :=
          \V{[\V{0}{\low},\V{0}{\low},\ldots,\V{0}{\low}]}{\lab}] \\
   \ii{rf\,}' = \ii{rf\,}[r_d := \V{(b,0)}{(\lab_n \join \lab')}]}
  {\RMACH{\V{\pc}{\lab_\pc}}{\ii{rf}}{\ii{cs}}{m} \step
   \RMACH{\V{(\pc\mathord{+}1)}{\lab_\pc}}{\ii{rf\,}'}{\ii{cs}}{m'}}
The returned pointer is protected by both $\lab_n$, the label of the
requested block size, and by $\lab'$, the label of the requested
block label.
If the requested block size is positive and there are still blocks left,
our \ii{Alloc} rule succeeds; in particular the block label
$\lab$ can be chosen {\em arbitrarily}.
This allows us to allocate a low block in high context, knowing that
at the end of the high context access to these blocks will only be
possible through high pointers; this invariant is a cornerstone of
\ifproof\chrev{our}\fi{} noninterference proof
(\autoref{sec:invariant}).
This is more permissive than the reference allocation rule of
\citet{BuirasSR14}, which can only use ``the current label'' (roughly
analogous to our \pc{} label) as ``the label on the reference label''
(analogous to our block label).
We return to the $\ii{fresh}~m~\ldots=b$ condition in \autoref{sec:invariant}.

Our \ii{Store} instruction arbitrarily changes the label of the
overwritten value. Inspired by \citet{BuirasSR14} we additionally
provide a \ii{Write} instruction that behaves the same as \ii{Store},
just that it keeps the label of the overwritten value unchanged:
\infrule[Write]
  {i(\pc) = \ii{Write}~r_p~r_s \quad
   \ii{rf\,}[r_p] = \V{(b, o)}{\lab_p} \quad
   \ii{rf\,}[r_s] = \V{v}{\lab_v} \quad
   m[b] = \V{\ii{vs}}{\lab_b} \quad
   \ii{vs}[o] {=} \V{v'}{\lab_v'} \\
   (\lab_\pc \join \lab_p \join \lab_v) \flowsto (\lab_b \join \lab_v') \quad
   \ii{vs}' {=} \ii{vs}[o := \V{v}{\lab_v'}] \quad
   m' {=} m[b := \V{\ii{vs}'}{\lab_b}]}
  {\RMACH{\V{\pc}{\lab_\pc}}{\ii{rf}}{\ii{cs}}{m} \step
   \RMACH{\V{(\pc\mathord{+}1)}{\lab_\pc}}{\ii{rf}}{\ii{cs}}{m'}}
The label check
is analogous to the one of \citet{BuirasSR14}; it can be broken into
two parts:
The first part,
$(\lab_\pc \join \lab_p) \flowsto (\lab_b \join \lab_v')$,
is more permissive than the
$(\lab_\pc \join \lab_p) \flowsto \lab_b$ check of \ii{Store}.
Because the write keeps the label $\lab_v'$ unchanged we do not need
to additionally protect this label; we only need to protect the new
value $v$ and for this $\lab_v'$ can help.
This allows for instance writing in a high context to a block with a
low label as long as we overwrite a value previously labeled high; a
\ii{Store} would be disallowed in this setting, because it could
potentially leak information via a label change.
The second part of the check,
$\lab_v \flowsto (\lab_b \join \lab_v')$,
ensures that the written value $v$ is at least as protected after the write
(by the block label $\lab_b$ and the preserved value label $\lab_v'$)
as it was before the write (by $\lab_v$).
This check was unnecessary for \ii{Store} because the label of the
stored value does not change.

\citet{BuirasSR14} also have an upgrade operation that can raise the
label of a value in memory before entering a high context. This
upgrade operation can in fact be faithfully encoded using our
\ii{Load} and \ii{Store} instructions (as well as judicious use of
\ii{Call} and \ii{Return}).\ifever\chfoot{TODO for later: do this
  encoding in full detail}\fi{}
For the purpose of stressing our testing
methodology we chose to include this as a primitive instruction, with
the following (otherwise derivable) operational semantics rule:
\infrule[Upgrade]
  {i(\pc) = \ii{Upgrade}~r_p~r_l \quad
   \ii{rf\,}[r_p] = \V{(b, o)}{\lab_p} \quad
   \ii{rf\,}[r_l] = \V{\lab}{\lab'} \quad
   \lab_\pc' = \lab_\pc \join \lab' \\
   m[b] = \V{\ii{vs}}{\lab_b} \andalso
   \ii{vs}[o] = \V{v'}{\lab_v'} \andalso
   \lab_v' \flowsto (\lab \join \lab_b) \\
   (\lab_\pc' \join \lab_p) \flowsto \lab_b \andalso
   \ii{vs}' = \ii{vs}[o := \V{v'}{\lab}] \andalso
   m' = m[b := \V{\ii{vs}'}{\lab_b}]}
  {\RMACH{\V{\pc}{\lab_\pc}}{\ii{rf}}{\ii{cs}}{m} \step
   \RMACH{\V{(\pc\mathord{+}1)}{\lab_\pc'}}{\ii{rf\,}'}{\ii{cs}}{m'}}
Perhaps surprisingly, this rule is more complex and more restrictive
than our \ii{Store} rule.
\ii{Store} does not have to deal with the label $\lab'$ protecting the
first-class label $\lab$, or with the label $\lab_v'$ of the value $v'$.
In particular, our \ii{Store} rule does not have the 
$\lab_v' \flowsto (\lab \join \lab_b)$ check, because for a \ii{Store} the value $v'$
is overwritten, and thus does not need to be protected in any way.
An important consequence of this is that \ii{Store} can change labels
arbitrarily, while \ii{Upgrade} can only change labels in a way that
does not diminish the total protection of the existing value in memory.
Using \ii{Store} to overwrite a memory location with $\V{0}{\lab}$ is
thus a better way to change the label of a location whose value is no
longer relevant to $\lab$.

The remaining instructions are much simpler. \ii{Eq} simply
illustrates that all values, including pointers, can be compared for
equality. From an IFC perspective, the rule for \ii{Eq} is the same as
the ones for for \ii{Add} and \ii{Mult}:
\infrule[Eq]
  {i(\pc) = \ii{Eq}~r_1~r_2~r_d \andalso
   \ii{rf\,}[r_1] = \V{v_1}{\lab_1} \andalso
   \ii{rf\,}[r_2] = \V{v_2}{\lab_2} \\
   \text{if } v_1 \mathrel{==} v_2 \text{ then } n = 1 \text{ else } n = 0 \andalso
   \ii{rf\,}' = \ii{rf\,}[r_d := \V{n}{(\lab_1\mathord\join \lab_2)}]}
  {\RMACH{\V{\pc}{\lab_\pc}}{\ii{rf}}{\ii{cs}}{m} \step
   \RMACH{\V{\pc\mathord{+}1}{\lab_\pc}}{\ii{rf\,}'}{\ii{cs}}{m}}

\ii{GetOffset} and \ii{SetOffset} allow direct access to the offset
of any pointer:
\infrule[GetOffset]
  {i(\pc) = \ii{GetOffset}~r_p~r_d \andalso
   \ii{rf\,}[r_p] = \V{(b, o)}{\lab_p} \andalso
   \ii{rf\,}' = \ii{rf\,}[r_d := \V{o}{\lab_p}]}
  {\RMACH{\V{\pc}{\lab_\pc}}{\ii{rf}}{\ii{cs}}{m} \step
   \RMACH{\V{(\pc\mathord{+}1)}{\lab_\pc}}{\ii{rf\,}'}{\ii{cs}}{m}}
\infrule[SetOffset]
  {i(\pc) = \ii{SetOffset}~r_p~r_o~r_d \andalso
   \ii{rf\,}[r_p] = \V{(b, o')}{\lab_p} \andalso
   \ii{rf\,}[r_o] = \V{o}{\lab_o} \\
   \ii{rf\,}' = \ii{rf\,}[r_d := \V{(b, o)}{(\lab_p \join \lab_o)}]}
  {\RMACH{\V{\pc}{\lab_\pc}}{\ii{rf}}{\ii{cs}}{m} \step
   \RMACH{\V{(\pc\mathord{+}1)}{\lab_\pc}}{\ii{rf\,}'}{\ii{cs}}{m}}
%

$\ii{GetBlockSize}~r_p~r_d$ returns the size of the block referenced
by the pointer in $r_p$:
\infrule[GetBlockSize]
  {i(\pc) = \ii{GetBlockSize}~r_p~r_d \andalso
   \ii{rf\,}[r_p] = \V{(b, o)}{\lab_p} \\
   m[b] = \V{\ii{vs}}{\lab_b} \andalso
   \ii{rf\,}' = \ii{rf\,}[r_d := \V{(\ii{length}~\ii{vs})}{\lab_b}]}
  {\RMACH{\V{\pc}{\lab_\pc}}{\ii{rf}}{\ii{cs}}{m} \step
   \RMACH{\V{(\pc\mathord{+}1)}{(\lab_\pc \join \lab_p)}}{\ii{rf\,}'}{\ii{cs}}{m}}
The result has to be protected by the block label $\lab_b$,
which in turn has to be protected by the pointer label $\lab_p$.
The latter is can only achieved by raising the \pc{} by $\lab_p$.
Finally, $\ii{GetBlockLabel}~r_p~r_d$ returns the label of the block
referenced by the pointer in $r_p$:
\infrule[GetBlockLabel]
  {i(\pc) = \ii{GetBlockLabel}~r_p~r_d \andalso
   \ii{rf\,}[r_p] = \V{(b, o)}{\lab_p} \\
   m[b] = \V{\ii{vs}}{\lab_b} \andalso
   \ii{rf\,}' = \ii{rf\,}[r_d := \V{\lab_b}{\lab_p}]}
  {\RMACH{\V{\pc}{\lab_\pc}}{\ii{rf}}{\ii{cs}}{m} \step
   \RMACH{\V{(\pc\mathord{+}1)}{\lab_\pc}}{\ii{rf\,}'}{\ii{cs}}{m}}

\ifever
\ch{It might be informative to have a look at the counterexample QC finds
  for joining $\lab_v$ on the RHS of the Store check. It's probably some
  kind of flow-sensitive leak through a label.}
\ch{In general there is still a lot to learn from all the counterexamples
  we get. Something to look at after the deadline I guess.}
\fi

\subsection{Per-level allocation, stamps, reachability, and
  noninterference}
\label{sec:invariant}


Dynamic allocation in high contexts can cause the values of the
pointers to differ between the two executions considered by
noninterference~\cite{BanerjeeN05}. We ensure soundness by
breaking up each pointer into a memory block identifier and an offset
into the memory block. While offsets are fully observable to the
program, block identifiers are opaque and can only be tested for
equality. To further simplify the technical development we allocate
block identifiers ``per level'', \IE we assume that we have a
separate allocator for each allocation context label. This
assumption ensures that, at the level of abstraction we consider here,
allocations in high contexts cannot influence the values 
of pointers allocated in low contexts, and we can
thus use syntactic equality to check indistinguishability of
pointers. While this assumption on allocation might seem unrealistic,
previous work has shown formally that because block identifiers are opaque, this
machine can be realized by a lower-level machine with a single
standard allocator~\cite{PicoCoq2013}.

To understand per-level allocation, one needs to understand the
structure of the block identifiers we have already used in the
previous subsection. Block identifiers are not opaque; they are pairs
of a label $\lab_\sigma$, which we will call a \emph{stamp}, and an
integer index $i$:
\[
b \mathrel{::=} (\lab_\sigma, i)
\]%
As mentioned in \autoref{sec:memory}, a memory $m$ is a partial map
between block identifiers and labeled lists of values.
One can also see the memory as a three-dimensional array indexed first
by stamps, then by indices, and finally by offsets.
Stamps record the level of the allocation, \IE the label of the
context in which the allocation occurred, and ensure that allocation
at one level cannot influence allocation at other levels.
The side-condition
\[
   \ii{fresh}~m~(\lab_\pc \join \lab_n \join \lab') = b
\]
in the \rn{Alloc} rule from \autoref{sec:memory} implements this
per-level allocation idea.
The function \ii{fresh} takes a memory $m$ and a stamp
$\lab_\sigma = \lab_\pc \join
\lab_n \join \lab'$, then uses $\lab_\sigma$ to index into the memory $m$,
then uses a deterministic strategy to find the first unallocated index
$i$ in $m[\lab_\sigma]$, and finally returns the block identifier
$b = (\lab_\sigma,i)$. Formally, we have:
\[
\ii{fresh}~m~\lab_\sigma =
(\lab_\sigma,\ \ii{find\_undefined\_index}~m[\lab_\sigma]).
\]

It turns out that stamps are not only a convenient mechanism for
implementing per-level allocation (thus simplifying the definition of
indistinguishability for pointers to just syntactic equality), but are
also a crucial ingredient in another complex invariant of our
noninterference proof.
The \rn{Alloc} rule from \autoref{sec:memory} allows choosing an
arbitrary block label, even in a high context.
This is only sound because at the end of the high context access to
the newly allocated blocks is only possible through high pointers.
The stamp in each block identifier captures precisely the label of the
context in which the allocation of that block occurred.
A key invariant
\ifproof\chrev{used in our noninterference proof}\fi{}
is that intuitively {\em an
  allocated block with identifier $(\lab_\sigma,i)$ can be reached
  from registers only via pointer paths that are protected by labels
  that are, when taken together, at least as secure as $\lab_\sigma$}.
In the following we will formalize this reachability invariant and use
it to define indistinguishability.

We start by defining the ``root set'' of our reachability invariant,
the memory blocks that are directly accessible at a certain label $\lab$.
Given a machine state $\RMACH{\V{n}{\lab_\pc}}{\ii{rf}}{\ii{cs}}{m}$,
the root set includes the blocks that can be directly accessed
by pointers $\V{(b,o)}{\lab_p}$ in the register file \ii{rf}
for which $(\lab_p \join \lab_\pc) \flowsto \lab$.
Because pointers are protected both by their explicit label $\lab_p$
and the \pc{} label $\lab_\pc$, if the machine is in a high state
(one for which $\lab_\pc \not\flowsto \lab$), then the current register file
does not contribute at all to the root set.
The saved register files on the call stack $\ii{cs}$ are added to the
root set or not depending on whether the label of the return address
in the same call frame is below $\lab$ or not.
The label of the return address becomes the new \pc{} on the
corresponding \ii{Return}, so even if the current \pc{} is high the
root set has to include all the low pointers in all register files
saved in low-saved-\ii{pc} call frames.

More formally, we define \ii{root-set}, a function from a label
and a machine state to a set of blocks, as follows:\chfoot{We use
  the contents of the result register $r$ to compute the root set,
  although that register will be overwritten on return. Is this correct
  but unnecessarily strong?}
\begin{align*}
\ii{root-set}~\lab~\RMACH{\V{n}{\lab_\pc}}{\ii{rf}}{\ii{cs}}{m} &=
  (\ii{root-set}'~\lab~\ii{cs}) \mathrel{\cup}
  \begin{cases}
   \ii{blocks}~\lab~\ii{rf\,}
    & \text{if } \lab_\pc \flowsto \lab \\
    \emptyset
    & \text{otherwise}
  \end{cases}
\\[\baselineskip]
\ii{root-set}'~\lab~((\V{n}{\lab_\pc},\ii{rf\,},r,\lab_\ii{res}) : \ii{cs}) &=
  (\ii{root-set}'~\lab~\ii{cs}) \mathrel{\cup}
  \begin{cases}
   \ii{blocks}~\lab~\ii{rf\,}
    & \text{if } \lab_\pc \flowsto \lab\\
   \emptyset
    & \text{otherwise}
  \end{cases}
\\
\ii{root-set}'~\lab~[\,] &= \emptyset\\[\baselineskip]
\ii{blocks}~\lab~\ii{vs} &= \{ b \mid
  \V{(b,i)}{\lab_v} \in \ii{vs} \land \lab_v \flowsto \lab \}
\end{align*}

Reachability with respect to a label and a state is a relation on
block identifiers defined as the reflexive transitive closure of a
direct $\ii{link}~\lab~m$ relation on block identifiers:
\begin{align*}
\ii{reachable}~\lab~\RMACH{\pc}{\ii{rf}}{\ii{cs}}{m} &=
  (\ii{link}~\lab~m)^*\\
\ii{link}~\lab~m &= \{ (b, b') \mid m[b] = \V{\ii{vs}}{\lab_{b}}
  \land \lab_b \flowsto \lab \land b' \in \ii{blocks}~\lab~\ii{vs} \}
\end{align*}

We formally state the reachability invariant as a well-formedness
property of stamps:
\begin{defn}\label{def:well-stamped}
We call a machine state $S$
{\em well-stamped} if
for all labels $\lab$ and for all block identifiers $b$ and $b'$, if
$b \in \ii{root-set}~\lab~S$ and
$(b,b') \in \ii{reachable}~\lab~S$ then
$b' = (\sigma,i)$ for some $\sigma \flowsto \lab$.
\end{defn}

We have discovered and refined the form of the well-stamped
property by testing. Subsequently we have also proved in Coq that it is
indeed an invariant of the execution of our register machine.\footnote{
  Lemma {\tt well\_stamped\_preservation} at\\
  \url{https://github.com/QuickChick/IFC/blob/master/NIProof.v}}
\begin{lemma}
If $S$ is well-stamped and $S \step S'$ then $S'$ is well-stamped.
\end{lemma}
The effort of proving this lemma was reduced by considering the
correct definitions and statement from the start.


\ifever
\ch{Should put some more effort into explaining {\bf why} things are
  defined the way they are for the parts that are non-obvious.}
\fi

The indistinguishability relation for register machine states
requires that both the considered states are well-stamped.
Like reachability, indistinguishability is defined with respect to an
observation level $\lab$.
\begin{defn}
  Machine states $S_1$ and $S_2$ are \emph{indistinguishable}
  given observer level $\lab$,
  written $S_1 \indistfullws^\lab S_2$, if $S_1$ and $S_2$ are both
  well-stamped and $S_1 \indistfull^\lab S_2$.
\end{defn}
The $S_1 \indistfull^\lab S_2$ relation is defined similarly to the
relation of the same name in \autoref{sec:ssni} (Definition~\ref{def:indistfull}):
\renewcommand{\labelof}[1]{\lab_{#1}}
\begin{defn}\label{defn:indistfull-regs}
  Machine states
  $S_1 = \RMACHI{\pc_1}{\ii{rf}_1}{\ii{cs}_1}{m_1}{i_1}$ and
  $S_2 = \RMACHI{\pc_2}{\ii{rf}_2}{\ii{cs}_2}{m_2}{i_2}$ are
  \emph{indistinguishable at level $\lab$ with respect to whole machine states},
  written $S_1 \indistfull^\lab S_2$,
  if $m_1 \indist^\lab m_2$, $i_1 \indist^\lab i_2$,
  and additionally
  \begin{itemize}
  \item if $\labelof{\pc_1} \flowsto \lab$ or $\labelof{\pc_2} \flowsto \lab$ then
    $\pc_1 = \pc_2$ and $\ii{rf}_1  \indist^\lab \ii{rf}_2$ and
    $\ii{cs}_1 \indist^\lab \ii{cs}_2$.
  \item otherwise
  $\ii{dropWhile}~(\ii{stack-frame-high}~\lab)~\ii{cs}_1 \indist^\lab
   \ii{dropWhile}~(\ii{stack-frame-high}~\lab)~\ii{cs}_2$,
   where $\ii{stack-frame-high}~\lab~(\V{n}{\lab_\pc},\ii{rf},r,\lab_\ii{res}) =
          \lab_\pc \not \flowsto \lab$
  \end{itemize}
\end{defn}
The differences with respect to Definition~\ref{def:indistfull} are
caused by moving from a 2-label lattice to a more general one.
In case one of the $\ii{pc}$s is high we still compare the stacks
after cropping all high elements, just that ``being high'' is now
defined as being protected by a label that does not flow to the
observation label $\lab$.
Moreover, if both \ii{pc}s are high
then the two \pc{} labels are not required to be
equal, while for the 2-label lattice from \autoref{sec:ssni}
any two high labels have to be equal.

The definition above relies on several auxiliary relations, most
interestingly on an indistinguishability relation for memories defined
as follows:
\[
m_1 \indist^\lab m_2 = \forall (\lab_\sigma,i).~ \lab_\sigma \flowsto \lab \Rightarrow
(m_1[(\lab_\sigma,i)]\Uparrow  \mathrel{\,\land\,}
 m_2[(\lab_\sigma,i)]\Uparrow) \mathrel{\,\lor\,}
m_1[(\lab_\sigma,i)] \indist^\lab m_2[(\lab_\sigma,i)]
\]
We require each observable block identifier $(\lab_\sigma,i)$
either to be undefined in both memories
or to point to respectively indistinguishable blocks.
Indistinguishability for labeled things (used both for labeled blocks
and the labeled values inside) is defined as follows:
\[
\V{y_1}{\lab_1} \indist^\lab \V{y_2}{\lab_2} =
  (\lab_1 = \lab_2 \land (\lab_1 \flowsto \lab \implies
                             y_1 \indist^\lab y_2))
\]
Indistinguishability for lists (of values or stack frames) is defined pointwise:
\[
(y_1 : \ii{ys}_1) \indist^\lab (y_2 : \ii{ys}_2) =
  y_1 \indist^\lab y_2 \land \ii{ys}_1 \indist^\lab \ii{ys}_2
  \quad\text{and}\quad [\,] \indist^\lab [\,]
\]
Because of per-level allocation, indistinguishability for
values (including for pointers) is defined simply as syntactic equality:
\[
v_1 \indist^\lab v_2 = (v_1 = v_2)
\]
%
%
Indistinguishability for (potentially cropped, see
Definition~\ref{defn:indistfull-regs}) stacks is defined 
using list indistinguishability and the following indistinguishability
relation on stack frames:
\[
\begin{array}{l}
(\pc_1,\ii{rf}_1,r_1,\lab_{{\ii{res}}_1}) \indist^\lab
(\pc_2,\ii{rf}_2,r_2,\lab_{{\ii{res}}_2}) = \\
\qquad
  (\lab_{\pc_1} \flowsto \lab \lor \lab_{\pc_2} \flowsto \lab) \implies
  (\pc_1 = \pc_2 \land  
   \ii{rf}_1 \indist^\lab \ii{rf}_2 \land
   r_1 = r_2 \land
   \lab_{{\ii{res}}_1} = \lab_{{\ii{res}}_2})
\end{array}
\]
If one of the stored $\ii{pc}$s is low, then the other has to be low
as well, and all elements of the stack frame have to be pointwise
related.
If both stored $\ii{pc}$s are high we do not impose any additional
constraints on the stack frame, in particular, the two high $\ii{pc}$s
can have different labels.
This mirrors the handing of $\ii{pc}$s and register files in
Definition~\ref{defn:indistfull-regs}.
While this definition seems natural in retrospect, it took us a while
to reach it; \autoref{app:regs-indist} presents the wrong
alternatives with which we started.

\ifproof\chrev{
  While we discovered the rules and the well-stamped invariant by
  testing, we finally proved in Coq that this IFC mechanism has
  noninterference with respect to $\indistfullws$.\footnote{
\url{https://github.com/QuickChick/IFC/blob/master/NIProof.v}}
\begin{theorem}
The register information-flow machine satisfies SSNI$_{\indistfullws}$.
\end{theorem}
While the proof of this theorem discovered no errors in the rules or
the well-stamped invariant, it did discover a serious flaw in the
indistinguishability relation for stacks, which was previously hidden
by an error in our stack generator (the wrong definition is described
in \autoref{app:regs-indist}).
This illustrates that keeping generators and checkers in sync is
challenging and brings further motivation to recent work on
domain-specific languages for generators~\cite{LampropoulosPHHPX15,
  ClaessenFLOPS14, FetscherCPHF15}.
}\fi

\subsection{Testing Results}\label{sec:regs-testing-results}

\ifever
\rjmh{I wonder would it be possible to include the size of the
  counterexample found in the table? It would be interesting to see
  how well counterexample size correlates with testing time to find
  the bugs.}
\ch{This won't work unless our shrinking is really good. And anyway,
  the tests in this table run with shrinking turned off. We did
  discuss about gathering shrinking statistics, but decided that it's
  not a priority for this (version of the) paper.}
\leo{I agree with Catalin gathering sizes for counterexamples generated is almost
  impossible to do until the deadline}
\fi

\begin{figure}[p]
  \centering
  \makebox[0pt]{
\begin{tabular}{@{}ml*8{mr}@{}}
\toprule
\multicolumn{1}{@{}l}{\em Tested property}
& \GTHeader{EENI}
& \GTHeader{EENI}
& \GTHeader{LLNI}
& \GTHeader{LLNI}
& \GTHeader{SSNI}
& \GTHeader{SSNI}
& \GTHeader{MSNI}
& \GTHeaderLast{MSNI} \\
\multicolumn{1}{@{}l}{\em Starting states}
& Init
& Any
& Any
& Any
& Tiny
& Tiny
& Any
& Any\\
\multicolumn{1}{@{}l}{\em Indistinguishability}
& \indist_\ii{ints-in-regs}
& \indistfullws
& \indistfullws
& \indistfullws
& \indistfullws
& \indistfullws
& \indistfullws
& \indistfullws\\
\multicolumn{1}{@{}l}{\em Generation strategy}
& ByExec
& ByExec
& ByExec
& ByExec
& Tiny
& Tiny
& ByExec
& ByExec\\
\multicolumn{1}{@{}l}{\em Variant}
& 
& 
& basic
& optimized
& basic
& optimized
& basic
& optimized\\
\midrule
\OpMOV{} & 17.29 & 224.26 & 13.34 & 13.11 & 16.39 & 22.25 & 12.66 & 16.10 \\
\OpLOAD{} & 5349.00 & 1423.70 & 70.57 & 57.30 & 116.54 & 121.63 & 67.73 & 71.49 \\
\OpLOAD{} & \GTNone & \GTNone & 144.64 & 119.82 & 77.62 & 83.60 & 34.02 & 36.49 \\
\OpLOAD{} & \GTNone & 80.00 & 111.28 & 95.12 & 127.10 & 131.50 & 104.07 & 110.92 \\
\OpSTORE{} & \GTNone & 1896.42 & 206.73 & 15.94 & 102.48 & 50.05 & 43.71 & 19.23 \\
\OpSTORE{} & \GTNone & \GTNone & 983.00 & 1077.07 & 47.56 & 22.04 & 22.47 & 12.17 \\
\OpSTORE{} & 5365.00 & 1027.42 & 123.57 & 41.57 & 187.58 & 87.10 & 121.79 & 50.74 \\
\OpBINOP{} & 55.15 & 2395.66 & 103.83 & 101.91 & 151.27 & 209.19 & 98.97 & 127.95 \\
\OpBINOP{} & 55.79 & 2683.88 & 101.93 & 102.05 & 154.52 & 210.68 & 98.08 & 123.61 \\
\OpNOOP{} & 286.07 & 123.00 & 1.57 & 9.91 & 8.02 & 11.07 & 8.51 & 11.85 \\
\OpJUMP{} & 1003.05 & 118.99 & 32.86 & 44.18 & 73.02 & 99.31 & 31.13 & 50.61 \\
\OpJUMP{} & 161.16 & 2731.78 & 11.90 & 11.82 & 20.93 & 29.19 & 11.31 & 14.29 \\
\OpBNZ{} & 2079.02 & \GTNone & 79.02 & 78.56 & 27.97 & 37.74 & 72.87 & 95.59 \\
\OpBNZ{} & 755.22 & 118.71 & 29.51 & 40.31 & 72.66 & 101.32 & 27.78 & 46.84 \\
\OpBCALL{} & 858.23 & 911.36 & 36.08 & 35.21 & 14.26 & 19.53 & 15.79 & 19.30 \\
\OpBCALL{} & 2128.50 & 117.68 & 93.99 & 148.48 & 10.83 & 14.71 & 4.18 & 6.67 \\
\OpBCALL{} & 249.53 & 4497.32 & 12.45 & 12.22 & 21.38 & 29.10 & 11.65 & 14.73 \\
\OpBRET{} & \GTNone & 452.99 & 1311.57 & 19.52 & 47.38 & 32.55 & 36.95 & 23.66 \\
\OpBRET{} & \GTNone & 1322.03 & 1346.00 & 119.56 & 178.14 & 123.99 & 215.00 & 152.14 \\
\OpBRET{} & 695.30 & 239.04 & 19.48 & 10.43 & 21.01 & 14.40 & 18.78 & 12.59 \\
\OpBRET{} & 3015.38 & 712.16 & 21.56 & 11.06 & 13.83 & 9.50 & 19.83 & 13.21 \\
\OpALLOC{} & 344.85 & 20.32 & 36.93 & 28.42 & 45.87 & 47.95 & 35.23 & 34.25 \\
\OpALLOC{} & 324.16 & 22.60 & 39.60 & 30.77 & 46.84 & 47.23 & 37.20 & 36.85 \\
\OpWRITE{} & \GTNone & \GTNone & 1340.00 & 773.40 & 183.35 & 85.13 & 76.82 & 40.87 \\
\OpWRITE{} & \GTNone & 4110.29 & 404.04 & 58.61 & 348.65 & 202.48 & 153.03 & 73.19 \\
\OpWRITE{} & \GTNone & \GTNone & 1104.73 & 69.98 & 288.42 & 152.07 & 197.57 & 83.56 \\
\OpWRITE{} & \GTNone & 757.69 & 69.92 & 25.22 & 28.96 & 13.24 & 14.31 & 7.55 \\
\OpUPGRADE{} & \GTNone & \GTNone & 790.67 & 144.17 & 399.17 & 338.26 & 352.94 & 184.00 \\
\OpUPGRADE{} & \GTNone & \GTNone & 1138.00 & 922.00 & 238.66 & 112.21 & 97.13 & 58.80 \\
\OpUPGRADE{} & \GTNone & \GTNone & 1843.50 & 84.70 & 267.86 & 132.53 & 223.79 & 101.73 \\
\OpUPGRADE{} & \GTNone & \GTNone & 860.49 & 90.22 & 407.16 & 295.01 & 242.51 & 109.57 \\
\OpUPGRADE{} & \GTNone & 2706.80 & 460.02 & 174.89 & 422.01 & 357.12 & 481.55 & 222.27 \\
\OpPGETOFF{} & 1236.10 & 513.52 & 17.49 & 17.89 & 27.55 & 38.86 & 16.46 & 21.25 \\
\OpPSETOFF{} & 3180.00 & 371.24 & 38.71 & 42.00 & 50.58 & 68.97 & 38.34 & 51.00 \\
\OpPSETOFF{} & 2188.00 & 744.37 & 30.16 & 29.25 & 49.54 & 69.20 & 28.91 & 35.26 \\
\OpMSIZE{} & \GTNone & 975.58 & 50.82 & 50.84 & 47.62 & 67.06 & 49.03 & 57.85 \\
\OpMSIZE{} & \GTNone & 1960.54 & 64.69 & 70.67 & 78.47 & 108.20 & 59.54 & 82.94 \\
\OpMLAB{} & \GTNone & 812.58 & 24.47 & 25.97 & 28.09 & 38.55 & 23.33 & 31.16 \\
\midrule
\GTMeanHeader{Arithmetic mean} & \GTMeanNone & \GTMeanNone & \GTMean{346.55} & \GTMean{126.42} & \GTMean{117.09} & \GTMean{95.65} & \GTMean{84.34} & \GTMean{59.53} \\
\GTMeanHeader{Geometric mean} & \GTMeanNone & \GTMeanNone & \GTMean{100.46} & \GTMean{53.44} & \GTMean{67.33} & \GTMean{62.29} & \GTMean{46.53} & \GTMean{40.65} \\
\bottomrule
\end{tabular}
}
  \caption{Experiments for the register machine. MTTF given in milliseconds.}
  \label{fig:giant-table}
\end{figure}

In order to evaluate how well our testing techniques scale, we apply
the best strategies from \autoref{sec:generation} and \autoref{sec:stronger}
to the register machine and devise an even stronger
property that is even better at finding bugs. In this subsection we
explain and discuss in detail the experimental results summarized in
\autoref{fig:giant-table}.

\ifever
\leo{TODO: Is there a better way to present the differences between these
rules and the original? Maybe grey out dropped parts? What about moved
parts?}\asz{I like grey dropped-out parts (greyed-out dropped parts?); maybe
treated moved parts as a dropped part and an insertion somehow?}\ch{I think
  sections 2 and 5 are full of such things, so I guess that by now the
  reader has a trained enough eye for such things.\asz{Agreed that we don't have
  time, but even trained eyes need time to grok inference rules; we want to help
  those eyes!}\ch{I went for explaining the difference more precisely in the
  text for now.}
\fi

For these experiments we
introduced bugs by dropping taints and checks and by moving taints from
the \pc{} to the result. 
A missing taint bug is formed by dropping some label in the result of the
correct IFC rule. For example, we can insert a bug in the \rn{Mult}
rule by only tainting the result with the label of one of its
arguments (we taint $n_1 \times n_2$ with $\lab_1$ instead or $\lab_1
\join \lab_2$):

\infrule[Mult*]
  {i(\pc) = \ii{Mult}~r_1~r_2~r_d \andalso
   \ii{rf\,}[r_1] = \V{n_1}{\lab_1} \andalso
   \ii{rf\,}[r_2] = \V{n_2}{\lab_2} \\
   \ii{rf\,}' = \ii{rf\,}[r_d := \V{(n_1 \times n_2)}{\lab_1}]}
  {\RMACH{\V{\pc}{\lab_\pc}}{\ii{rf}}{\ii{cs}}{m} \step
   \RMACH{\V{(\pc\mathord{+}1)}{\lab_\pc}}{\ii{rf\,}'}{\ii{cs}}{m}}

A missing check bug is formed by dropping some part of the requirements of 
the IFC rule. For example, to insert a bug in the \rn{Store} rule we turn
the $\lab_\pc \join \lab_p \flowsto \lab_b$ check into just $\lab_p
\flowsto \lab_b$:

\infrule[Store*]
  {i(\pc) = \ii{Store}~r_p~r_s \andalso
   \ii{rf\,}[r_p] = \V{(b, o)}{\lab_p} \andalso
   \ii{rf\,}[r_s] = \V{v}{\lab_v} \\
   m[b] = \V{\ii{vs}}{\lab_b} \quad
   \lab_p \flowsto \lab_b \quad
   \ii{vs}' = \ii{vs}[o := \V{v}{\lab_v}] \quad
   m' = m[b := \ii{vs}']}
  {\RMACH{\V{\pc}{\lab_\pc}}{\ii{rf}}{\ii{cs}}{m} \step
   \RMACH{\V{(\pc\mathord{+}1)}{\lab_\pc}}{\ii{rf}}{\ii{cs}}{m'}}

A final and more subtle class of bugs is moving the taint from the \pc{} to the
result. For some rules, like the one for \ii{Load}, it is imperative that the
\pc{} is tainted instead of the result so that the labels involved are protected.
The following incorrect rule, in which we taint the value with
the block label $\lab_b$ instead of the \pc{},
yields a counterexample:

\infrule[Load*]
  {i(\pc) = \ii{Load}~r_p~r_d \andalso
   \ii{rf\,}[r_p] = \V{(b, o)}{\lab_p} \\
   m[b] = \V{\ii{vs}}{\lab_b} \andalso
   \ii{vs}[o] = \V{v}{\lab_v} \andalso
   \ii{rf\,}' = \ii{rf\,}[r_d := \V{v}{(\lab_v\join \lab_b)}]}
  {\RMACH{\V{\pc}{\lab_\pc}}{\ii{rf}}{\ii{cs}}{m} \step
   \RMACH{\V{(\pc\mathord{+}1)}{(\lab_\pc \join \lab_p)}}
         {\ii{rf\,}'}{\ii{cs}}{m}}
%


The baseline for our comparison is generation by execution and
EENI$_{\ii{Init}, \Halted\cap\Low, \indist_\ii{ints-in-regs}}$,
a basic instantiation of EENI (as
defined in \autoref{sec:eeni} and Appendix~\ref{app:strengthening}),
stating that starting from empty initial states and
executing the same program, if both machines reach a low halting state
then their register files need to contain low integers at the same
positions and these integers need to be pairwise equal.
Formally, we define indistinguishability as follows:
\begin{defn}
  $S_1 = \RMACH{\pc_1}{\ii{rf}_1}{\ii{cs}_1}{m_1}$ and
  $S_2 = \RMACH{\pc_2}{\ii{rf}_2}{\ii{cs}_2}{m_2}$ are indistinguishable
  with respect to integers stored in registers,
  written $S_1 \indist_\ii{ints-in-regs} S_2$,
  if $\ii{rf}_1 \indist_\ii{ints} \ii{rf}_2$, which is the pointwise
  extension of the following indistinguishability relation on values:
\[
\V{v_1}{\lab_1} \indist^\lab_\ii{ints} \V{v_2}{\lab_2} =
  (\lab_1 = \lab_2 \land (\lab_1 \flowsto \lab \implies
                          (v_1 = n \Leftrightarrow v_2 = n)))\\
\]
\end{defn}
We choose this property as the baseline because it is simple; in
particular it does not compare pointers or memories or stacks, which
as we saw in the previous subsection is very involved.
The results for this property appear in
  the first column of \autoref{fig:giant-table} and as expected are not
  satisfactory: most of the bugs are not found at all even after 5 minutes of
  testing.  For the rest of the experiments, we use the indistinguishability
  relation $\indistfullws$ described in the previous subsection.
Moreover, we start execution from arbitrary states,
which also significantly improves testing. 

\sloppy
The next two columns show the result of using the
EENI$_{\ii{Any},\Halted\cap\Low,{\indistfullws}}$ and
LLNI$_{\ii{Any},{\indistfullws}}$
properties (LLNI is defined generically in \autoref{sec:llni} and
Appendix~\ref{app:strengthening}).
These properties do not actually use the full invariants
shown in \autoref{sec:invariant}, just the parts of it that pertain to low states,
since both EENI and LLNI will only compare such states.
The generation strategy is again
generation by execution---simpler strategies result in very poor performance. 
The results show that the extended machine is too complex for EENI
to discover all injected bugs, while LLNI does find all of them.
\par\fussy

Two simple observations allow us to improve LLNI even further.
The first one is that our implementation of generation by execution is
``naive'' about generating each next instruction. To be precise, we
continuously update the uninitialized instruction memory with new
random instructions by repeatedly indexing into the list. This is
clearly the source of some overhead, which can be alleviated by using
a random-access data structure like a map or---as we chose to
use---a \emph{zipper}; a zipper provides even faster average-case
performance since most of the time the program counter is only
incremented by one. This improvement yields a small performance
boost.

The second and more important observation is that some instructions
have very restrictive IFC checks (\ii{Store}, \ii{Return}, \ii{Write}
and \ii{Upgrade}), which often lead to the machine failing as soon as
they are encountered.
This causes these instructions to be underrepresented in the machine
states produced by generation by execution, which only chooses an
instruction if this instruction can execute for at least a step.
Adjusting the frequency of instruction generation experimentally, so
that each instruction ends up being equally frequent among the ones
that can successfully take a step, leads to the optimized LLNI column
of \autoref{fig:giant-table}. This strategy successfully discovers all bugs
relatively quickly. There is, however, a trade-off: the bugs that were
easier to find before become slightly harder to find.

We also consider an instance of the SSNI property
(\autoref{sec:ssni}), SSNI$_{\indistfullws}$, which we expect to take
advantage of all our invariants. Similarly to LLNI, IFC-check-heavy
instructions cause a lot of failures; in this case, the failures lead
to many discarded tests, since only one instruction is run.
In \autoref{fig:giant-table} we show the performance of
SSNI${_{\indistfullws}}$ with uniform and weighted instruction
generation so that instructions empirically appear uniform in the
non-discarded tests.\ch{Can we have discards in the table
please?}\leo{Probably too late for that now, maybe during the
reviews.}  The required weights turn out to be very
similar to the ones required for LLNI.  Moreover, the same trade-off appears
here: we can find the hard-to-find bugs faster by sacrificing a bit of
speed for the easy-to-find ones.

As was the case for the basic machine, when optimizing the generation for
SSNI, we must be extremely cautious to avoid ruling out
useful parts of the state space. Since SSNI operates by executing a machine
state for a {\em single} step to check the invariant, being able to generate the
entire state space of pairs of indistinguishable machines becomes very
important. For example, a reasonable assumption might seem to be that the
stacks are ``monotonic'', as described in the end of the previous
section. However, if we use the incorrect indistinguishability
relation in \autoref{eq:wrong-indist} and generate only ``monotonic''
stacks for
the starting states, SSNI does not uncover the bug in
\autoref{fig:counter-extended-stack}, whereas LLNI does.

Comparing LLNI and SSNI with respect to their efficiency in testing, we can
spot an interesting tradeoff.  On the one hand, a significant limitation of
LLNI is that bugs that appear when the $\pc$ is high are not detected
immediately, but only after the $\pc$ goes back low, if ever. One example is the
 {\sc Store*} buggy rule above, where we do not check whether
the $\pc$ label flows to the label of the memory cell.
On such bugs LLNI has orders of magnitude worse results.
On the other hand, SSNI is significantly less robust with respect to starting
state generation.  If we do not generate every valid starting state,
then SSNI will not
test executions starting in the missing states, since it only executes one
instruction. LLNI avoids this problem as long as all valid states are
eventually reachable from the generated starting states.

These observations lead us to formulate a new property: {\em
multi-step noninterference (MSNI)}, that combines the advantages of
both LLNI and SSNI. The formal definition of MSNI is given in
Appendix~\ref{app:strengthening}.  Informally, we start from an
arbitrary pair of indistinguishable machine states and we check the
SSNI {\em unwinding conditions} along a whole execution trace. Using
generation by execution with uniform and adjusted instruction
frequencies for this property yields the last two columns
of \autoref{fig:giant-table}. MSNI performs on most bugs on par with the
better of SSNI or LLNI by uncovering IFC violations as
soon as they appear; at the same time, unlike SSNI, MSNI is robust
against faulty generation.

\ifever{
\leo{The following should go to some discussion/conclusion paragraph
somewhere, but I am putting notes here for now to avoid conflicts. Still in draft
form! }. 

\begin{itemize}
\item In future attempts to design IFC artifacts we believe testing can be
extremely useful
\item Early in the design process, one can still use LLNI with a
not-fully-formed invariant to get a lot of benefit out of it
\item One can even use LLNI to slowly state the unwinding conditions.
\item When one has candidates for unwinding conditions, MSNI should be the
go-to property since it combines speed and robustness.\ch{Again, all this
  assumes that LLNI doesn't already use the invariant, although I think
  in your experiments it very much does. At the risk of
  repeating myself, the smarts don't go into the once and for all properties
  like EENI, LLNI, SSNI, MSNI. It goes into the indistinguishability!
  LLNI can indeed be instantiated with a naive indistinguishability
  relation that ignores pointers, and that would be a great start
  for testing, but as far as I understand it's not what we did.}
\end{itemize}
}
\fi

\ifever
\ch{Testing only SSNI but not well-stampedness preservation misses the
  alloc mutants that well-stampedness would catch. Q: What does this
  mean?}
\fi

\FloatBarrier

\section{Related Work}
\label{sec:related}

Generating random inputs for testing is a large research area,
but the particular
sub-area of testing language implementations by generating random \emph{programs}
is less well studied.
\iffull Redex~\cite{Klein09,Klein09thesis,Klein12,FetscherCPHF15} (\textit{n\'e} PLT Redex) \else PLT Redex~\cite{Klein12} \fi
is a domain-specific language for defining operational semantics within
\iffull Racket (\textit{n\'e} PLT Scheme), \else PLT Scheme, \fi
which includes a property-based random testing framework inspired by
QuickCheck.
This framework uses a formalized language definition to automatically generate
simple test-cases.  To generate better test cases, however, Klein~\ETAL find
that the generation strategy needs to be tuned for the particular language; this
agrees with our observation that fine-tuned strategies are required to obtain
the best results.  They argue that the effort required to find bugs
using \iffull Redex \else PLT Redex \fi is less than the effort required for a
formal proof of correctness, and that random testing is sometimes viable in
cases where full proof seems infeasible.

\iffull
\citet{KleinRacketMachine} use PLT Redex's
QuickCheck-inspired random testing framework to asses the safety of
the bytecode verification algorithm for the Racket virtual machine.
They observe that naively generated programs only rarely pass bytecode
verification (88\% discard rate), and that many programs fail
verification because of a few common violations that can be easily
remedied in a post-generation pass that for instance replaces
out-of-bounds indices with random in-bounds ones.
These simple changes to the generator are enough for reducing the
discard rate (to 42\%) and for finding more than two dozen bugs in the
virtual machine model, as well as a few in the Racket machine
implementation, but three known bugs were missed by this naive
generator.
The authors conjecture that a more sophisticated test generation
technique could probably find these bugs.
\fi

CSmith~\cite{YangCER11} is a C compiler testing tool that
generates random C programs, avoiding ones whose behavior is undefined by the
C99 standard.  When generating programs, CSmith does not attempt to model the
current state of the machine; instead, it chooses program fragments that are
correct with respect to some static safety analysis (including type-,
pointer-, array-, and initializer-safety, \ETC).  We found that modeling the
actual state of our (much simpler) machine to check that generated programs were
hopefully well-formed, as in our generation by execution strategy, made our
test-case generation far more effective at finding noninterference bugs.
In order to get smaller counterexamples, Regehr~\ETAL present
C-Reduce~\cite{regehr2012test}, a tool for reducing test-case C programs such as
those produced by CSmith. They note that conventional shrinking methods
usually introduce test cases with undefined behavior; thus, they put a great
deal of effort and domain specific knowledge into shrinking well-defined
programs only to programs that remain well-defined.  To do this, they use a
variety of search techniques to find better reduction steps and to couple
smaller ones together. Our use of QuickCheck's \emph{double shrinking} is
similar to their simultaneous reductions, although we observed no need in our
setting for more sophisticated searching methods than the greedy one that is
guaranteed to produce a local minimum.
Regehr~\ETAL's work on reduction is partly based on Zeller and
Hildebrandt's formalization of the delta debugging algorithm
\textit{ddmin}~\cite{zeller2002simplifying}, a non-domain-specific algorithm for
simplifying and isolating failure-inducing program inputs with an extension of
binary search. In our work, as in Regehr~\ETAL's, domain-specific knowledge is
crucial for successful shrinking.
%
In recent work, \citet{KoopmanAP13} propose a technique for model-based
shrinking.

\sloppy

Another relevant example of testing programs by generating random input is
Randoop~\cite{Pacheco:2007}, which generates random sequences of calls to Java
APIs.  Noting that many generated sequences crash after only a few calls, before
any interesting bugs are discovered, Randoop performs \emph{feedback directed}
random testing, in which previously found sequences that did not crash are
randomly extended.  This enables Randoop to generate tests that run much longer
before crashing, which are much more effective at revealing bugs. Our
generation by execution  
strategy is similar in spirit, and likewise results in a substantial improvement
in bug detection rates.

\fussy



A state-machine modeling library for (an Erlang version of)
QuickCheck has been developed by Quviq \cite{Hughes07}%
. It generates sequences of API calls to a stateful system satisfying
preconditions formulated in terms of a model of the system state,
associating a (model) state transition function with each API
call. API call generators also use the model state to avoid generating
calls whose preconditions cannot be satisfied. Our
generation-by-execution strategy works in a similar way for
straightline code.

A powerful and widely used approach to testing is symbolic execution---in
particular, \emph{concolic testing} and related dynamic symbolic execution
techniques~\cite{ConcolicAssessment:ICSE11\iffull,
HybridConcolic:2007:Majumdar\fi}.  The idea is to mix symbolic and concrete
execution in order to achieve higher code coverage.  The choice of which
concrete executions to generate is guided by a constraint solver and path
conditions obtained from the symbolic executions.  Originating with
DART~\cite{DART:2005} and PathCrawler~\cite{WilliamsMM04}, a variety of
tools and methods have appeared%
\iffull%
; some of the state-of-the-art tools include
CUTE~\cite{CUTE:2005}, CREST~\cite{CREST:2008}, and KLEE~\cite{KLEE:2008:OSDI}
(which evolved from EXE~\cite{EXE:2006})
\fi.
We wondered whether dynamic symbolic execution could be used instead of
random testing for finding noninterference bugs.  As a first step, we
implemented a simulator for a version of our abstract machine in C and
tested it with
\iffull
KLEE.
\else
KLEE~\cite{KLEE:2008:OSDI}, a state-of-the-art symbolic execution tool.
\fi
Using KLEE out of the box and without any expert knowledge in the area, we
attempted to invalidate various assertions of noninterference.  Unfortunately,
we were only able to find a counterexample for \IfcBugPushNoTaint, the simplest
possible bug, in addition to a few implementation errors (\EG out-of-bound
pointers for invalid machine configurations).  The main problem seems to be that
the state space we need to explore is too large\iffull~\cite{Cadar:ACM:2013}\fi, so we
don't cover enough of it to reach the particular IFC-violating configurations.
More recently, \citet{TorlakB14} have used our information-flow stack
machine and its bugs with respect to EENI as a case study for their
symbolic virtual machine, and report better results.


\iffull 
\sloppy

\citet{BalliuDG12} created \textsc{ENCoVer}, an extension of
Java PathFinder, to verify information-flow properties of Java programs by means
of concolic testing. In their work, concolic testing is used to extract an
abstract model of a program so that security properties can be verified by an
SMT solver.
%
%
While \textsc{ENCoVer} tests the security of individual programs, we
use testing to check the soundness of an entire enforcement mechanism.
Similarly, \citet{MilushevBC12} have used KLEE for testing the
noninterference of individual programs, as opposed to our focus on
testing dynamic IFC mechanisms that are meant to provide
noninterference for all programs.

\fussy
\fi


\sloppy

In recent work, \citet{LampropoulosPHHPX15} introduce a
domain-specific language for random generators that puts together
random instantiation \cite{Antoy94aneeded, ClaessenFLOPS14,
  FetscherCPHF15} and constraint solving \cite{MohrH86}. As a case
study, they tests noninterference for the register machine from
\autoref{sec:regs} using our generators for indistinguishable machine
states (including generation by execution) and the SSNI and LLNI
properties.

In interactive theorem provers, automatically generating
counterexamples for false conjectures can prevent wasting time and
effort on proof attempts doomed to fail~\cite{GroceHJ07}.
\citet{DybjerHT03} propose a QuickCheck-like tool
for the Agda/Alfa proof assistant.
\citet{BerghoferN04} proposed a QuickCheck-like
tool for Isabelle/HOL. This was recently extended by
\citet{Bulwahn12} to also support exhaustive and
narrowing-based symbolic
testing\iffull~\cite{Lindblad07,RuncimanNL08,ChristiansenF08}\fi.
Moreover, Bulwahn's tool uses Horn clause data flow analysis to
automatically devise generators that only produce data that satisfies
the precondition of the tested conjecture~\cite{Bulwahn12smartgen}.
\citet{DoubleCheck} implemented DoubleCheck,
an adaption of QuickCheck for ACL2.
\citet{ChamarthiDKM11} later proposed a more
advanced counterexample finding tool for ACL2s,
which uses the full power of the theorem
prover and libraries to simplify conjectures so that they
are easier to falsify.
While all these tools are general and only require the statement of
the conjecture to be in a special form (\EG executable specification),
so they could in principle be applied to test noninterference, our
experience with QuickCheck suggests that for the best results one
has to incorporate domain knowledge about the machine and the
property being tested.
%
We hope to compare our work against these tools in the future and
provide experimental evidence for this intuition.
Recently, \citet{itp2015} introduced a port of Haskell QuickCheck to
Coq together with a foundational verification framework for testing
code and use our testing noninterference techniques as their main case
study, proving our generator for ``Tiny'' indistinguishable states
used to test SSNI for the register machine
(\autoref{sec:regs-testing-results}) sound and complete with respect
to indistinguishability.

\fussy

\iffull\ifunfinished
\asz{This will need to be expanded for the full version.}
Reich~\ETAL's ``Lazy Generation of Canonical Test Programs''~\cite{reichlazy}
is about how to constrain an exhaustive generation of \emph{all} programs of a
certain depth.  We do unconstrained generation of programs of all depths.  They
don't do the same sort of direct evaluation we do, so it's not clear how related
it is.
\fi\fi

\sloppy

On the dynamic IFC side \citet{BirgissonHS12} have a good overview
of related work.
Our correct rule for \ii{Store} for the stack machine
is called the \emph{no-sensitive-upgrades}
policy in the literature and was first proposed by
\citet{zdancewic02:thesis} and later adapted to the dynamic IFC
setting by \citet{AustinF09}.
\iffull
To improve precision, \citet{AustinF:2010}
later introduced a different \emph{permissive-upgrade} policy, where
public locations can be written in a high context as long as branching
on these locations is later prohibited, and they discuss adding
\emph{privatization operations} that would even permit this kind of
branching safely.
\fi
\citet{HedinS12} improve the precision of the
no-sensitive-upgrades policy by explicit \emph{upgrade annotations},
which raise the level of a location before branching on secrets.  They
apply their technique to a core calculus of JavaScript that includes
objects, higher-order functions, exceptions, and dynamic code
evaluation.\ifever\bcp{The reader may wonder whether we've experimented with
  QCing these variants, or whether there would be any technical impediments
  to doing so.}\fi{}
\citet{BirgissonHS12} show that random testing with
QuickCheck can be used to infer upgrade
instructions in this setting. The main idea is that whenever a random
test causes the program to be stopped by the IFC monitor because it
attempts a sensitive upgrade, the program can be rewritten by
introducing an upgrade annotation that prevents the upgrade from being
deemed sensitive on the next run of the program.
In recent work, \citet{BichhawatRGH14b} generalize the
permissive-upgrade check to arbitrary IFC lattices. They present
involved counterexamples, apparently discovered manually while doing
proofs. We believe that our testing techniques are well-suited at
automatically discovering such counterexamples.

\iffull
\citet{TerauchiA05} and later
\citet{BartheDR11} propose a technique for statically
verifying the noninterference of individual programs using the idea of
self-composition.
This reduces the problem of verifying secure information flow for a
program $P$ to a safety property for a program $\hat{P}$ derived from
$P$, by composing $P$ with a renaming of itself.
Self-composition enables the use of standard (\IE not
relational~\cite{Benton04,BartheCK11}) program logics and model
checking for showing noninterference.
The problem we address in this paper is different: we test the
soundness of dynamic IFC mechanisms by randomly generating (a large
number of) pairs of related programs.
One could imagine extending our technique in the future to testing the
soundness of static IFC mechanisms such as
type systems~\cite{sabelfeld03:lang_based_security},
relational program logics~\cite{Benton04,BartheCK11},
and self-composition based tools~\cite{BartheDR11}.

\fussy

In recent work \citet{OchoaCPH15} discuss a preliminary model-checking
based technique for discovering unwanted information flows in
specifications expressed as extended finite state machines. They also
discuss about testing systems for unwanted flows using unwinding-based
coverage criteria and mutation testing.
%
In a recent position paper, \citet{Kinder15} discusses testing of
hyperproperties~\cite{ClarksonS10}.

\ifunfinished
\ch{Is \emph{low-lockstep
    noninterference} a standard property? If so who invented it?}

Some references from Domagoj Babic (possibly worth including, but even if so,
only in the full version):
\begin{itemize}
  \item Emek~\ETAL: ``X-Gen: a random test-case generator for systems and
SoCs'', HLDVT '02
\url{http://dx.doi.org/10.1109/HLDVT.2002.1224444}
--- X-Gen was developed roughly at the same time as Verisity developed
the Specman tool and the E verification language.  Cadence later
bought Verisity, while Synopsys developed SystemVerilog.  Both
languages (and the corresponding tools), as well as X-Gen essentially
have the same idea---special language for defining the environment
constraints, coverage, distributions,\ldots

  \item Dorit Baras, Shai Fine, Laurent Fournier, Dan Geiger, Avi Ziv:
``Automatic Boosting of Cross-Product Coverage Using Bayesian
Networks'', STTT 2011
I couldn't find the link to the journal version, this is the
conference version:
\url{http://link.springer.com/chapter/10.1007\%2F978-3-642-01702-5_10?LI=true}

  \item Chockler~\ETAL: ``Cross-Entropy Based Testing'', FMCAD'07
\url{http://dx.doi.org/10.1109/FAMCAD.2007.19}
\end{itemize}

\ch{A good survey of related work on random program generation can be
  found in the related work section of Micha\l\ Pa{\l}ka's thesis,
  ``Testing an Optimising Compiler by Generating Random Lambda
  Terms''~\cite{Palka12}.}

\ch{Other people observed that the rates at which bugs manifest
  themselves with random testing can vary by many orders of magnitude
  \cite{Taming2013,LiblitNZAJ05} -- in realistic scenarios (not ours!)
  this leads to the fuzzer taming problem. We could use this in the
  text to motivate using the geometric mean?}

\ch{\cite{Taming2013} also makes the claim that large test cases are
  better at finding bugs, but small ones are more useful afterwards
  (thus justifying shrinking): ``Randomly generated test cases are
  more effective at finding bugs when they are large [1]. There are
  several reasons for this. First, large tests are more likely to bump
  into implementation limits and software aging effects in the system
  under test. Second, large tests amortize start-up costs. Third,
  undesirable feature interactions in the system under test are more
  likely to occur when a test case triggers more behaviors. The
  obvious drawback of large random test cases is that they contain
  much content that is probably unrelated to the bug.''}
\ch{\cite{AndrewsGWX08} could directly motivate generation by
  execution}

\ch{TODO: Have a look at \cite{PLTExp2013}}

\fi 
\fi 

\section{Conclusions and Outlook}
\label{sec:conclusions}

We have shown how random testing can be used to discover counter\-examples to
noninterference in a simple information-flow machine and how to shrink
counterexamples discovered in this way to simpler, more comprehensible
ones.  
The techniques we present bring many orders of magnitude improvement
in the rate at which bugs are found, and for the hardest-to-find bugs
(to EENI) the minimal counterexamples are 10-15 instructions long --
well beyond the scope of naive exhaustive testing.
Even if we ultimately care about full security proofs~\cite{PicoCoq2013},
using random
testing should greatly speed the initial design process and allow us to
concentrate more of our energy on proving things that are correct or nearly
correct.

We are hopeful that we can scale the methodology introduced in this
paper to test noninterference and other properties~\cite{micropolicies2015} for
abstract machines built on top of real-life instruction set
architectures.
The results in \autoref{sec:regs} are particularly encouraging in this
respect.
For a real-life architecture, even if were to find bugs $100\times$
slower than for the machine in \autoref{sec:regs}, that would still
only be a matter of seconds.

We expect that our techniques are flexible enough to be applied to checking
other relational properties of programs (i.e., properties of pairs of
related runs\iffull~\cite{Benton04,BartheCK11,ClarksonS10}\fi)---in particular,
the many variants and generalizations of
noninterference, for instance taking into account
declassification~\cite{sabelfeld05:_dimensions_declass}.
Beyond noninterference properties, preliminary
experiments with checking correspondence between concrete and abstract
versions of our current stack machine suggest that many of our techniques
can also be adapted for this purpose.  For example, the
generate-by-execution strategy and many of the shrinking tricks apply just
as well to single programs as to pairs of related programs.  This gives us
hope that they may be useful for checking yet further properties of abstract
machines.

\ifever\finish{One issue that we have not begun to address is concurrency...}\fi

\ifever\finish{Maybe say this observation more explicitly: Two labels are
  not enough! To show this we can add a labelOf instruction, which is
  unsound for the current setting ... just that every counterexample needs
  at least 3 labels, so QC doesn't find anything!}  \fi

\ifever\finish{
An important point that we have not yet carefully addressed is {\em coverage
  checking}---i.e., gathering execution statistics to increase confidence
that our test-case generation is not missing relevant parts of the execution
space.  Our execution-length and reason-for-halting statistics are small
steps in this direction, but we could imagine going much further.  One
possible idea (suggested by Regehr) is to check that every rule is actually
being executed under all possible combinations of high and low labels on
each of its inputs.}
\ch{A related idea is to be exhaustive and automatic about removing
  individual checks and taints in rules. While this wouldn't say
  anything about getting the noninterference property +
  indistinguishability relation right, it would provide some
  confidence without turning the interpreter into spaghetti.}
\fi


\ifever
\feedback{Deian: I'm not sure if this helps your story but in LIO I've
  been doing something similar, but much simpler (I didn't go beyond
  your first few optimizations), in generating random
  "instructions"/actions and testing properties against it (ref:
  \url{https://github.com/scslab/lio/blob/master/quickcheck-lio-instances/LIO/Instances.hs}).}
\ch{Your point 3 below is particularly interesting, we could try to
  apply our ideas to LIO and see how they fly. It's a pity we probably
  won't have time to do that before ICFP. Still, I'll keep it in mind
  for later and mention it as future work}
\fi

\paragraph*{Acknowledgments}
We thank the participants in the discussion at the IFIP WG 2.8
meeting in Storulv\accent'27an that originated this work:
    Ulf Norell,
    Rishiyur S. Nikhil,
    Micha\l{} Pa\l{}ka,
    Nick Smallbone, and
    Meng Wang.
%
%
We are grateful to
  Johannes Borgstr\"{o}m,
  Cristian Cadar,
  Delphine Demange,
  Matthias Felleisen,
  Robby Findler,
  Alex Groce,
  Casey Klein,
  Ben Karel,
  Scott Moore,
  Micha\l{} Pa\l{}ka,
  John Regehr,
  Howard Reubenstein,
  Alejandro Russo,
  Deian Stefan,
  Greg Sullivan, and
  Andrew Tolmach,
for providing feedback on a draft,
and to the members of the CRASH{\slash}SAFE team
and to Manolis Papadakis for fruitful discussions.
Finally, we thank the anonymous reviewers for their suggestions
and Andreas Haeberlen for kindly providing us computing time
on his cluster.
This material is based upon work supported by the DARPA CRASH
program through the US Air Force Research Laboratory (AFRL) under Contract
No. FA8650-10-C-7090, and NSF award 1421243,
{\em Random Testing for Language Design}.
The views expressed are those of the authors and do
not reflect the official policy or position of the Department of Defense or
the U.S. Government. The work is also partially funded under the
Swedish Foundation for Strategic Research grant RAWFP.

\appendix
\iftechrep
\renewcommand{\thesection}{\Alph{section}}
\fi

\section{Varying the Indistinguishability Relation}
\label{app:indist}

\subsection{Labels being high or low needs to be observable}
\label{sec:counter-observable}

As seen in \autoref{sec:basic-ifc-rules},
our definition of
indistinguishability of values (Definition~\ref{def:indist}) allows
the observer to distinguish between final memory states that differ
only in their \emph{labels}.
One might imagine changing the definition of indistinguishability so
that labels are not observable.
There are at least two ways one can imagine doing this; however, both are wrong.
First, one could try defining indistinguishability of values so that
$\V{x}{\low} \indist \V{y}{\high}$ for any $x$ and $y$. QuickCheck
easily finds a counterexample to this
(\autoref{fig:counter-high-equiv-everything}).
Second, one could try refining this so that only $\V{x}{\low} \indist
\V{x}{\high}$, \IE a high value is equivalent with a low one only when
the payloads are equal.
QuickCheck also disproves this alternative
(\autoref{fig:counter-not-observable}), and the counterexample
produced by QuickCheck illustrates how, even with the correct rules, a
difference in the labels of two values can be turned into a difference
in the values of two values.
This counterexample is reminiscent of a well-known ``flow-sensitivity
attack'' (Figure~1 in \citet{RussoS10}; attributed to \citet{Fenton74}).
This counterexample relies on \ii{Call} and \ii{Return} as introduced in
\autoref{sec:cally}.

\counterexample{high-equiv-everything}{A counterexample showing that it is
  wrong to make high values be equivalent to all other values.
}

\counterexample{not-observable}{A counterexample showing that it is also
  wrong to make high values equivalent to low values with the same payload.
\vspace{3.5ex} 
}


\subsection{Weakening EENI when adding calls and returns}
\label{sec:counter-eeni-low}

The counterexample in \autoref{fig:counter-eeni-no-low} shows that once we have
a way to restore the \pc{} label, we can no longer expect all pairs of halting
states to be indistinguishable in EENI.
In particular, as the counterexample shows, one machine can halt in a
high state, while the other can return to low, and only then halt.
Since our indistinguishability relation only equates states with
the same \pc{} label, these two halting states are distinguishable.
The solution we use in \autoref{sec:calls-and-returns} is to weaken
the EENI instance, by considering only ending states that are both
halting and low (\IE we change to
EENI$_{\Init,\Halted\cap\Low,{\indistmem}}$).

\counterexample{eeni-no-low}{A counterexample justifying the change to
  EENI$_{\Init,\Halted\cap\Low,{\indistmem}}$ in \autoref{sec:calls-and-returns}.
\vspace{3.5ex} 
}

\subsection{Indistinguishability for stack elements when adding calls and returns}
\label{sec:counter-stack}

In \autoref{sec:calls-and-returns} we defined the indistinguishability relation
on stack elements so that return addresses are only equivalent to other return
addresses and (as for values) $\badret{x_1}{\lab_1} \indist \badret{x_2}{\lab_2}$ if
either $\lab_1 = \lab_2 = \high$ or $x_1 = x_2$ and $\lab_1 = \lab_2 = \low$.
If instead we considered high return addresses and high values to be
indistinguishable, QuickCheck would find a counterexample.
This counterexample requires quasi-initial states (and $\indistlow$)
and is listed in \autoref{fig:counter-stk-elt-equiv-label-on-top}.
The first machine performs only one \ii{Return} that throws away
two elements from the stack and then halts.
The second machine returns twice: the first time to the same \ii{Return},
unwinding the stack and raising the \pc{}; and the second time to the \ii{Halt}
instruction, labeling the return value high in the process.
The final states are distinguishable because the elements on the stack have
different labels.
As we saw earlier, such a counterexample can be extended to one in which values
also differ.

\ifever
\ch{This also shows the fact that quasi-initial states
  are not necessarily reachable states -- although this
  is not really needed for this counterexample.}
\fi

\counterexample[ht]{stk-elt-equiv-label-on-top}{A counterexample that motivates
  the indistinguishability of stack elements for the machine with
  calls and returns.}

\iffull
\subsection{Counterexamples justifying indistinguishability for SSNI}
\label{sec:counter-ssni}

The indistinguishability relation high states used for SSNI needs to
be strong enough to ensure that when both machines return to low
states, those low states are also indistinguishable.
Since $\indistlow$ is too weak, QuickCheck can find counterexamples to
condition~\ref{ssni:return-step} in Definition~\ref{def:SSNI}
(see \autoref{fig:counter-ssni-equivlow-too-weak}).

\counterexample{ssni-equivlow-too-weak}{A counterexample showing that
  $\indistlow$ is too weak for SSNI. Since the \pc{} is initially
  high, $\indistlow$ does not require the initial stacks to be related
  in any way, which means the two machines can jump to two different
  addresses while still both lowering the \pc{}. The two resulting
  states are, however, distinguishable, since they have different
  \pc{}s.}

On the other hand, treating high states exactly like low states in the
indistinguishability relation is too strong, since that would prevent
the stacks to change between successive high states. In this case
QuickCheck finds counterexamples to condition~\ref{ssni:high-step} in
Definition~\ref{def:SSNI} (see
\autoref{fig:counter-ssni-equivwrongfull}). This motivates comparing
stacks for high state only below the first low return, while allowing
the tops of the stacks to vary arbitrarily, as done in the definition
of $\indistfull$ (Definition~\ref{def:indistfull}).
These two counterexamples guide our search for the correct
indistinguishability relation---\IE one that correctly captures
the invariant that the machine can only alter stack frames below
the current one by using the \ii{Return} instruction.

\counterexample{ssni-equivwrongfull}{A counterexample that shows that
  treating high states exactly like low states in the
  indistinguishability relation over machine states is too strong and
  breaks condition~\ref{ssni:high-step} in
  Definition~\ref{def:SSNI}. When a machine steps from a high state to
  another high state the contents of the stack can change.}

\ch{Could it be that some of Antal's(?)  LaTeX hacks are biting back?
  These definition references don't work the way they should}
\fi

\subsection{Wrong alternatives for indistinguishability of register
  machine states}
\label{app:regs-indist}

The definition of indistinguishability for the register machine
(Definition~\ref{defn:indistfull-regs}) might seem natural in
retrospect, but it took us a while to reach it. Here we presents two
wrong alternatives with which we started. The handing of call stacks
differs from Definition~\ref{defn:indistfull-regs}.

In the first wrong alternative we required a very strong matching
between stack frames:
\begin{multline}\label{eq:wrong-indist}
(\V{n_1}{\lab_{\pc_1}},\ii{rf}_1,r_1,\lab_{{\ii{res}}_1}) \indist^\lab
(\V{n_2}{\lab_{\pc_2}},\ii{rf}_2,r_2,\lab_{{\ii{res}}_2}) = \\
  n_1 = n_2 \land \lab_{\pc_1} = \lab_{\pc_2} \land
  \ii{rf}_1 \indist^\lab \ii{rf}_2 \land
  r_1 = r_2 \land \lab_{{\ii{res}}_1} = \lab_{{\ii{res}}_2}
\end{multline}

However, the above indistinguishability relation assumes that the
stacks are ``monotonic'', in the sense that the program counters
stored in the stack are only decreasing with respect to the label
flows-to relation. While this was true for the stack machines with the
two-label lattice, this is the case for the register machine with the
more complex diamond lattice, as can be seen
in \autoref{fig:counter-extended-stack}.
In this counterexample even after cropping the top high part of the
stack, a high stack element frame remains on the stack, which varies
in the return label, causing our indistinguishability relation to fail
when it shouldn't. Since the return $\pc$s are high, this difference
in labels is not observable, and therefore does not break
noninterference.

\counterexample{extended-stack}{An example of an execution trace that produces
  a non-monotonic stack}

In the second wrong alternative we tried to deal with this observation
by filtering out {\em all} high elements of the stack, leaving only
low stack elements to compare pairwise. This required us to change
Definition~\ref{defn:indistfull-regs} as follows:

\renewcommand{\labelof}[1]{\lab_{#1}}
\begin{defn}\label{defn:indistfull-regs-wrong}
  Machine states
  $S_1 = \RMACHI{\pc_1}{\ii{rf}_1}{\ii{cs}_1}{m_1}{i_1}$ and
  $S_2 = \RMACHI{\pc_2}{\ii{rf}_2}{\ii{cs}_2}{m_2}{i_2}$ are
  \emph{indistinguishable at level $\lab$ with respect to whole machine states},
  written $S_1 \indistfull^\lab S_2$,
  if $m_1 \indist^\lab m_2$, $i_1 \indist^\lab i_2$,
  $\ii{cs}_1 \indist^\lab \ii{cs}_2$,
  and additionally
  \begin{itemize}
  \item if $\labelof{\pc_1} \flowsto \lab$ or $\labelof{\pc_2} \flowsto \lab$ then
    $\pc_1 = \pc_2$ and $\ii{rf}_1  \indist^\lab \ii{rf}_2$.\ifever\asz{I don't like a
      bulleted list with just one bullet.}\ch{how else should we
      enforce proper parsing of complex logical formulas (in Coq I
      would use parentheses, but they don't have the same meaning in
      English)?}\fi
  \end{itemize}
\end{defn}
We then defined indistinguishability of stacks as follows.
\[
\begin{array}{c}
\ii{cs}_1 \indist^\lab \ii{cs}_2 =
\ii{filter}~(\ii{stack-frame-below}~\lab)~\ii{cs}_1 \indist^\lab
\ii{filter}~(\ii{stack-frame-below}~\lab)~\ii{cs}_2\\
\ii{stack-frame-below}~\lab~(\V{n}{\lab_\pc},\ii{rf},r,\lab_\ii{res}) =
  \lab_\pc \flowsto \lab
\end{array}
\]

\counterexample{extended-stack-2}{Counterexample for the filtering
stack-indistinguishability relation}

Unfortunately, this is now too weak and leads to an execution trace
that breaks noninterference (\autoref{fig:counter-extended-stack-2}).
After taking a step in the first machine, we get an distinguishable pair
of machines where one has an observable $\pc$ while the other one
does not.

\section{Theorems for Strengthening IFC Properties}
\label{app:strengthening}

We have proved in Coq%
\footnote{\url{https://github.com/QuickChick/IFC/blob/master/NotionsOfNI.v}}
that, under some reasonable assumptions, MSNI implies SSNI,
SSNI implies LLNI,
and LLNI implies EENI.
All these are generic properties of information-flow abstract machines;
a machine $M$ is composed of:
\begin{itemize}
\item an arbitrary type of machine states,
\item a partial step function on states written $\step$
  (reduction is deterministic), and
\item a set of observation levels $o$ (\EG labels).
\end{itemize}

As in \autoref{sec:basic}, we write $\manystep$ for the reflexive,
transitive closure of $\step$.
When $S \manystep S'$ and
\chrev{$S'$ is stuck ($\not\exists S''.~S' \step S''$)},
we write $S \stuckswith S'$.
We write $S \manystep_t$ when an execution ($\manystep$) from $S$
produces trace $t$ (a list of states).

\begin{defn}
A machine $M$ has {\em end-to-end noninterference (EENI)} with respect to
\begin{itemize}
  \item a predicate on states $\Start$ (initial states), and
  \item a predicate on states $\End$ (successful ending states),
  \item an observation-level-indexed
    indistinguishability relation on states $\indist$,
\end{itemize}
written EENI$_{\Start,\End,{\indist}}\,M$, when 
\begin{itemize}
  \item
for all states $S_1, S_2 \in \Init$, if
$S_1 \indist_o S_2$,
$S_1 \stuckswith S_1'$,
$S_2 \stuckswith S_2'$, and
$S_1', S_2' \in \End$ then
$S_1' \indist_o S_2'$.
\end{itemize}
\end{defn}

\begin{defn}
A machine $M$ has {\em low-lockstep noninterference (LLNI)} with respect to
\begin{itemize}
  \item a predicate on states $\Start$,
  \item an indistinguishability relation $\indist$, and
  \item an observation-level-indexed predicate on states $\Low$
   (\EG in which only data
    labeled below a certain label has influenced control flow),
\end{itemize}
written LLNI$_{\Start,\Low,{\indist}}\,M$, when for all $S_1,S_2 \in \Start$ 
\let\oldLow\Low%
\renewcommand{\Low}{\ensuremath{\oldLow_o}}%
\let\oldindist\indist%
\renewcommand{\indist}{\ensuremath{\oldindist_o}}%
\llnibody{}
\end{defn}

\begin{theorem} LLNI$_{\Start,\Low,{\indist}}\,M$ implies
 EENI$_{\Start,\End,{\indist}}\,M$ provided that
\begin{itemize}
\item $\indist$ is symmetric,
\item $\End{} \subset \Low{}$,
\item $S \in \End{}$ implies that $S$ is stuck ($\not\exists S'.~S \step S'$),
\item $S_1 \indist S_2$ implies $S_1 \in \End \Leftrightarrow S_2 \in \End$,
\end{itemize}
\end{theorem}

\begin{defn}
A machine $M$ has {\em single-step noninterference (SSNI)} with respect to
\begin{itemize}
  \item an indistinguishability relation $\indist$, and
  \item an observation-level-indexed predicate on states $\Low$,
\end{itemize}
written SSNI$_{\Low,{\indist}}\,M$, when
\let\oldLow\Low%
\renewcommand{\Low}{\ensuremath{\oldLow_o}}%
\let\oldindist\indist%
\renewcommand{\indist}{\ensuremath{\oldindist_o}}%
\renewcommand{\ssnilabel}{something} 
\ssnibody{$o$ and }
\end{defn}

\begin{theorem} SSNI$_{\Low,{\indist}}\,M$ implies LLNI$_{\Start,\Low,{\indist}}\,M$
under the following assumptions:
\begin{itemize}
\item $\indist$ is a partial equivalence relation (symmetric and transitive),
\item $S_1 \indist S_2$ implies $S_1 \in \Low \Leftrightarrow S_2 \in \Low$.
\end{itemize}
\end{theorem}

\ch{\strikeout{These theorems are weak; they require all common arguments to be
  equal (and that's a strong assumption), when usually it is enough
  that they are in some inclusion relationship. We need that at least
  for the indist relation in SSNI-implies-LLNI, but we could actually
  try to find the weakest assumptions everywhere.}}
\ch{Actually, it seems a better idea to prove weakening lemmas for
  each property. SSNI-implies-LLNI + a weakening of indistinguishability
  lemma for LLNI would give us what we want.}\ch{In fact we claim two
  other such weakening/strengthening lemmas at the end of 6.1 and 6.2}

\ch{Duplicate in the main text? (using macro)}%
\begin{defn}
A machine $M$ has {\em multi-step noninterference (MSNI)} with respect to
\begin{itemize}
  \item an indistinguishability relation $\indist$, and
  \item an observation-level-indexed predicate on states $\Low$,
\end{itemize}
written MSNI$_{\Low,{\indist}}\,M$, when 
\let\oldLow\Low%
\renewcommand{\Low}{\ensuremath{\oldLow_o}}%
\let\oldindist\indist%
\renewcommand{\indist}{\ensuremath{\oldindist_o}}%
  $S_1$ and $S_2$
  with $S_1 \indist S_2$, $S_1 \manystep_{t_1}$, $S_2 \manystep_{t_2}$
  we have $t_1 \indist^* t_2$,
  where indistinguishability on traces $\indist^*$
  is defined inductively by the following rules:
\typicallabel{High to High Step}
\infrule[Low Steps]
  {S_1, S_2 \in \Low \quad S_1' \indist S_2'
   \quad (S_1' : t_1) \indist^* (S_2' :t_2)}
  {(S_1 : S_1' : t_1) \indist^* (S_2 : S_2' : t_2)}
\infrule[High to Low Steps]
  {S_1, S_2 \not\in \Low \quad
   S_1', S_2' \in \Low \quad
   S_1' \indist S_2'
   \quad (S_1' : t_1) \indist^* (S_2' :t_2)}
  {(S_1 : S_1' : t_1) \indist^* (S_2 : S_2' : t_2)}
\infrule[High to High Step]
  {S_1, S_1' \not\in \Low \quad S_1 \indist S_1' \quad 
   (S_1' : t_1) \indist^* (S_2 : t_2)}
  {(S_1 : S_1' : t_1) \indist^* (S_2 : t_2)}
\infrule[Low Step End]
  {(S_1 \in \Low \lor S_1' \in \Low) \quad (S_1' : t_1) \indist^* (S_2 : [\,])}
  {(S_1 : S_1' : t_1) \indist^* (S_2 : [\,])}
\infrule[Both End]
  {}
  {(S_1 : [\,]) \indist^* (S_2 : [\,])}
\infrule[Symmetry]
  {t_1 \indist^* t_2}
  {t_2 \indist^* t_1}
\end{defn}

\begin{theorem} MSNI$_{\Low,{\indist}}\,M$ implies SSNI$_{\Low,{\indist}}\,M$
under the following assumptions:
\begin{itemize}
\item $\indist$ is reflexive and symmetric\ch{special name for this?},
\item $S_1 \indist S_2$ implies $S_1 \in \Low \Leftrightarrow S_2 \in \Low$.
\end{itemize}
\end{theorem}

\ch{Leo believes the other direction is also true; \IE MSNI and SSNI
  are equivalent. No time to try that.}

\ch{We were using a couple of bibtex hacks that interact badly
  with the jfp and abbrvnat styles. Will need to clean up.}

\let\section\OLDsection
\bibliographystyle{abbrvnaturl} 
\bibliography{quick-chick,local,safe}


\ifscr
\clearpage
\appendix

\section*{Scratchpad}

\section{Graph}
\bugtablegraph

\clearpage

\section{Comparing ByExec, ByExec2, and ByExec4}

\bugtable[h]{Cally-EENI-Mem-Initial-ByExec-True}{}
\bugtable[h]{Cally-EENI-Mem-Initial-ByExec2-True}{}
\bugtable[h]{Cally-EENI-Mem-Initial-ByExec4-True}{}
\clearpage

\bugtable[h]{Cally-EENI-Low-Initial-ByExec-True}{}
\bugtable[h]{Cally-EENI-Low-Initial-ByExec2-True}{\ch{Warning: 1sec!}}
\bugtable[h]{Cally-EENI-Low-Initial-ByExec4-True}{}
\clearpage

\bugtable[h]{Cally-EENI-Low-QuasiInitial-ByExec-True}{}
\bugtable[h]{Cally-EENI-Low-QuasiInitial-ByExec2-True}{\ch{Warning: 1sec!}}
\bugtable[h]{Cally-EENI-Low-QuasiInitial-ByExec4-True}{}
\clearpage

\bugtable[h]{Cally-LLNI-Low-QuasiInitial-ByExec-True}{}
\bugtable[h]{Cally-LLNI-Low-QuasiInitial-ByExec2-True}{}
\bugtable[h]{Cally-LLNI-Low-QuasiInitial-ByExec4-True}{}
\clearpage

\section{Profiling}
\proftable[h]{Initial-Naive-False}{}
\profvartable[h]{Initial-Naive-False}{}
\proftable[h]{Initial-Naive-True}{}
\profvartable[h]{Initial-Naive-True}{}
\proftable[h]{Initial-Weighted-False}{}
\profvartable[h]{Initial-Weighted-False}{}
\proftable[h]{Initial-Weighted-True}{}
\profvartable[h]{Initial-Weighted-True}{}
\proftable[h]{Initial-Sequence-False}{}
\profvartable[h]{Initial-Sequence-False}{}
\proftable[h]{Initial-Sequence-True}{}
\profvartable[h]{Initial-Sequence-True}{}
\proftable[h]{Initial-ByExec-False}{}
\profvartable[h]{Initial-ByExec-False}{}
\proftable[h]{Initial-ByExec-True}{}
\profvartable[h]{Initial-ByExec-True}{}
\proftable[h]{QuasiInitial-Naive-False}{}
\profvartable[h]{QuasiInitial-Naive-False}{}
\proftable[h]{QuasiInitial-Naive-True}{}
\profvartable[h]{QuasiInitial-Naive-True}{}
\proftable[h]{QuasiInitial-Weighted-False}{}
\profvartable[h]{QuasiInitial-Weighted-False}{}
\proftable[h]{QuasiInitial-Weighted-True}{}
\profvartable[h]{QuasiInitial-Weighted-True}{}
\proftable[h]{QuasiInitial-Sequence-False}{}
\profvartable[h]{QuasiInitial-Sequence-False}{}
\proftable[h]{QuasiInitial-Sequence-True}{}
\profvartable[h]{QuasiInitial-Sequence-True}{}
\proftable[h]{QuasiInitial-ByExec-False}{}
\profvartable[h]{QuasiInitial-ByExec-False}{}
\proftable[h]{QuasiInitial-ByExec-True}{}
\profvartable[h]{QuasiInitial-ByExec-True}{}

\clearpage
\section{Looking at individual bugs}

\subsection{Why is {\IfcBugStoreNoPointerTaint}
  so hard to find?}

\begingroup\tiny
\begin{verbatim}
[Push 0@\lab,Push 0@H,Push 1@\lab,Store,Push 0@H,Push 0@\lab,Store,{Push 0@H/Push 1@H},Store,Halt]
--- Common execution prefix:
0@\lab	M=[0@\lab,0@\lab]	S=[]	next=Push 0@\lab
1@\lab	M=[0@\lab,0@\lab]	S=[AData 0@\lab]	next=Push 0@H
2@\lab	M=[0@\lab,0@\lab]	S=[AData 0@H,AData 0@\lab]	next=Push 1@\lab
3@\lab	M=[0@\lab,0@\lab]	S=[AData 1@\lab,AData 0@H,AData 0@\lab]	next=Store
4@\lab	M=[0@\lab,0@H]	S=[AData 0@\lab]	next=Push 0@H
5@\lab	M=[0@\lab,0@H]	S=[AData 0@H,AData 0@\lab]	next=Push 0@\lab
6@\lab	M=[0@\lab,0@H]	S=[AData 0@\lab,AData 0@H,AData 0@\lab]	next=Store
7@\lab	M=[0@H,0@H]	S=[AData 0@\lab]	next={Push 0@H/Push 1@H}
8@\lab	M=[0@H,0@H]	S=[{AData 0@H/AData 1@H},AData 0@\lab]	next=Store
9@\lab	M=[{0@\lab/0@H},{0@H/0@\lab}]	S=[]	next=Halt
--- Machine 1 continues...
--- Machine 2 continues...
\end{verbatim}%
\endgroup


\begin{tabular}{mlmlmlml}
  \multicolumn{4}{mc}{i: \left[\begin{array}{l}\ii{Push}\;0@\high, \ii{Push}\;0@\high, \ii{Push}\;1@\low, \ii{Store}, \ii{Push}\;0@\low, \\ \ii{Store}, \ii{Push}\;0@\low, \ii{Push}\;\variation{0}{1}@\high, \ii{Store}, \ii{Halt},\end{array}\right]} \\
  \addlinespace\toprule
  \pc & m & s & i(\pc) \\
  \midrule
  0@\low & \left[0@\low,0@\low\right] & \left[\right] & \ii{Push}\;0@\high \\
  1@\low & \left[0@\low,0@\low\right] & \left[0@\high\right] & \ii{Push}\;0@\high \\
  2@\low & \left[0@\low,0@\low\right] & \left[0@\high,0@\high\right] & \ii{Push}\;1@\low \\
  3@\low & \left[0@\low,0@\low\right] & \left[1@\low,0@\high,0@\high\right] & \ii{Store} \\
  4@\low & \left[0@\low,0@\high\right] & \left[0@\high\right] & \ii{Push}\;0@\low \\
  5@\low & \left[0@\low,0@\high\right] & \left[0@\low,0@\high\right] & \ii{Store} \\
  6@\low & \left[0@\high,0@\high\right] & \left[\right] & \ii{Push}\;0@\low \\
  7@\low & \left[0@\high,0@\high\right] & \left[0@\low\right] & \ii{Push}\;\variation{0}{1}@\high \\
  8@\low & \left[0@\high,0@\high\right] & \left[\variation{0}{1}@\high,0@\low\right] & \ii{Store} \\
  9@\low & \left[0@\variation{\low}{\high},0@\variation{\high}{\low}\right] & \left[\right] & \ii{Halt} \\
  \bottomrule
\end{tabular}
\medskip

\begin{tabular}{mlmlmlml}
  \multicolumn{4}{mc}{i: \left[\begin{array}{l}\ii{Push}\;1@\low, \ii{Push}\;\variation{9}{7}@\high, \ii{Jump}, \ii{Push}\;6@\low, \ii{Jump}, \\ \ii{Halt}, \ii{Store}, \ii{Push}\;5@\low, \ii{Jump}, \ii{Push}\;0@\low, \\ \ii{Push}\;3@\low, \ii{Jump}\end{array}\right]} \\
  \addlinespace\toprule
  \pc & m & s & i(\pc) \\
  \midrule
  0@\low & \left[0@\low\right] & \left[\right] & \ii{Push}\;1@\low \\
  1@\low & \left[0@\low\right] & \left[1@\low\right] & \ii{Push}\;\variation{9}{7}@\high \\
  2@\low & \left[0@\low\right] & \left[\variation{9}{7}@\high,1@\low\right] & \ii{Jump} \\
  \midrule
  \multicolumn{4}{l}{Machine 1 continues\ldots} \\
  9@\high & \left[0@\low\right] & \left[1@\low\right] & \ii{Push}\;0@\low \\
  10@\high & \left[0@\low\right] & \left[0@\low,1@\low\right] & \ii{Push}\;3@\low \\
  11@\high & \left[0@\low\right] & \left[3@\low,0@\low,1@\low\right] & \ii{Jump} \\
  \midrule
  \multicolumn{4}{l}{Machine 2 continues\ldots} \\
  7@\high & \left[0@\low\right] & \left[1@\low\right] & \ii{Push}\;5@\low \\
  8@\high & \left[0@\low\right] & \left[5@\low,1@\low\right] & \ii{Jump} \\
  5@\low & \left[0@\low\right] & \left[1@\low\right] & \ii{Halt} \\
  \bottomrule
\end{tabular}
\medskip

\begin{tabular}{mlmlmlml}
  \multicolumn{4}{mc}{i: \left[\begin{array}{l}\ii{Push}\;\variation{0}{1}@\high, \ii{Push}\;0@\low, \ii{Store}, \ii{Halt}\end{array}\right]} \\
  \addlinespace\toprule
  \pc & m & s & i(\pc) \\
  \midrule
  \counterexampleline 0@\low & \left[0@\low\right] & \left[\right] & \ii{Push}\;\variation{0}{1}@\high \\
  \counterexampleline 1@\low & \left[0@\low\right] & \left[\variation{0}{1}@\low\right] & \ii{Push}\;0@\low \\
  \counterexampleline 2@\low & \left[0@\low\right] & \left[0@\low,\variation{0}{1}@\low\right] & \ii{Store} \\
  \counterexampleline 3@\low & \left[\variation{0}{1}@\low\right] & \left[\right] & \ii{Halt} \\
  \bottomrule
\end{tabular}
\medskip

\begingroup
\[
  i:\quad
  \begin{array}{l}
  \ii{Push}\;0@\low,\ii{Push}\;0@\high,\ii{Push}\;1@\low,\ii{Store},\ii{Push}\;0@\high, \\
  \ii{Push}\;0@\low,\ii{Store},\ii{Push}\;\variation{0@\high}{1@\high},\ii{Store},\ii{Halt}
  \end{array}
\]
\begin{tabular}{llll}
\toprule
$\pc$ & $m$ & $s$ & $i(pc)$ \\
\midrule
$0@\low$ & $[0@\low,0@\low]$                          & $[]$                        & $\ii{Push}\;0@\low$ \\
\multicolumn{4}{c}{$\vdots$} \\
$7@\low$ & $[0@\high,0@\high]$                        & $[0@\low]$                  & $\ii{Push}\;\variation{0@\high}{1@\high}$ \\
$8@\low$ & $[0@\high,0@\high]$                        & $\left[\variation{0@\high}{1@\high},
                                                          0@\low\right]$                  & $\ii{Store}$ \\
$9@\low$ & $\left[\variation{0@\low}{0@\high}
                 ,\variation{0@\high}{0@\low}\right]$ & $[]$                        & $\ii{Halt}$ \\
\bottomrule
\end{tabular}
\endgroup

In \IfcBugStoreNoPointerTaint{} we do the no-sensitive-upgrade check,
but we don't taint the value we store with the label on the pointer.

This means that for such a store through a high pointer to succeed 2
locations in memory have to be labeled high, which is very hard to
obtain when starting from an initially low memory.
{\tiny
\begin{verbatim}
[Push 0@\lab,Push 0@H,Push 7@\lab,Store,Push 0@H,Push 8@\lab,Store,{Push 8@H/Push 7@H},Store]
--- Common execution prefix:
2666@\lab  M=[0@\lab,0@\lab]     S=[]    next=Push 0@\lab
2667@\lab  M=[0@\lab,0@\lab]     S=[AData 0@\lab]   next=Push 0@H
2668@\lab  M=[0@\lab,0@\lab]     S=[AData 0@H,AData 0@\lab] next=Push 7@\lab
2669@\lab  M=[0@\lab,0@\lab]     S=[AData 7@\lab,AData 0@H,AData 0@\lab]       next=Store
2670@\lab  M=[0@H,0@\lab]     S=[AData 0@\lab]   next=Push 0@H
2671@\lab  M=[0@H,0@\lab]     S=[AData 0@H,AData 0@\lab] next=Push 8@\lab
2672@\lab  M=[0@H,0@\lab]     S=[AData 8@\lab,AData 0@H,AData 0@\lab]       next=Store
2673@\lab  M=[0@H,0@H]     S=[AData 0@\lab]   next={Push 8@H/Push 7@H}
2674@\lab  M=[0@H,0@H]     S=[{AData 8@H/AData 7@H},AData 0@\lab]     next=Store
2675@\lab  M=[{0@H/0@\lab},{0@\lab/0@H}] S=[]    next=<eof>
--- Machine 1 continues...
--- Machine 2 continues...
*** Falsifiable, numTests = 4351, numShrinks = 52
\extrabuginfo{InstrsBasic}{\PropLLNI}{EquivLow}{StartInitial}
{\GenByExec}{True}{1sec}{110}{Wed Feb  6 17:14:40 EST 2013}
\end{verbatim}
}

Counterexamples for StateQuasiInitial can directly start with a high
memory, so the shortest ones look like this:
{\tiny
\begin{verbatim}
[{Push 7@H/Push 8@H},Store]
--- Common execution prefix:
2666@\lab  M=[0@H,0@H]     S=[AData 0@\lab]   next={Push 7@H/Push 8@H}
2667@\lab  M=[0@H,0@H]     S=[{AData 7@H/AData 8@H},AData 0@\lab]     next=Store
2668@\lab  M=[{0@\lab/0@H},{0@H/0@\lab}] S=[]    next=<eof>
--- Machine 1 continues...
--- Machine 2 continues...
*** Falsifiable, numTests = 52, numShrinks = 167
\extrabuginfo{InstrsBasic}{\PropLLNI}{EquivLow}{StartQuasiInitial}
{\GenByExec}{True}{1sec}{110}{Wed Feb  6 17:08:02 EST 2013}
\end{verbatim}
}

\subsection{Why is {\IfcBugValueOrVoidOnReturn}
  hard to find?}

It's because the counterexample is really involved?

The best LLNI strategy finds this, and yes, it seems indeed complicated
to find.
{\tiny
\begin{verbatim}
[{Push 2669@H/Push 2668@H},Call 1 False,Return False,Return True]
--- Common execution prefix:
2666@\lab  M=[]    S=[AData 0@\lab]   next={Push 2669@H/Push 2668@H}
2667@\lab  M=[]    S=[{AData 2669@H/AData 2668@H},AData 0@\lab]       next=Call 1 False
--- Machine 1 continues...
2669@H  M=[]    S=[AData 0@\lab,ARet (2668,False)@\lab]       next=Return True
2668@\lab  M=[]    S=[AData 0@H]   next=Return False
--- Machine 2 continues...
2668@H  M=[]    S=[AData 0@\lab,ARet (2668,False)@\lab]       next=Return False
2668@\lab  M=[]    S=[]    next=Return False
*** Falsifiable, numTests = 580, numShrinks = 168
\extrabuginfo{InstrsCally}{\PropLLNI}{EquivLow}{StartQuasiInitial}
{\GenWeighted}{True}{1sec}{110}{Thu Feb  7 10:28:11 EST 2013}
\end{verbatim}
}

The best SSNI strategy finds this:
{\tiny
\begin{verbatim}
*** Failed! Falsifiable (after 202 tests and 23 shrinks): 
Shrink2 {AS { amem = [], aimem = [ Return True, Return False ]
   , astk = [ AData 0@\lab, ARet (0,False)@\lab ], apc = 2666@H }
/AS { amem = [], aimem = [ Return True, Return False ]
   , astk = [ ARet (0,False)@\lab ], apc = 2667@H }}
[...]
*** Falsifiable, numTests = 202, numShrinks = 23
\extrabuginfo{InstrsCally}{\PropSSNI}{EquivFull}{StartArbitrary}
{\GenTinySSNI}{True}{1sec}{110}{Thu Feb  7 10:42:26 EST 2013}
\end{verbatim}
}
It originally seemed to me that this shouldn't be so hard to find, but
it is.  If each return has 1/20 chance, then the pair has 1/400
chance.  And we additionally need a high pc, and two low return
addresses on the stack. So it's no surprise that SSNI was finding this
is only once in 1700 tests. Improved this (to 1 in 300) by making the
second generated instruction be often a variation of the first.

\subsection{Why is {\IfcBugReturnNoTaint} hard to find?}

For SSNI the pc needs to be H, and both machines have to run returns.

{\tiny
\begin{verbatim}
*** Failed! Falsifiable (after 319 tests and 13 shrinks):     
Shrink2 {AS { amem = [], aimem = [ Return False, Return False ]
   , astk = [ AData 0@\lab, ARet (0,True)@\lab ], apc = 2666@H }
/AS { amem = [], aimem = [ Return False, Return False ]
   , astk = [ AData 1@\lab, ARet (0,True)@\lab ], apc = 2667@H }}
[...]
*** Falsifiable, numTests = 319, numShrinks = 13
\extrabuginfo{InstrsCally}{\PropSSNI}{EquivFull}{StartArbitrary}
{\GenTinySSNI}{True}{60sec}{700}{Thu Feb  7 12:30:06 EST 2013}
\end{verbatim}
}

{\tiny
\begin{verbatim}
Shrink2 {AS { amem = [], aimem = [ Return True ]
   , astk = [ AData 0@\lab, ARet (0,True)@\lab ], apc = 2666@H }
/AS { amem = [], aimem = [ Return True ]
   , astk = [ AData 1@\lab, ARet (0,True)@\lab ], apc = 2666@H }}
[...]
*** Falsifiable, numTests = 522, numShrinks = 19
\extrabuginfo{InstrsCally}{\PropSSNI}{EquivFull}{StartArbitrary}
{\GenTinySSNI}{True}{60sec}{700}{Thu Feb  7 12:32:43 EST 2013}
\end{verbatim}
}

\clearpage

\textbf{DUMP OF ALL TABLES FROM GenerateMakeFiles }

\bugtable[h]{Basic-EENI-Mem-Initial-Weighted-False}{}
\bugtable[h]{Basic-EENI-Mem-Initial-Sequence-False}{}
\bugtable[h]{Basic-EENI-Mem-Initial-Sequence-True}{}
\bugtable[h]{Basic-EENI-Mem-Initial-Naive-False}{}
\bugtable[h]{Basic-EENI-Mem-Initial-ByExec-True}{}
\clearpage
\bugtable[h]{Cally-EENI-Mem-Initial-ByExec2-True}{}
\bugtable[h]{Cally-EENI-Low-Initial-ByExec2-True}{}
\bugtable[h]{Cally-EENI-Low-QuasiInitial-ByExec2-True}{}
\clearpage
\bugtable[h]{Cally-LLNI-Low-QuasiInitial-ByExec2-True}{}
\clearpage
\bugtable[h]{Cally-SSNI-Full-Arbitrary-Naive-True}{}
\bugtable[h]{Cally-SSNI-Full-Arbitrary-TinySSNI-True}{}
\clearpage

\fi

\end{document}